\documentclass[12pt,a4paper]{article}
\usepackage{graphicx,epsfig}
\usepackage{color}
\usepackage[colorlinks,citecolor=red,linkcolor=blue]{hyperref}
\usepackage{epsfig,float,amsmath,amsfonts,amssymb,graphicx,bm,bbm,array,dcolumn,cite}
\usepackage[small]{caption}
\usepackage[caption=false]{subfig}

\usepackage{geometry}
\def\beq{\begin{equation}}
\def\eeq{\end{equation}}
\def\bea{\begin{eqnarray}}
\def\eea{\end{eqnarray}}
\def\bmat{\begin{pmatrix}}
\def\emat{\end{pmatrix}}
\newcommand{ \slashchar }[1]{\setbox0=\hbox{$#1$}   % set a box for #1
   \dimen0=\wd0                                     % and get its size
   \setbox1=\hbox{/} \dimen1=\wd1                   % get size of /
   \ifdim\dimen0>\dimen1                            % #1 is bigger
      \rlap{\hbox to \dimen0{\hfil/\hfil}}          % so center / in box
      #1                                            % and print #1
   \else                                            % / is bigger
      \rlap{\hbox to \dimen1{\hfil$#1$\hfil}}       % so center #1
      /                                             % and print / 
   \fi}                                             %
\def\to{\rightarrow}

\begin{document}

\title{LHC Signatures of Warped-space Vectorlike Quarks}

\author{Shrihari~Gopalakrishna$^{a}$\thanks{shri@imsc.res.in}~, Tanumoy~Mandal$^a$\thanks{tanumoy@imsc.res.in}~, Subhadip~Mitra$^b$\thanks{subhadip.mitra@th.u-psud.fr}~, \\  Gr\'{e}gory~Moreau$^b$\thanks{gregory.moreau@th.u-psud.fr} \\
$^a$~\small{Institute of Mathematical Sciences (IMSc),} \\ 
\small{C.I.T Campus, Taramani, Chennai 600113, India.} \\
$^b$~\small{Laboratoire de Physique Th\'{e}orique,} \\ 
\small{CNRS-UMR 8627, Universit\'{e} Paris-Sud 11, F-91405 Orsay Cedex, France.}
}

\maketitle

\begin{abstract}
We study the LHC signatures of TeV scale vectorlike quarks  
$b'$, $t'$ and $\chi$ with electromagnetic charges $-1/3$, $2/3$ and $5/3$ that appear in many 
beyond the standard model (BSM) extensions.  
We consider warped extra-dimensional models and analyze the phenomenology of such vectorlike quarks that are the
custodial partners of third generation quarks. 
In addition to the usually studied pair-production channels which depend on the strong coupling, we put equal emphasis on single production channels 
that depend on electroweak couplings and on electroweak symmetry breaking induced mixing effects between the heavy 
vectorlike quarks and standard model quarks. 
We identify new promising $gg$-initiated pair and single production channels and 
find the luminosity required for discovering these states at the LHC. 
For these channels, we propose a cut that allows one to extract the relevant electroweak couplings. 
Although the motivation is from warped models, we present many of our results model-independently.  
\end{abstract}

%%%%%%%%%%%%%%%%%%%%%%%%%%%%%%%%%%%%%%%%%%%%%
\section{Introduction}
The standard model (SM) of particle physics suffers from the gauge hierarchy and flavor hierarchy problems and
many beyond the standard model (BSM) extensions have been proposed to solve these problems.
The extra particles in these BSM extensions are being searched for at the CERN Large Hadron Collider (LHC). 
Some BSM extensions contain vectorlike colored fermions, for example, 
having electromagnetic (EM) charges $5/3$, $2/3$, and $-1/3$, which 
we denote as $\chi$, $t'$ and $b'$ respectively. 
For instance, warped-space extra-dimensional models with bulk fermions contain these
vectorlike fermions. 

In this work we consider the LHC signatures of the $\chi$, $t'$ and $b'$. 
We present a few warped-space extra-dimensional models that contain these states, specify realistic parameter values, 
and extract the couplings of these vectorlike fermion states with SM states. 
We identify promising pair and single production channels at the LHC and find the luminosity required for discovering 
these states at the LHC. 
We emphasize that the signatures we identify and the search strategies 
are common to many other BSM theories that contain such vectorlike quarks.   
A particular emphasis is the single-production of the heavy colored fermions in addition to their pair-production, 
since single-production couplings depend 
more directly on their electroweak quantum numbers, while pair-production is dominated by its 
coupling to the gluon which is given by the $SU(3)_C$ gauge coupling $g_s$, and thus hides its electroweak 
aspects.
Although measuring the branching ratios using pair-production channels gives information on the electroweak couplings, 
it is only the ratios of couplings that is determined and not their actual values. 
But single-production can fix the actual values of the couplings.
Moreover, compared to pair-production, single production also have less complications from combinatorics. Depending on the coupling, some single production channel can
even be the dominant production channel for heavy vectorlike quarks due to the phase-space suppression in pair-production.

In Ref.~\cite{Gopalakrishna:2011ef}, we analyzed the LHC signatures of a vectorlike $b'$  
in a model-independent fashion. We highlighted there many general aspects of vectorlike fermions and 
contrasted them with chiral (4th generation) fermions in how they decay and their resulting signatures at the LHC. 
In this work we extend this to include the $\chi$ and $t'$ also.  
%%%%%%%%%%%%%%%%%%%%%%%%%%%%%%%%%%%%%%%%%%%%%%%%%%%%%%%%%%%%%%%%%%%%

Other than in extra dimensional theories,
vectorlike quarks appear in many new-physics models such as   
composite Higgs models~\cite{Contino:2006qr,Anastasiou:2009rv,Vignaroli:2012sf,DeSimone:2012fs}, 
little Higgs models~\cite{Han:2003wu,Carena:2006jx,Matsumoto:2008fq,Berger:2012ec}, 
some supersymmetric extensions~\cite{Kang:2007ib,Graham:2009gy,Martin:2010dc}, quark-lepton
unification models~\cite{Li:2012zy} etc. 
Extensive studies on vectorlike fermions are available in the literature. 
Here we briefly survey some references that are relevant to our study.
Vectorlike fermions in the context of Higgs boson production have been considered in 
Refs.~\cite{delAguila:1989rq,delAguila:1989ba,Djouadi:2007fm,Azatov:2012rj,Bouchart:2009vq}.
Based on the recent discovery of a Higgs boson at the LHC~\cite{Aad:2012tfa,Chatrchyan:2012ufa}, 
Refs.~\cite{Moreau:2012da,Bonne:2012im} constrain vectorlike fermion masses and couplings from the recent data. 
It has been pointed out~\cite{Djouadi:2011aj,Djouadi:2009nb,Djouadi:2006rk,Bouchart:2008vp} that vectorlike fermions can address the forward-backward
asymmetry in top quark pair production at the Tevatron. 
Refs.~\cite{delAguila:1998tp,delAguila:2000rc,AguilarSaavedra:2002kr,Cacciapaglia:2010vn,Cacciapaglia:2011fx,Okada:2012gy,Aguilar-Saavedra:2013qpa} 
analyze vectorlike fermion representations and mixing of the new fermions with the SM quarks and the relevant experimental bounds. 
Refs.~\cite{Contino:2008hi,Mrazek:2009yu,Dennis:2007tv,Cacciapaglia:2012dd,AguilarSaavedra:2005pv,AguilarSaavedra:2006gw,Atre:2013ap,Harigaya:2012ir,AguilarSaavedra:2009es} 
study the LHC signatures of $b^{\prime}$, $t^{\prime}$ and $\chi$ vectorlike quarks. 
Ref.~\cite{Dennis:2007tv} studies the LHC signatures of vectorlike $b^{\prime}$ and $\chi$ in the 4-$W$ channel.
Ref.~\cite{Harigaya:2012ir} studies multi-$b$ signals for $t^\prime$ quarks. 
The LHC signatures of vectorlike $t'$ and $b'$ decaying to a Higgs boson are discussed in Ref.~\cite{Atre:2013ap}. 
Ref.~\cite{AguilarSaavedra:2009es} studies pair-production of the vectorlike quarks followed by decays into single and multi-lepton channels 
and the pair-production of the Kaluza-Klein (KK) top is explored in Ref.~\cite{Carena:2007tn}.
Ref.~\cite{Barcelo:2011wu} studies the signatures of vectorlike quarks resulting from the decay of a KK gluon.    
Ref.~\cite{Bini:2011zb} analyzes the single production of $t'$ and $b'$ via KK gluon and finds that these channels could be 
competitive with the direct electroweak single production channels of these heavy quarks.
Model independent LHC searches of vectorlike fermions have been discussed in Refs.~\cite{Atre:2008iu,Han:2010rf,Atre:2011ae,Buchkremer:2013bha}.
Many important pair and single production channels for probing a vectorlike $b'$ at the LHC in the context of a warped extra-dimension were explored in 
Ref.~\cite{Brooijmans:2010tn}. 
Mixing of the SM $b$-quark with a heavy vectorlike $b'$ and partial decay widths were worked out in Ref.~\cite{Ghosh:2011zz}. 
In Ref.~\cite{Alves:2013dga}, the LHC phenomenology of new heavy chiral quarks with electric charges $-4/3$ and $5/3$ are discussed.
Exploiting same-sign dileptons signal to beat the SM background, Refs.~\cite{Contino:2008hi,Mrazek:2009yu} show that the pair-production at the 14~TeV LHC 
can discover charge $-1/3$ and $5/3$ vectorlike quarks with a mass up to 1~TeV (1.5~TeV) with about 10~fb$^{-1}$ (200~fb$^{-1}$) integrated luminosity.
Ref.~\cite{Cacciapaglia:2012dd} considers pair production of charge $5/3$ vectorlike 
quarks  and shows that with the search for same sign dilepton the discovery reach of the 7 TeV LHC is about 700~GeV with 5~fb$^{-1}$ integrated luminosity. 
The LHC signatures of $t^{\prime}$ vectorlike quarks have been discussed in~\cite{AguilarSaavedra:2005pv} 
using $pp\to t^{\prime}\bar{t}^{\prime}\to bW^+\bar{b}W^-$ channel with the semileptonic decay of the $W$'s and the reach is found to be about 1~TeV 
with 100~fb$^{-1}$ integrated luminosity at the 14~TeV LHC.  
With 14.3~fb$^{-1}$ of integrated luminosity at the 8~TeV LHC, ATLAS has excluded a weak-isospin singlet $b'$ quark with mass below 645~GeV, 
while for the doublet representation the limit is 725~GeV~\cite{ATLAS:2013TB}.
With 4.64~fb$^{-1}$ luminosity, using single production channels with charged and neutral current interactions, 
vectorlike $b'$, $t'$ and $\chi$ quarks up to masses about 1.1~TeV, 1~TeV and 1.4~TeV respectively have been excluded~\cite{ATLAS:2012apa}, 
for couplings taken to be $v/M$, where $v$ is the Higgs vacuum expectation value (VEV), and $M$ the mass of the vectorlike quark.  
In Ref.~\cite{CMS:2012cya} the CMS collaboration presents the results for the search of a charge 5/3 quark at the 7 TeV LHC. 
With 5~fb$^{-1}$ luminosity and assuming 100\% branching ratio (BR) for the $\chi\rightarrow tW$ channel a $\chi$ quark with mass below 645 GeV is excluded. 
With the 8 TeV LHC, the CMS collaboration has improved their limit on the $\chi$ quark to 770 GeV~\cite{CMS:vwa}. 
In Ref.~\cite{ATLAS8TeV} the ATLAS collaboration shows the exclusion limits for a $t'$ quark in the  
BR($t'\rightarrow Wb$) versus BR($t'\rightarrow t h$) plane.

In this work, we detail some warped models with different $SU(2)_L \otimes SU(2)_R$ fermion representations 
that have been proposed earlier in the literature. 
For each of these we carefully work out the couplings induced by electroweak symmetry breaking (EWSB)
relevant for single production of the vectorlike quarks after diagonalizing the mass matrices including the EWSB contributions. 
We show what sizes of the relevant couplings are realistic by varying the parameters of the theory. 
For these warped models with the couplings above, and for vectorlike quark masses of about a TeV,  
the direct single production channels that most of the studies above focus on have 
too small cross-sections and therefore extraction of the electroweak couplings from these are difficult. 
Typically these quark initiated processes have small rates.  
In this work, we identify channels which are $gg$ initiated but yet sensitive to electroweak couplings after our cuts.
For vectorlike quark masses of about a TeV, the channels are signal rate limited and the backgrounds under control after cuts.
We show that these channels can be observed above background. 
These are our main contributions.   

This paper is organized as follows: 
In Sec.~\ref{ThFrmWrk.SEC} we give the details of the warped models both with and without custodial 
protection of the $Zb\bar b$ coupling, show the mass mixing terms and their diagonalization, and
work out the couplings in the mass basis relevant to the phenomenology we consider. 
In \ref{fermProf.APP} we give the fermion profiles that we use to compute the couplings, 
and the dependence of the mass eigenvalue on the $c$-parameter that parametrizes the fermion bulk masses in units of the curvature scale of the extra-dimension. 
In \ref{tR3AnaDiag.APP} we give some analytical results of the diagonalization and the resulting couplings 
in the small mixing limit for the model with custodial protection of $Zb\bar b$. 
In Sec.~\ref{parCoup.SEC} we give details of the parameter choices we make in the warped models and  
show the vectorlike fermion couplings and their dependence on the $c$-parameter.
Readers not wanting to know all the details of the warped models can go directly to the next section,
although the above sections will guide which channels we consider in later sections. 
In Sec.~\ref{DecBR.SEC} we give the decay partial widths and the branching ratios into the various decay modes. 
In Sec.~\ref{LHCSign.SEC} we discuss some promising discovery channels for the vectorlike quarks and 
present the reach for the 8 and 14~TeV LHC.
We offer our conclusions in Sec.~\ref{CONCL.SEC}.

%%%%%%%%%%%%%%%%%%%%%%%%%%%%%%%%%%%%%%%%%%%%%%%%%%%%%%%%
\section{Warped models}
\label{ThFrmWrk.SEC}

The Randall-Sundrum model~\cite{rs1} is a theory defined on a slice of AdS space which 
solves the gauge hierarchy problem.
Due to the AdS/CFT duality conjecture~\cite{Maldacena:1997re}, this construction 
may be dual to a spontaneously broken conformal four dimensional strongly coupled theory. 
By letting SM fields propagate in the bulk, the fermion mass hierarchy of the SM can 
also be addressed~\cite{Grossman:1999ra,Gherghetta:2000qt} without badly violating the flavor-changing neutral current (FCNC) constraints. 
The bulk mass parameters $c_\psi$ are chosen so that the SM fermion masses match the measured values. 

Precision electroweak constraints place strong bounds on such extensions of the SM. 
Gauging $SU(2)_R$ in the bulk offers a custodial symmetry that protects~\cite{Agashe:2003zs} the $T$-parameter from receiving large tree-level shifts,
but can still lead to problems due to an excessive shift to the $Zb\bar{b}$ coupling. This can also be protected~\cite{Agashe:2006at} by  
taking the third generation $Q_L$ as a bi-doublet under $SU(2)_L \times SU(2)_R$, {\it i.e.} $Q_L=(2,2)$.

An equivalent $4D$ theory can be written down by performing a KK expansion.
For LHC phenomenology, it is sufficient to keep only the zero-mode and the 1st KK excitation with mass $M_{KK}$.  
EWSB makes some zero-modes massive like in the SM, and mixes various KK modes, and after diagonalization the 
light eigenmodes are identified with the SM states. 
In this work, we ignore mixings between zero-mode and 1st KK modes in the gauge sector as this mixing is of order 
$\sqrt{k\pi R}\, v^2/M^2_{KK}$ and will be a few percent effect.  
We keep the $(0)-(1)$ mixing in the fermion sector to fermions with Dirichlet-Neumann $(-,+)$ 
boundary conditions (BC) as these can be bigger owing to the smaller mass of the 
$(-,+)$ custodians. 

The $SU(2)_R$ symmetry implies extra exotic 5D fermions not present in the SM, and
the light zero-modes of these which are not observed in Nature are ``projected-out'' by imposing $(-,+)$ BC on the bulk fields.
The first KK excitation of such $(-,+)$ fermions, {\it i.e.} the custodial partners, especially of third generation quarks 
can be significantly lighter~\cite{Choi:2002ps,Agashe:2003zs,Agashe:2004bm,Agashe:2004cp} than the gauge KK excitations, 
leading to measurable signals at the LHC.

In warped space extra-dimensional theories, 
in order to relax electroweak constraints, $SU(2)_R$ is gauged
in the bulk~\cite{Agashe:2003zs} to provide a custodial symmetry in the gauge-Higgs sector that protects the $T$ parameter.
We therefore take the bulk gauge group as ${\rm SU(2)_L\! \times\! SU(2)_R\! \times\! U(1)_X}$. 
We start with the simplest realization of this in Sec.~\ref{ThNoZbb.SEC} although the constraint coming from the shift 
of the $Z b \bar b$ coupling is quite strong. 
In order to avoid this constraint, the $SU(2)_R$ can also be used to protect this coupling~\cite{Agashe:2006at}, and we present this model 
in Sec.~\ref{ThZbb.SEC}. 
The most important aspect, as already pointed out, is that the new heavy fermions (the first KK fermion modes in particular)
are vectorlike with respect to the gauge group. 
In this work we focus on the LHC signatures of three such custodial vectorlike quarks, namely the 
$\chi$, $t'$ and $b'$.  
This complements other studies of warped KK states at the LHC, for example, 
KK graviton in Ref.~\cite{Davoudiasl:1999jd} and KK Gauge bosons in Refs.~\cite{KKgluon,Agashe:2007ki,Djouadi:2007eg,Agashe:2008jb,Ledroit:2007ik}. 

Following usual practice, we denote the field representations as $(l,r)_X$ where $l$, $r$ denote the $SU(2)_L$ and $SU(2)_R$ representations respectively,
and $X$ denotes the $U(1)_X$ charge.
In all the Lagrangian terms in the following, we will not show terms that are the same as in the SM, but will only show 
the terms either new to this BSM theory, or SM couplings that are shifted.

%%%%%%%%%%%%%%%%%%%%%%%%%%%%%%%
\subsection{Model without $Zb\bar b$ protection (DT model)}
\label{ThNoZbb.SEC}
We start with the quark representations  
$$ Q_{L} \equiv ({\bf 2},{\bf 1})_{1/6} =(t_L,b_L), \ \ \ \ $$
$$
\Psi_{b_R} \equiv ({\bf 1},{\bf 2})_{1/6} =(t',b_R), \ \ \ \
\Psi_{t_R} \equiv ({\bf 1},{\bf 2})_{1/6} =(t_R,b') \ .
$$
The representation for the Higgs field, responsible for the EWSB, is
$$
\Sigma \equiv ({\bf 2},{\bf 2})_{0}.
$$ 
We refer to this model as the doublet-top (DT) model. 
The extra fields $t'$ and $b'$ (the ``custodians'') are ensured to be without zero-modes by applying Dirichlet-Neumann $(-,+)$ 
BC on the extra dimensional interval $[0,\pi R]$, 
and their KK excitations are vectorlike with respect to the SM gauge group, 
while the SM particles are the zero-modes of fields with Neumann-Neumann $(+,+)$ BC, and are chiral.
As mentioned above, the $(-,+)$ fields are most likely the lowest mass KK excitation, and, among them
the $b'$ couplings to SM states are larger due to a larger mixing angle. This is because the mixing angle is 
inversely proportional to $M_{b'}$ which is smaller due to the $c_{t_R}$ choice required for the correct top-quark mass. 
Therefore, the $b'$ promises to have the best observability at the LHC, and we will only study its phenomenology 
and will not comment further on the $t'$ for this model. 
Elsewhere in the literature, sometimes the $L,R$ subscripts on fermion fields denote the gauge-group, but in our notation here, 
the $b'_{L,R}$ will mean the two Lorentz chiralities of the vectorlike $b'$. 

Electroweak symmetry is broken by $\left< \Sigma \right> = {\rm diag}(v, v)/\sqrt{2}$
(the Higgs boson VEV is $v \approx 246$ GeV). 
The Goldstone bosons of electroweak symmetry breaking ($\phi$) are contained in 
$\Sigma = (v/\sqrt{2}) e^{2i\phi^a T^a/v}$, written in the nonlinear realization,
where $T^a$ are the generators of $SU(2)_L$. 
We work here in the unitary gauge
for which we absorb the Goldstone bosons as the longitudinal polarization of the
gauge bosons. 
Nevertheless, for completeness and to have a clear understanding of the 
couplings involved, a derivation of the couplings using Goldstone boson equivalence is
presented in Appendix~A of Ref.~\cite{Brooijmans:2010tn}.
The theory as written above has also been presented before in Ref.~\cite{Brooijmans:2010tn}. 

The Yukawa couplings are given by 
\beq
{\cal L}_{5D}  \supset - \lambda_t \bar{Q}_L \Sigma \Psi_{t_R} 
                 - \lambda_b \bar{Q}_L \Sigma \Psi_{b_R} \ ,
\label{yuk.EQ}
\eeq
where $\lambda_{t,b}$ are the 5D Yukawa coupling constants. 
We write down an equivalent $4D$ theory by a Kaluza-Klein expansion. 
After EWSB, the zero-mode $b$ mixes with the $b'$ due to
off-diagonal terms in the following mass matrix: 
\beq
{\cal L}_{4D} \supset - \bmat \bar{b_L} & \bar{{b'}_L} \emat 
\bmat \lambda_{Q_L b_R} v/\sqrt{2} & \lambda_{Q_L b'_R} v/\sqrt{2} \\
      \lambda_{b_R b'_L} v/\sqrt{2}&         M_{b'}             \emat 
\bmat b_R \\ {b'}_R \emat + {\rm h.c.} \ ,
\label{LintM.EQ}
\eeq
where $\lambda_{Q_L b_R}$ is the zero-mode b-quark Yukawa coupling, the $M_{b'}$ is the vectorlike mass, 
and $\lambda_{ij} v/\sqrt{2}$ terms are induced after EWSB. 
In this work we set $\lambda_{b_R b'_L}$ to zero since this will always be the case we are interested in. 

The above mass matrix written in the $(b, b')$ basis is diagonalized by bi-orthogonal rotations, 
and we denote the sine (cosine) of the mixing angles by $s_\theta^{L,R}$ ($c_\theta^{L,R}$).
We denote the corresponding mass eigenstates as $(b_1, b_2)$. 
We define the off-diagonal mass $\tilde{m} \equiv \lambda_{Q_L b'_R} v/\sqrt{2}$ for notational ease. 
The mixing angles are 
\beq
\tan{(2\theta_L)} = -\frac{2\tilde{x}}{(1-\tilde{x}^2-x_b^2)} \ ; \quad 
\tan{(2\theta_R)} = -\frac{2 x_b \tilde{x}}{(1+\tilde{x}^2-x_b^2)} \ ,
\eeq
where $x_b \equiv (\lambda_{Q_L b_R} v/\sqrt{2})/M_{b'}$ and $\tilde{x} \equiv \tilde{m}/M_{b'}$. 
The mass eigenvalues to leading order in $x_b$ are: \\*
$(\lambda_{Q_L b_R} v/\sqrt{2})/\sqrt{1+\tilde{x}^2}$ and 
$M_{b'}\sqrt{(1+\tilde{x}^2)(1+x_b^2 \tilde{x}^2/(1+\tilde{x}^2)^2)}$. 
Although we do not show the mass matrix for the top sector, 
analogously, the top mass is given by $m_t \approx \lambda_{Q_L t_R} v/\sqrt{2}$.

The $b'$ mixes with the zero-mode $b$ due to off-diagonal terms in the mass matrix induced by EWSB as shown in Eq.~(\ref{LintM.EQ}).  
Diagonalizing this, we go from the $(b,b')$ basis to the $(b_1,b_2)$ mass-basis
and write an effective Lagrangian relevant for this model in the mass-basis as~\cite{Gopalakrishna:2011ef}
\bea
{\cal L}_{4D} \supset 
&-&\frac{e}{3} \bar{b_1} \gamma^\mu b_1 A_\mu 
-\frac{e}{3} \bar{b}_{2} \gamma^\mu b_{2} A_\mu
+ g_s \bar{b_1} \gamma^\mu T^\alpha b_1 g^\alpha_\mu  
+ g_s \bar{b}_{2}  \gamma^\mu T^\alpha b_{2} g^\alpha_\mu \nonumber \\
&-& \left( \kappa^L_{b t W} \bar{t_L} \gamma^\mu {b_1}_L W_\mu^+ 
+ \kappa^L_{b_2 t W} \bar{t}_{1L}  \gamma^\mu {b_2}_L W_\mu^+ + {\rm h.c.} \right) \nonumber \\
&+& \kappa_{b b Z}^L \bar{b_1}_L \gamma^\mu {b_1}_L Z_\mu
+ \kappa_{b_2 b_2 Z}^L \bar{b}_{2L} \gamma^\mu b_{2L} Z_\mu \nonumber \\
&+& \left( \kappa_{b_2 b Z}^L \bar{b}_{1L} \gamma^\mu b_{2L}  Z_\mu + {\rm h.c.} \right)  \nonumber \\
&+& \kappa_{b b Z}^R \bar{b_1}_R \gamma^\mu {b_1}_R Z_\mu 
+ \kappa_{b_2 b_2 Z}^R \bar{b}_{2R}  \gamma^\mu b_{2R}  Z_\mu \ ,
\label{b'uni.EQ}
\eea
and the Higgs interactions as~\cite{Gopalakrishna:2011ef}\footnote{Our convention of the 
Higgs coupling $\kappa$'s here differ by a factor of $\sqrt{2}$ compared to that in Ref.~\cite{Gopalakrishna:2011ef}.}
\bea
{\cal L}_{4D} \supset - h \left[ 
\kappa_{h b_{L} b_{R}} \bar{b}_{1L} {b}_{1R} + 
\kappa_{h b_{2L} b_{2R}} \bar{b}_{2L} {b}_{2R} \right. \nonumber \\
\left. + \kappa_{h b_{L} b_{2R}} \bar{b}_{1L} {b_2}_R +
\kappa_{h b_{2L} b_{R}} \bar{b}_{2L} {b_1}_R  \right] + {\rm h.c.} \ . 
\label{Lagb'H.EQ}
\eea
We have not introduced $\kappa_{b_2 b Z}^{R} \bar{b}_{2R}  \gamma^\mu b_{1R}  Z_\mu + {\rm h.c.} $
or $ \kappa^R_{b_2 t W} \bar{t}_{1R}  \gamma^\mu {b_2}_R W_\mu^+ + {\rm h.c.}$
in Eq.~(\ref{b'uni.EQ}) since these couplings will not arise in this model. 
For convenience, we use $b$ and $b_1$ interchangeably, and also $b'$ and $b_2$ interchangeably, but it should be clear from the context which one we mean. 

For mixing with a single $b'$, the effective couplings $\kappa$ as defined in 
Eqs.~(\ref{b'uni.EQ}) and (\ref{Lagb'H.EQ}) are given by
\bea
\kappa^L_{b t W} &=& \frac{g c^L_\theta}{\sqrt{2}} \ ; \ \  \kappa^L_{b_2 t W} = \frac{g s^L_\theta}{\sqrt{2}} \ ; \nonumber\\ 
\kappa_{b b Z}^L &=& g_Z \left( -\frac{1}{2}{c_\theta^L}^2 + \frac{1}{3} s_W^2 \right) \ ; \ \  
\kappa_{b_2 b_2 Z}^L = g_Z \left( -\frac{1}{2}{s_\theta^L}^2 + \frac{1}{3} s_W^2 \right) \ ; \nonumber \\
\kappa_{b_2 b Z}^L &=& g_Z c_\theta^L s_\theta^L \left(\frac{1}{2}\right) \ ; \ \  
\kappa_{b b Z}^R = g_Z \left(\frac{1}{3}s_W^2\right)\ ; \ \ 
\kappa_{b_2 b_2 Z}^R = g_Z \left(\frac{1}{3}s_W^2\right) \ ;  \label{kapNoZbb.EQ} \\
\kappa_{h b_{L} b_{R}} &=& \frac{1}{\sqrt{2}} (c^L_\theta c^R_\theta \lambda_{Q_L b_R} + c^L_\theta s^R_\theta \lambda_{Q_L b'_R}) \ ; \ \ 
\kappa_{h b_{2L} b_{2R}} = \frac{1}{\sqrt{2}} (s^L_\theta s^R_\theta \lambda_{Q_L b_R} - s^L_\theta c^R_\theta \lambda_{Q_L b'_R}) \ ; \nonumber \\
\kappa_{h b_{L} b_{2R}} &=& \frac{1}{\sqrt{2}} (-c^L_\theta s^R_\theta \lambda_{Q_L b_R} + c^L_\theta c^R_\theta \lambda_{Q_L b'_R}) \ ; \ \ 
\kappa_{h b_{2L} b_{R}} = \frac{1}{\sqrt{2}} (-s^L_\theta c^R_\theta \lambda_{Q_L b_R} - s^L_\theta s^R_\theta \lambda_{Q_L b'_R}). \nonumber 
\eea
From Eq.~(\ref{kapNoZbb.EQ}) we see that $\kappa_{b b Z}^L$ is shifted, and experimental constraints require that this shift be less than about $1~\%$,
roughly implying $s^L_\theta < 0.1$, {\it i.e.} $M_{b'} \gtrsim 10\,\tilde m \approx 3~$TeV. 
But as we have mentioned, since we have in mind application to the model in 
Ref.~\cite{Agashe:2006at} where this coupling is protected by the custodial symmetry, 
we consider much lighter $M_{b'}$ when we discuss the phenomenology.

For this model, the effective $4D$ Yukawa couplings parametrized in Eq.~(\ref{LintM.EQ}) are given by
\bea
\lambda_{b_R b'_L} &=& 0 \ ; \ \ \lambda_{Q_L b_R} = \frac{\tilde\lambda_b}{k\pi R} f_{Q_L}(\pi R) f_{b_R}(\pi R) e^{k\pi R} \ ; \\ 
\lambda_{Q_L b'_R} &=& \frac{\tilde\lambda_t}{k\pi R} f_{Q_L}(\pi R) f_{b'_R}(\pi R) e^{k\pi R} \ ; \ \
\lambda_{Q_L t_R} = \frac{\tilde\lambda_t}{k\pi R} f_{Q_L}(\pi R) f_{t_R}(\pi R) e^{k\pi R} \ , \nonumber
\eea
where $\lambda_{Q_L b_R}$ is the b-quark Yukawa coupling,
$\lambda_{Q_L t_R}$ is the top-quark Yukawa coupling,
$\tilde\lambda_{b,t}$ are the (dimensionless) $5D$ Yukawa couplings $\tilde \lambda_{b,t} \equiv k \lambda_{b,t}$, and 
$f_\psi$ are the fermion wavefunctions which depend on the fermion bulk mass parameters $c_\psi$~\cite{Gherghetta:2000qt}. 
We present the fermion profiles in \ref{fermProf.APP}. 

The mixing in the gauge boson sector {\it i.e.}, $V_\mu^{(0)} \leftrightarrow V_\mu^{(1)}$ mixing, 
where $V_\mu = \{W_\mu,\, Z_\mu\}$, also induces the $W t b'$ coupling,
and this mixing is of order $(v/M_{KK})^2$ with an additional $\sqrt{k\pi R}$ enhancement
for an IR-brane-peaked Higgs.
The contribution to the $b'$ decay rate due to $b'\leftrightarrow b$ mixing is proportional to $(\tilde{m}/M_{b'})^2$, 
while due to $W_L^{(0)}\leftrightarrow W_R^{(1)}$ mixing it is proportional to 
$\left(\sqrt{k\pi R} (g_R/g_L) m_W^2/M^2_{W'_R}\right)^2$, 
and it should be noted that
the gauge KK boson mass ({\it i.e.} $M_{W'_R}$) is constrained to be $\gtrsim 2~{\rm TeV}$ by precision electroweak 
constraints (see Ref.~\cite{Davoudiasl:2009cd} and references therein). 
Thus, the contribution due to gauge KK mixing is about $1.3\,$\% of the fermion KK mixing contribution
for $M_{b'} = M_{W'_R} = 2~$TeV, and even smaller for lighter $b'$ masses.   
We thus do not include the $W^{(0)}\leftrightarrow W^{(1)}$ mixing contribution in our study.
See Ref.~\cite{Dennis:2007tv} for another discussion of the $W^{(0)}\leftrightarrow W^{(1)}$ mixing contribution.
For the model without custodial protection of $Zbb$, we ignore $t \leftrightarrow t'$ mixing 
since this mixing angle is small, being suppressed by the larger $M_{t'}$ (above $3$~TeV) due to the choice of the 
$c_{b_R}$ required for the correct $b$-quark mass.
We also ignore mixings to the heavier KK modes in both the gauge and fermion sectors.  

%%%%%%%%%%%%%%%%%%%%%%%%%%%%%%%%%%%%%%%%%%%%%%%%%%%%%%%%%%%%%%%%%
\subsection{Model with $Zb\bar b$ custodial protection}
\label{ThZbb.SEC}
In order to ease precision electroweak constraints on warped models, the custodial symmetry can
be used to protect the $Zb\bar b$ coupling as proposed in Ref.~\cite{Agashe:2006at}. 
One way to achieve this is to complete the 3rd generation left-handed quarks into the 
$Q_L = (2,2)_{2/3}$ bi-doublet representation and the theory made invariant under
a discrete $L \leftrightarrow R$ symmetry defined as $P_{LR}$. 
The kinetic energy (KE) term for $Q_L$ is 
\beq
{\cal L}_{\rm KE} \supset {\rm Tr}\! \left[ \bar{Q}_L i\gamma^\mu D_\mu Q_L \right]  \, \label{LKEQL.EQ} \ ,
\eeq
and $\Sigma = (2,2)_0$ is the bidoublet Higgs, and their component fields are  
\beq
Q_L = \begin{pmatrix} t_L & \chi \\ b_L & t'  \end{pmatrix} \ , \qquad \Sigma = \begin{pmatrix} \phi_0^* & \phi^+ \\ -\phi^- & \phi_0  \end{pmatrix} \ .
\eeq
EWSB is due to $\left< \phi_0 \right> = v/\sqrt{2}$ and ${\rm Im}{(\phi^a)}$ are the Goldstone bosons. 
Note that to complete the bidoublet representation, two new fermions have been introduced, namely
$\chi$ and $t'$, with electromagnetic charge $5/3$ and $2/3$ respectively. 
The extra-fields $\chi$ and $t'$ (the ``custodians'') are ensured to be without zero-modes by applying Dirichlet-Neumann $(-,+)$ 
boundary conditions (BC) on the extra dimensional interval $[0,\pi R]$, 
and their KK excitations are vectorlike with respect to the SM gauge group,
while the SM particles are the zero-modes of fields with Neumann-Neumann $(+,+)$ BC, and are chiral.
Elsewhere in the literature, sometimes the $L,R$ subscripts on fermion fields denote the gauge-group, but in our notation, 
the subscripts $L,R$ on the fields denote the left and right (Lorentz) chiralities. 

The above ${\cal L}$ in Eqs.~(\ref{LKEQL.EQ}) implies the following couplings of the component fields 
\bea
{\cal L}_{\rm KE} &\supset& \frac{g_L}{2} W_{L\mu}^3 \left[ \bar{t}_L \gamma^\mu t_L - \bar{b}_L \gamma^\mu b_L + \bar{\chi}\gamma^\mu \chi - \bar{t'}\gamma^\mu t' \right] %\hspace{7cm}
+ \frac{g_L}{\sqrt{2}} \left[ W_{L\mu}^+ \left( \bar{t}_L \gamma^\mu b_L + \bar{\chi} \gamma^\mu t' \right) + h.c. \right]
 \nonumber \\
                    &&+ g^\prime B_\mu \left[\frac{1}{6} \bar{t}_L \gamma^\mu t_L + \frac{1}{6} \bar{b}_L \gamma^\mu b_L + \frac{7}{6} \bar{\chi} \gamma^\mu \chi + \frac{7}{6} \bar{t'} \gamma^\mu t' \right]  \ .
\label{LKEQQV.EQ}
\eea
The $SU(3)_c$ QCD interaction of the colored fermions are standard and are not shown. 

We go to the electroweak gauge boson mass basis by the usual orthogonal rotation
\beq
\bmat B \\ W_L^3 \emat = \bmat c_W & -s_W \\ s_W & c_W \emat \bmat A \\ Z  \emat \ , 
\eeq
defined by the weak mixing angle $c_W \equiv  \cos{(\theta_W)} = g_L/\sqrt{g_L^2 + {g^\prime}^2}$, $s_W \equiv \sin{(\theta_W)} = g^\prime/\sqrt{g_L^2 + {g^\prime}^2}$,
and the electric charge as $e \equiv g_L g^\prime /\sqrt{g_L^2 + {g^\prime}^2}$. 

After KK reduction, we obtain, in addition to the SM neutral current (NC) and charge current (CC) interactions, 
the following new interactions
\bea
{\cal L}^{4D}_{\rm NC} &\supset& {\cal L}_{\rm NC}^{SM} + \left[ e \, {\cal I}_{\chi\chi A} \left(\frac{5}{3}\right) A_\mu + g_Z \, {\cal I}_{\chi\chi Z} \left( \frac{1}{2} - s_W^2 \frac{5}{3} \right) Z_\mu \right] \bar{\chi}\gamma^\mu \chi + \nonumber \\
                          & &\quad \left[ e \, {\cal I}_{t't'A} \left(\frac{2}{3}\right) A_\mu + g_Z \, {\cal I}_{t't'Z} \left(-\frac{1}{2} - s_W^2 \frac{2}{3} \right) Z_\mu \right] \bar{t'} \gamma^\mu t' \ , \\
{\cal L}^{4D}_{\rm CC} &\supset& {\cal L}_{\rm CC}^{SM} + \frac{g_L}{\sqrt{2}}\, {\cal I}_{\chi t' W} {W^+_L}_{\! \mu} ~ \bar{\chi} \gamma^\mu t' + h.c. \ ,
\label{ZbbWLCC.EQ}
\eea
where $g_Z = \sqrt{g_L^2 + {g^\prime}^2}$, and the overlap integrals are given by 
$$
{\cal I}_{\psi\psi V} \equiv \frac{1}{\pi R} \int_0^{\pi R} dy \, e^{ky} f_\psi(y) f_\psi(y) f_V(y) \ .
$$
Since $U(1)_{EM}$ is unbroken ${\cal I}_{\psi \psi A} = 1$;
${\cal I}_{\psi\psi Z}$ and ${\cal I}_{\psi\psi W}$ differ from unity by a few percent due to EWSB $(0)-(1)$ gauge boson mixing effects, 
and since we are neglecting this small effect, we take all the ${\cal I} = 1$. 

It is possible to write down an invariant top quark Yukawa coupling with either the 
$t_R = (1,1)_{2/3}$ or with $t_R \subset (1,3)_{2/3} \oplus (3,1)_{2/3}$. 
We refer to these possibilities as the singlet top (ST) and the triplet top (TT) models respectively, 
and will elaborate on both these possibilities in the following subsections. 
We will show the couplings relevant to the phenomenology we are interested in, 
controlled by the (diagonal) coupling of the new heavy fermions to the gluon (set by $g_s$),
and a model dependent (off-diagonal) coupling of one heavy fermion, a ``light'' SM fermion
and a gauge boson or the Higgs boson. 
The off-diagonal couplings are induced by mass mixings between the zero-mode and 1st KK mode fermions,
which in turn is governed by the Yukawa couplings.
We will elaborate on these couplings below. 

%%%%%%%%%%%%%%%%%%%%%%%%%
\subsubsection{Model with $t_R = (1,1)_{2/3}$ (ST model)}
\label{tR1Th.SEC}
For the case of $t_R = (1,1)_{2/3}$ the kinetic-energy term is  
\beq
{\cal L}_{\rm K.E.}^{t_R} \supset \bar{t}_R i \gamma^\mu D_\mu t_R \label{LKEtR.EQ} \ ,
\eeq
and the top Yukawa coupling is the invariant combination $\overline{(2,2)}_{2/3} (2,2)_0 (1,1)_{2/3}$ written as 
\beq
{\cal L}_{\rm Yuk} \supset \lambda_t ~ {\rm Tr}\! \left[ \bar{Q}_L \Sigma \right] t_R + h.c. \ , \label{LYukt.EQ}
\eeq
The above ${\cal L}$ in Eqs.~(\ref{LKEtR.EQ})~and~(\ref{LYukt.EQ}) adds in addition to Eq.~(\ref{LKEQQV.EQ})
the following couplings of the component fields 
\bea
{\cal L}_{\rm KE} &\supset& 
                    g^\prime B_\mu \left[ \frac{2}{3} \bar{t}_R \gamma^\mu t_R \right] \ , \\ 
{\cal L}_{\rm Yuk} &\supset& \lambda_t \left( \bar{t}_L t_R \phi_0^* - \bar{b}_L t_R \phi^- + \bar{\chi} t_R \phi^+  + \bar{t'} t_R \phi_0 \right) + h.c. \ .
\label{tR1-LYuk.EQ}
\eea

In the fermion sector, the mass matrix including zero-mode and (light) KK mixing but neglecting the smaller mixings to heavier KK states is 
\beq
{\cal L}_{\rm mass} \supset \bmat \bar{t}_L & {\bar{t'}}_L \emat \bmat m_t  & 0 \\ \tilde{m} & M_{t'} \emat \bmat t_R \\ {t'}_R \emat 
+ \bar{b}_L \left( \lambda_b \frac{v}{\sqrt{2}} \right) b_R + h.c. \ , 
\label{MtTmat.EQ}
\eeq
where $m_t,\tilde{m} = \tilde\lambda_t (v/\sqrt{2}) f^{(n)}_{t_R}(\pi R) f^{(m)}_{t_L,t'_L}(\pi R) e^{k\pi R}/(k\pi R)$,
$\tilde\lambda_t \equiv k \lambda_t$ is the dimensionless $5D$ Yukawa coupling, 
and we have not shown mixing terms 
in the b-quark sector since in this model the new heavy charge $-1/3$ vectorlike fermions could only arise as
the partners of the $b_R$ but we ignore them since they are very heavy.  
The above mass matrix is diagonalized by
\beq
\bmat t_L \\ t'_L \emat = \bmat c_L & - s_L \\ s_L & c_L \emat \bmat {t_1}_L \\ {t_2}_L \emat \ ; \quad 
\bmat t_R \\ t'_R \emat = \bmat c_R & - s_R \\ s_R & c_R \emat \bmat {t_1}_R \\ {t_2}_R \emat \ ,
\label{tTrot.EQ}
\eeq
where $\{t_1,t_2\}$ are the mass eigenstates (ignoring mixings to higher KK states),
with the mixing angles given by
\beq 
\tan{(2\theta_L)} = \frac{-2 m_t \tilde{m}}{\left( M_t'^2 - m_t^2 + \tilde{m}^2 \right)}  \ ; \quad 
\tan{(2\theta_R)} = \frac{-2 \tilde{m} M_t'}{\left( M_t'^2 - m_t^2 - \tilde{m}^2 \right)} \ .
\eeq 
The mass eigenvalues $m_{1,2}$ are given by
\beq
m_{1,2}^2 = \frac{M_t'^2}{2} \left[ (1 + x_t^2 + \tilde{x}^2) \mp \sqrt{(1 + x_t^2 + \tilde{x}^2)^2 - 4 x_t^2} \right] \ , 
\eeq
where $x_t \equiv m_t/M_t'$ and $\tilde{x} \equiv \tilde{m}/M_t'$. 
In the limit of large $M_t'$, {\it i.e.}, $x_t,\tilde{x} \ll 1$, we have
\beq
m_1 = m_t \left[ 1 + O(x^4) \right] \quad ; \quad m_2 = M_t' \left[ 1 + \frac{\tilde{x}^2}{2} + O(x^4) \right] \ .
\eeq

In the mass basis the final interactions we obtain are as below. 
The charged current interaction is
\bea
{\cal L}_{CC} \supset \frac{g_L}{\sqrt{2}} \left( c_L \bar{t}_{1L} \gamma^\mu b_L - s_L \bar{t}_{2L} \gamma^\mu b_L
+ s_L \bar{\chi}_L \gamma^\mu t_{1L} + c_L {\bar{\chi}}_L \gamma^\mu t_{2L} \right.  \nonumber \\
\left. + s_R \bar\chi_R t_{1R} + c_R \bar\chi_R t_{2R}  \right) {W^+_L}_\mu + h.c. \ .
\label{tR1-CC.EQ}
\eea
The neutral current interaction is 
\bea
{\cal L}_{NC} \supset && e \left[ \bar{\chi} \gamma^\mu \left( \frac{5}{3} \right) \chi + \bar{t}_1 \gamma^\mu \left( \frac{2}{3} \right) t_1
+ \bar{t}_2 \gamma^\mu \left( \frac{2}{3} \right) t_2 + \bar{b} \gamma^\mu \left( -\frac{1}{3} \right) b  \right] A_\mu \nonumber \\ 
&+& g_Z \left\{ \bar{t}_{1L} \gamma^\mu \left[ \frac{1}{2} \cos{2\theta_L} - \frac{2}{3} s_W^2 \right] t_{1L} 
+ \bar{t}_{2L} \gamma^\mu \left[ -\frac{1}{2} \cos{2\theta_L} - \frac{2}{3} s_W^2 \right] t_{2L} \right. \nonumber \\
&+& \left. \left[ \bar{t}_{2L} \gamma^\mu \left( -\frac{1}{2} \sin{2\theta_L} \right) t_{1L} + h.c. \right] \right. \nonumber \\
&+& \left. 
\bar{t}_{1R} \gamma^\mu \left[ -\frac{1}{2} s_R^2 - \frac{2}{3} s_W^2 \right] t_{1R} 
+ \bar{t}_{2R} \gamma^\mu \left[ -\frac{1}{2} c_R^2 - \frac{2}{3} s_W^2 \right] t_{2R} \right. \nonumber \\
&+& \left. \left[ \bar{t}_{2R} \gamma^\mu \left( -\frac{1}{2} s_R c_R \right) t_{1R} + h.c. \right] \right. \nonumber \\
&+& \left. \bar{b}_{L} \gamma^\mu \left[ -\frac{1}{2} - s_W^2 \left(-\frac{1}{3}\right) \right] b_{L}
+ \bar{\chi} \gamma^\mu \left[ \frac{1}{2} - s_W^2 \left(\frac{5}{3}\right) \right] \chi
\right\} Z_\mu \ ,
\eea
where $g_Z \equiv \sqrt{g_L^2 + {g^\prime}^2}$.
The $\chi$ interactions above include both the $L$ and $R$ chiralities. 
The Higgs interactions are got by replacing $v \rightarrow h$ in Eq.~(\ref{MtTmat.EQ}), after which going to the mass basis using
Eq.~(\ref{tTrot.EQ}) we get
\bea
{\cal L}_h &\supset& \frac{\tilde\lambda_t}{\sqrt{2}} h 
\left[ \left(c_L f_{t_L} + s_L f_{t'_L} \right) \bar{t}_{1L} + \left(c_L f_{t'_L} - s_L f_{t_L} \right) \bar{t}_{2L}  \right] 
\left(c_R t_{1R} - s_R t_{2R} \right) f_{t_R} \frac{e^{k\pi R}}{k\pi R} + h.c. \ , \nonumber \\
&=& h \left[ \left(c_L \frac{m_t}{v} + s_L \frac{\tilde{m}}{v} \right) \bar{t}_{1L} + \left(c_L \frac{\tilde{m}}{v} - s_L \frac{m_t}{v} \right) \bar{t}_{2L}  \right] 
\left(c_R t_{1R} - s_R t_{2R} \right) + h.c. \ ,
\eea
where the wavefunctions are evaluated at $\pi R$, {\it i.e.}, $f_{t_L,t'_L}(\pi R)$ is implied in the first line above, and
in the second line above we have written the Higgs couplings in terms of $m_t,\tilde{m}$ defined below 
Eq.~(\ref{MtTmat.EQ}).

%%%%%%%%%%%%%%%%%%%%%%%%%%
\subsubsection{Model with $t_R \subset (1,3)_{2/3} \oplus (3,1)_{2/3} $ (TT model)}
\label{tR3Defn.SEC}
Here we pursue another option detailed in Ref.~\cite{Agashe:2006at}
in which the $t_R$ can be embedded into a $(1,3)_{2/3}$ representation, 
and as explained there, due to the required $P_{LR}$ invariance, a $(3,1)_{2/3}$ must also be added. 
Thus, the multiplet containing the $t_R$ is
\beq
\psi_{t_R} = \psi_{t_R}^\prime \oplus \psi_{t_R}^{\prime\prime} =
\begin{pmatrix} t_R/\sqrt{2} & \chi^\prime \\ b^\prime & -t_R/\sqrt{2} \end{pmatrix} \oplus
\begin{pmatrix} t^{\prime\prime}/\sqrt{2} & \chi^{\prime\prime} \\ b^{\prime\prime} & -t^{\prime\prime}/\sqrt{2} \end{pmatrix} \ ,
\eeq
where $\psi_{t_R}^\prime = (1,3)_{2/3}$ and $\psi_{t_R}^{\prime\prime} = (3,1)_{2/3}$.  
The top Yukawa coupling is obtained from
\beq
{\cal L}_{Yuk}^{t_R} \supset - \sqrt{2} \lambda_t^\prime {\rm Tr} \left[ \bar{Q}_L \Sigma \psi_{t_R}^\prime \right] 
-\sqrt{2} \lambda_t^{\prime\prime} {\rm Tr} \left[ \bar{Q}_L \psi_{t_R}^{\prime\prime} \Sigma \right] + h.c. \ , 
\eeq
and $P_{LR}$ invariance requires  $\lambda_t^\prime = \lambda_t^{\prime\prime}$ (which we will just denote
as $\lambda_t$ henceforth),
and also $c_{\psi^\prime_{t_R}} = c_{\psi^{\prime\prime}_{t_R}}$.   

After EWSB due to $\left<\phi_0 \right> = v/\sqrt{2}$, 
with the restrictions mentioned in the previous paragraph,
the mass matrix is
\bea
{\cal L}_{mass} \supset 
&-& \bmat \bar{b}_L & \bar{b^{\prime}_L} & \bar{b^{\prime\prime}_L} \emat 
\bmat 0  & \sqrt{2} m_{bb'} & \sqrt{2} m_{bb''} \\ 0 & M_{b^{\prime}} & 0 \\ 0 & 0 & M_{b''} \emat 
\bmat b_R \\ b^\prime_R \\ b^{\prime\prime}_R \emat  \nonumber \\
& & \quad - \bmat \bar{t}_L & \bar{t'}_L & \bar{t^{\prime\prime}_L} \emat 
\bmat     m_{tt}  &   0            &  m_{tt''} \\ 
       -m_{t't}    &  M_{t'}         &  -m_{t't''} \\ 
         0       & -m_{t't''}      & M_{t^{\prime\prime}} \emat 
\bmat t_R \\ {t'}_R \\ t''_R \emat 
\label{tR3Mmat.EQ} \\ 
 & & \quad - \bmat \bar{\chi}_L & \bar{\chi^\prime}_L & \bar{\chi^{\prime\prime}_L} \emat 
\bmat M_\chi  & \sqrt{2} m_{\chi\chi'} & \sqrt{2} m_{\chi\chi''} \\
        \sqrt{2} m_{\chi\chi'} & M_{\chi^{\prime}} & 0 \\ 
        \sqrt{2} m_{\chi\chi''} & 0 & M_{\chi^{\prime\prime}} \\ \emat 
\bmat \chi_R \\ \chi^\prime_R \\ \chi^{\prime\prime}_R \emat + h.c. \ , \nonumber 
\eea
where the $M_i$ are the vectorlike masses, and the EWSB generated masses $m_{ij}$ are given by
\beq
m_{ij} = \tilde\lambda_t \frac{v}{\sqrt{2}} \frac{1}{k\pi R} f^{(n)}_{\psi_L^i}(\pi R) f^{(m)}_{\psi_R^j}(\pi R) e^{k\pi R} \ ,
\label{mij.EQ}
\eeq
$\tilde\lambda_t \equiv k \lambda_t$ is the dimensionless $5D$ Yukawa coupling. 

We will work out next the couplings in the  mass basis.
We write $\psi^\alpha \equiv (\psi\ \psi^\prime\ \psi^{\prime\prime})^T$ and the mass eignestates as 
$\psi^i \equiv (\psi_1\ \psi_2\ \psi_3)^T$ for each of the $\psi = \{b, t, \chi \}$ sectors 
(here $\psi'$ for the $t$-sector is really what we have called $t'$). 
We perform a bi-orthogonal rotation (we take the masses to be real for simplicity)
$\psi_L^\alpha = R_{\psi L}^{\alpha i} \psi_L^i$ and $\psi_R^\alpha = R_{\psi R}^{\alpha i} \psi_R^i$ to diagonalize 
each of the mass matrices in Eq.~(\ref{tR3Mmat.EQ}).
 
The $Z$ couplings for the $\psi = \{b,t,\chi \}$-sectors in unitary gauge in the mass basis are
\bea
{\cal L} \supset g_Z \bar{\psi}^i_{L,R}  \left[ R_{\psi_{L,R}}^{\alpha i^*} \left( q^{3L}_{\psi^\alpha_{L,R}} - Q_\psi s_W^2 \right)  R_{\psi_{L,R}}^{\alpha j} \right] \gamma^\mu 
          {\cal I}_{\psi\psi Z} \psi_{L,R}^j Z_\mu \ ,
\eea
where the $q^{3L}$ are the $W^3_L$ charges and $Q_\psi$ are EM charges as given below and 
we ignore differences in the overlap integrals and take ${\cal I} = 1$.
The $W_L^3$ charges are $q^{3L}_{b^\alpha_{L}} = \{-1/2, 0, -1 \}$, $q^{3L}_{b^\alpha_{R}} = \{0, 0, -1 \}$, $q^{3L}_{t^\alpha_{L}} = \{1/2, -1/2, 0 \}$, $q^{3L}_{t^\alpha_{R}} = \{0, -1/2, 0 \}$,
$q^{3L}_{\chi^\alpha_{L}} = \{1/2, 0, 1 \}$, $q^{3L}_{\chi^\alpha_{R}} = \{1/2, 0, 1 \}$.
The EM chages are $Q_b = -1/3$, $Q_t = 2/3$ and $Q_\chi = 5/3$.

The Higgs couplings in the mass basis are
\beq
{\cal L} \supset - \bar{\psi}_L^i {R_{\psi_L}^{\alpha i^*}} \frac{m^{\alpha \beta}}{v}  R_{\psi_R}^{\beta j} \psi_R^j h  + h.c. \ ,
\eeq
where $m^{\alpha \beta}$ are the off-diagonal EWSB induced masses in Eq.~(\ref{mij.EQ}). 

The charged current $W^\pm$ interactions, in addition to those in Eq.~(\ref{ZbbWLCC.EQ}), are  
\beq
{\cal L}_{CC}^W \supset 
%{\cal L}_{CC}^{\rm Eq.(\ref{ZbbWLCC.EQ})} + 
g_L \left( \bar{t^{\prime\prime}} \gamma^\mu b^{\prime\prime} {\cal I}_{t'' b'' W} 
- \bar{\chi^{\prime\prime}} \gamma^\mu t^{\prime\prime} {\cal I}_{\chi'' t'' W} \right) {W_L^+}_\mu + h.c. \ ,
\eeq
which in the mass basis in unitary gauge are
\bea
{\cal L}_{CC}^W \supset \frac{g_L}{\sqrt 2} \left[ 
\bar{t}^i_L R^{1i^*}_{t_L} R^{1j}_{b_L} \gamma^\mu b_L^j + 
\left(
\bar{\chi}^i_L R^{1i^*}_{\chi_L} R^{2j}_{t_L} \gamma^\mu t_L^j 
\phantom{\sqrt{1}} \hspace{4cm} \right. \right.  \nonumber \\  \left. \left. \hspace{2cm} 
+ \sqrt{2} \bar{t}^i_R R^{3i^*}_{t_R} R^{3j}_{b_R} \gamma^\mu b_R^j - \sqrt{2} \bar{\chi}^i_R R^{3i^*}_{\chi_R} R^{3j}_{t_R} \gamma^\mu t_R^j
\right) + (L \leftrightarrow R) \right] {W^+_L}_\mu + h.c. \ ,
\label{tR3-CC.EQ}
\eea
and again we ignore differences in the overlap integrals and take ${\cal I} = 1$.

In \ref{tR3AnaDiag.APP} we present analytical expressions for the 
mixing matrices in the $b$-quark sector in the limit of $m_{ij}/M_{\psi^\prime} \ll 1$,
and the resulting couplings in the mass basis. 
We present this for illustration only and have used exact numerical diagonalization in all our results.

One way to generate the bottom-quark mass is to have a Yukawa coupling that respects the custodial symmetry. 
With $Q_L = (2,2)_{2/3}$, the $b_R$ can be embedded into the representation 
$\psi_{b_R}^\prime = (1,3)_{2/3}$ 
and the $b$-quark Yukawa coupling obtained from, 
${\cal L}_{Yuk}^{b_R} \supset - \lambda_b^\prime {\rm Tr} \left[ \bar{Q}_L \Sigma \psi_{b_R}^\prime \right] 
+ h.c.
$
This breaks the $P_{LR}$ symmetry but the resulting shifts are acceptable since the 
$c_{b_R}$ choice required to get the correct $b$-quark mass makes the new vectorlike fermions in 
the $\psi_{b_R}^\prime$ multiplet all very heavy ($> 3~$TeV). 
In our analysis we have therefore ignored the mixing effects and the signatures of these heavy fermions.
Many more possibilities for $b_R$ representations are discussed in Ref.~\cite{Agashe:2006at}. 

%%%%%%%%%%%%%%%%%%%%%%%%%%%%%%%%%%%%%%%%%%%%%%%%%
\section{Parameters and Couplings}
\label{parCoup.SEC}

The vectorlike fermions can mix among themselves and with SM fermions. 
We take this into account and denote the mass eigenstates by a subscript, {\it i.e.,} $X_n$ denotes the
$n^{\rm th}$ mass eigenstate of X type quark except for the SM quarks where we use 
$t$ or $t_1$ and $b$ or $b_1$ interchangeably.

We parametrize the relevant vectorlike quark couplings model-independently as
\bea
\mathcal{L}_\chi &\supset& \kappa_{\chi_{1L}t_{1L}W}~ \bar{\chi}_{1L}\gamma^{\mu}t_{1L}W^{+}_{\mu} + \kappa_{\chi_{1R}t_{1R}W}~ \bar{\chi}_{1R}\gamma^{\mu}t_{1R}W^{+}_{\mu} + {\rm h.c.}
\label{kapChiMI.EQ} \\
\mathcal{L}_{t'} & \supset & \kappa_{t_{2L}t_{1R}h}~\bar{t}_{2L}t_{1R}h + \kappa_{t_{1L}t_{2R}h} ~\bar{t}_{1L}t_{2R}h \nonumber \\
 &+&\kappa_{t_{1L}t_{2L}Z}~\bar{t}_{1L}\gamma^{\mu}t_{2L}Z_{\mu} + \kappa_{t_{1R}t_{2R}Z} ~\bar{t}_{1R}\gamma^{\mu}t_{2R}Z_{\mu} \nonumber\\
 &+&\kappa_{t_{2L}\chi_{1L}W}~\bar{t}_{2L}\gamma^{\mu}\chi_{1L}W_{\mu} + \kappa_{t_{2R}\chi_{1R}W} ~\bar{t}_{2R}\gamma^{\mu}\chi_{1R}W_{\mu} + {\rm h.c.}
\label{t'kapMI.EQ} \\
\mathcal{L}_{b'} &\supset & \kappa_{b_{2L}b_{1L}Z}~\bar{b}_{2L}\gamma^{\mu}b_{1L}Z_{\mu} + \kappa_{b_{2R}b_{1R}Z} ~\bar{b}_{2R}\gamma^{\mu}b_{1R}Z_{\mu} \nonumber\\
&+& \kappa_{b_{2L}b_{1R}h}~\bar{b}_{2L} b_{1R} h + \kappa_{b_{2R}b_{1L}h} ~\bar{b}_{2R} b_{1L} h + {\rm h.c.} \ .
\label{b'kapMI.EQ}
\eea
Wherever possible we show results model-independently as functions of the $\kappa$'s defined above.

In the following, we present the parameter choices we make for the different warped models 
discussed in Sec.~\ref{ThFrmWrk.SEC} for which we present numerical results.  
The analytical expressions for the fermion mass eigenvalue, the fermion profiles along the extra dimensions,
and their dependence on the $c$-parameters are given in \ref{fermProf.APP}.~\footnote{We find 
that after mixing the couplings relevant for our study are largely insensitive to the  choice of $k \pi R$ and $\tilde{\lambda}_{b,t}$; 
for instance, for $M_{KK} = 3~$TeV, varying $k/M_{Pl}$ between 0.1 and 1 changes the couplings by at most 1\,\% and varying $\tilde{\lambda}_{b,t}$ between 1 and 2 changes couplings only about a few percent.}
Using these profiles, we compute the overlap integrals and determine the couplings of the vectorlike fermions 
to the SM states relevant to our study. 
Various choices of the three relevant $c$-parameters, namely $c_{q_L}$, $c_{t_R}$ and $c_{b_R}$, are possible that reproduce the measured masses and couplings (see for {\it e.g.} Refs.~\cite{Moreau:2006np,Moreau:2005kz}
and references therein). 
Furthermore, there is freedom to choose the 5D Yukawa couplings $\tilde\lambda$ which we set to $1$, 
and $M_{KK}$ which we take to be $3$~TeV.
After these choices and imposing the constraint that the lightest eigenvalues in the top and bottom quark sectors 
correspond to the measured top mass ($172$~GeV) and bottom mass ($4.2$~GeV) respectively, there is one free parameter remaining 
which we take to be $c_{q_L}$.  
In the following, we show some representative benchmark points for the various warped models
detailed in Sec.~\ref{ThFrmWrk.SEC}, 
for each of the $\chi$, $t'$ and $b'$.

%%%%%%%%%%%%%%%%%
\subsection{$\chi$ Parameters and Couplings}

\begin{figure}[]
\begin{center}
\includegraphics[width=0.49\textwidth]{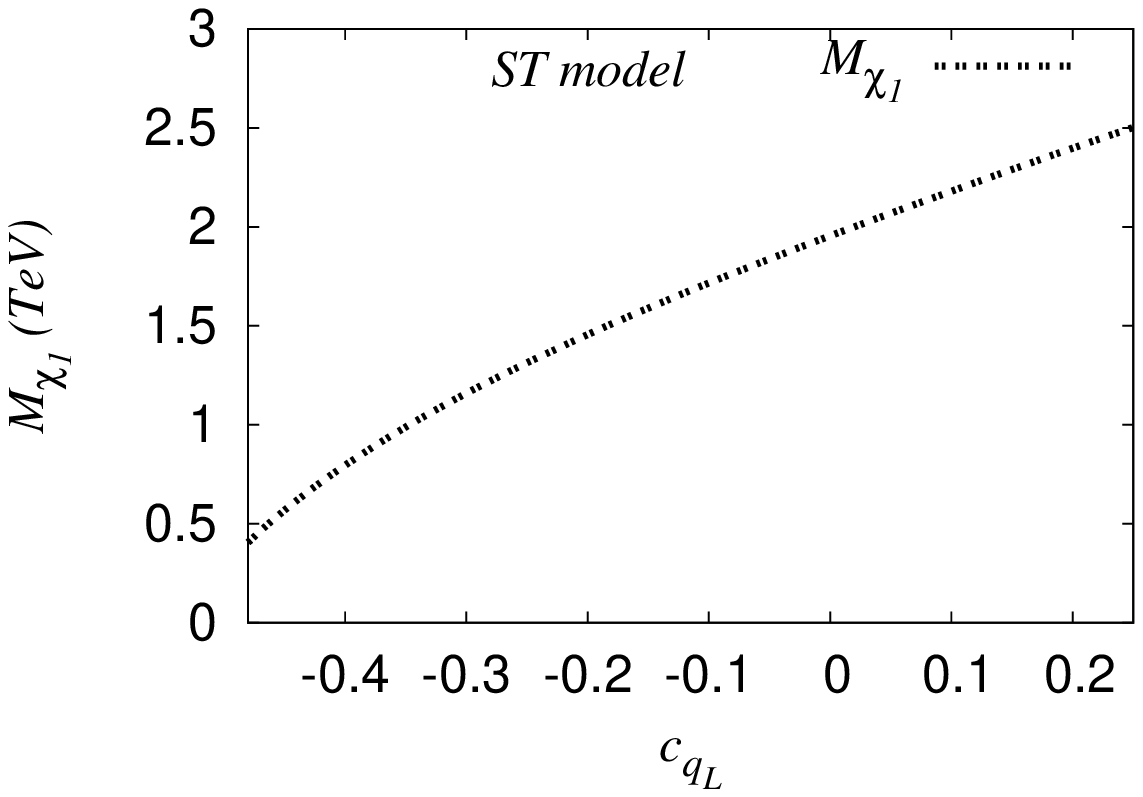}
\includegraphics[width=0.49\textwidth]{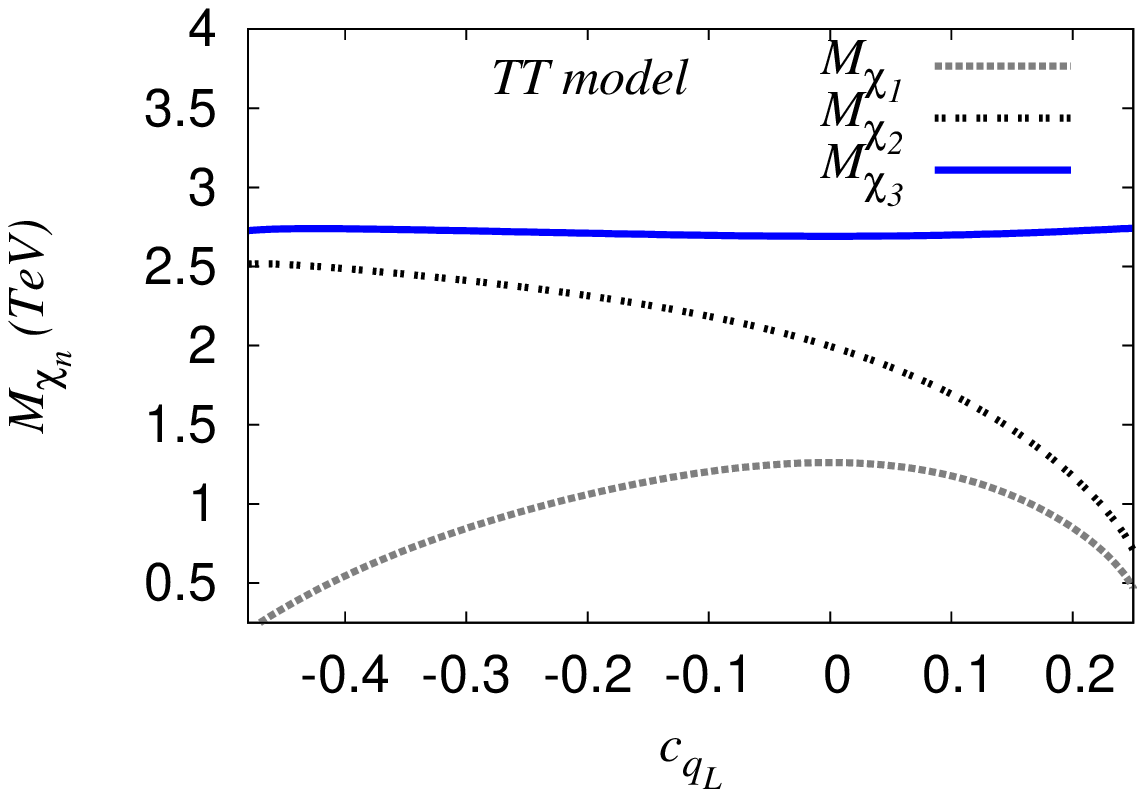}
\caption{$M_{\chi_1,\chi_2,\chi_3}$ as functions of $c_{q_L}$ in the ST and TT models with
$\tilde{\lambda}_t=1$, $\tilde{\lambda}_b=1$ and $M_{KK}=3$~TeV.
\label{chimass}}
\end{center}
\end{figure}

\begin{figure}[]
\begin{center}
\includegraphics[width=0.49\textwidth]{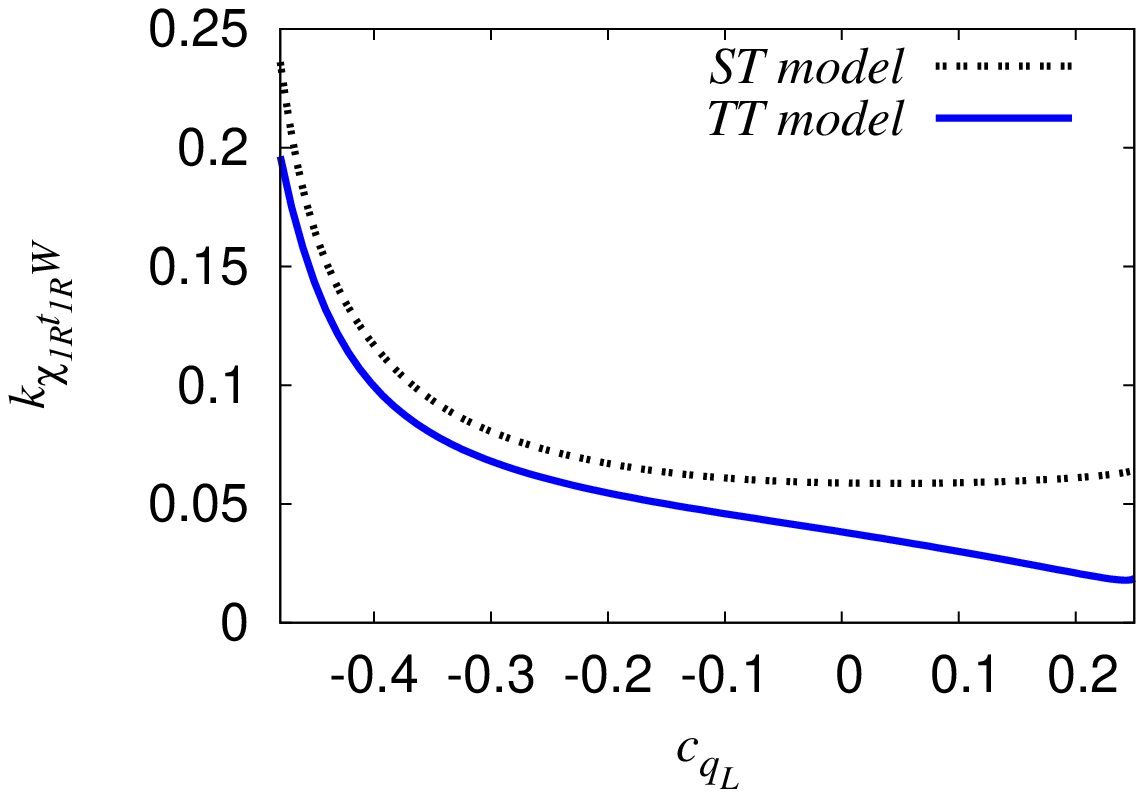}
\includegraphics[width=0.49\textwidth]{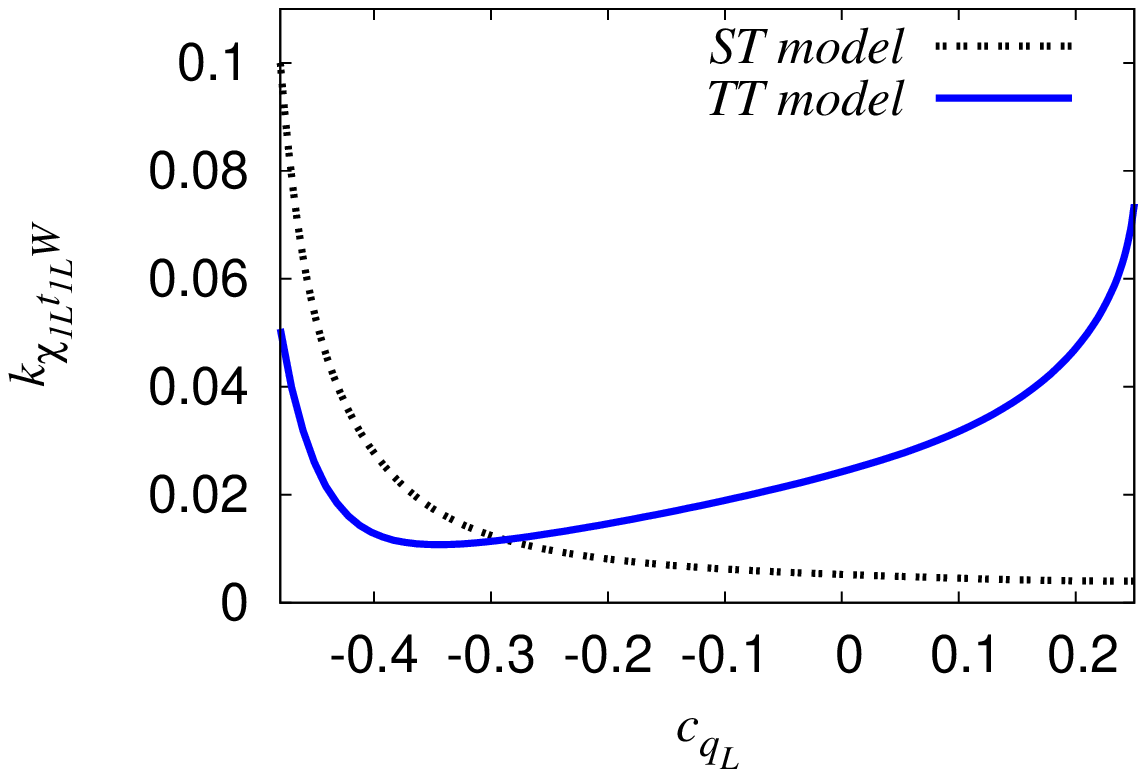}
\caption{$\kappa_{\chi_1tW}$'s as functions of $c_{q_L}$ in the ST and TT models with
$\tilde{\lambda}_t=1$, $\tilde{\lambda}_b=1$ and $M_{KK}=3$ TeV.
\label{chikappa}}
\end{center}
\end{figure}

The $\kappa$ for the warped model are as detailed in Sec.~\ref{ThFrmWrk.SEC}. 
In Figs.~\ref{chimass}~and~\ref{chikappa}, we show $M_{\chi_1}$, $\kappa_{\chi_{1L}t_{1L}W}$ and $\kappa_{\chi_{1R}t_{1R}W}$ 
as functions of $c_{q_L}$ for the $Zb\bar b$ protected ST and TT models. 
There is no $\chi$ state in the DT model. 
In the TT model, after $\chi$-$\chi^{\prime}$-$\chi^{\prime\prime}$ mixing, the $\chi_2$, $\chi_3$ becomes much heavier than $\chi_1$
because the appearance of the large off-diagonal term in
the $\chi$ mass matrix causes a significant split between $M_{\chi_1}$ and $M_{\chi_2,\chi_3}$. 
Therefore, for both ST and TT models, we focus only on the phenomenology of $\chi_1$. 
In the TT model $M_{\chi_1}$ in Fig.~\ref{chimass} shows an unusual behavior --  
with increasing $c_{Q_L}$, it first increases and then decreases. 
This is an effect of the diagonalization of Eq.~(\ref{tR3Mmat.EQ}), 
with $c_{q_L}\lesssim 0$ having $M_{\chi} < M_{\chi^{\prime}}$ while $c_{q_L}\gtrsim 0$ has $M_{\chi} > M_{\chi^{\prime}}$, 
and the maximum of the eigenvalue is attained when $M_{\chi} = M_{\chi^{\prime}}$.
In Table~\ref{chiSTpara} we explicitly display the benchmark parameters and couplings in the ST model that we use 
for our numerical computations when we discuss $\chi_1$ phenomenology. 
\begin{table}[]
\centering
\begin{tabular}{|c|c|c|c|c|c|}
\hline $\mathcal X$ & $c_{q_L}$ & $c_{t_R}$  & $c_{b_R}$ & $\sin\theta_L$ & $\sin\theta_R$ \\ 
\hline 
 $\mathcal X_1$ & -0.463 & 0.206 & 0.586 & -0.136 & -0.394 \\
 $\mathcal X_2$ & -0.414 & 0.216 & 0.585 & -0.058 & -0.253 \\
 $\mathcal X_3$ & -0.350 & 0.202 & 0.584 & -0.033 & -0.192 \\
 $\mathcal X_4$ & -0.274 & 0.177 & 0.583 & -0.022 & -0.159 \\
 $\mathcal X_5$ & -0.186 & 0.137 & 0.581 & -0.016 & -0.140 \\
 $\mathcal X_6$ & -0.088 & 0.078 & 0.578 & -0.013 & -0.129 \\
\hline 
\hline $\mathcal X$ & $M_{\chi}$ (GeV)& $\kappa_{\chi_{1R}t_{1R}W}$ & $\kappa_{\chi_{1L}t_{1L}W}$ & $\kappa_{\chi_{1R}t_{2R}W}$ & $\kappa_{\chi_{1L}t_{2L}W}$ \\ 
\hline 
 $\mathcal X_1$ &  500 & 0.182 & 0.063 & 0.424 & 0.458 \\
 $\mathcal X_2$ &  750 & 0.117 & 0.027 & 0.447 & 0.461 \\
 $\mathcal X_3$ & 1000 & 0.089 & 0.015 & 0.453 & 0.462 \\
 $\mathcal X_4$ & 1250 & 0.074 & 0.010 & 0.456 & 0.462 \\
 $\mathcal X_5$ & 1500 & 0.065 & 0.007 & 0.457 & 0.462 \\
 $\mathcal X_6$ & 1750 & 0.060 & 0.006 & 0.458 & 0.462 \\
 \hline 
\end{tabular} 
\caption{$\chi$ benchmark parameters (parameter set denoted by $\mathcal X$) and couplings obtained using $\tilde{\lambda}_t=1$, $\tilde{\lambda}_b=1$ and $M_{KK}=3$ TeV in the
ST model.  
The $c$ values for all the benchmark parameter sets reproduce correct top and bottom quark masses after mixing.
\label{chiSTpara}}
\end{table}
In the ST model, we restrict ourselves to $c_{q_L} < 0$, {\it i.e.} with the $q_L$ partners peaked towards the IR brane,
since otherwise the partners become very heavy and this may be out of reach at the LHC.

In the TT model we have $M_{\chi^{\prime}}=M_{\chi^{\prime\prime}}$ due to the $P_{LR}$ symmetry of the theory
and we find the $\chi_2 \chi_1 h$ couplings (both $L$ and $R$) to be zero as a consequence of this.
The $\chi_2 \chi_3 h$ coupling is also zero.
Furthermore, the $P_{LR}$ symmetry also constrains $m_{\chi\chi'} =  m_{\chi\chi''}$ and as a result we find 
$\chi_3 \chi_1 Z$ (both $L$ and $R$) couplings to be zero.

%%%%%%%%%%%%%%%%%%%%%%%%%%%
\subsection{$t'$ Parameters and Couplings}
\label{tpParamCoup.SEC}

\begin{figure}[]
\begin{center}
\includegraphics[width=0.49\textwidth]{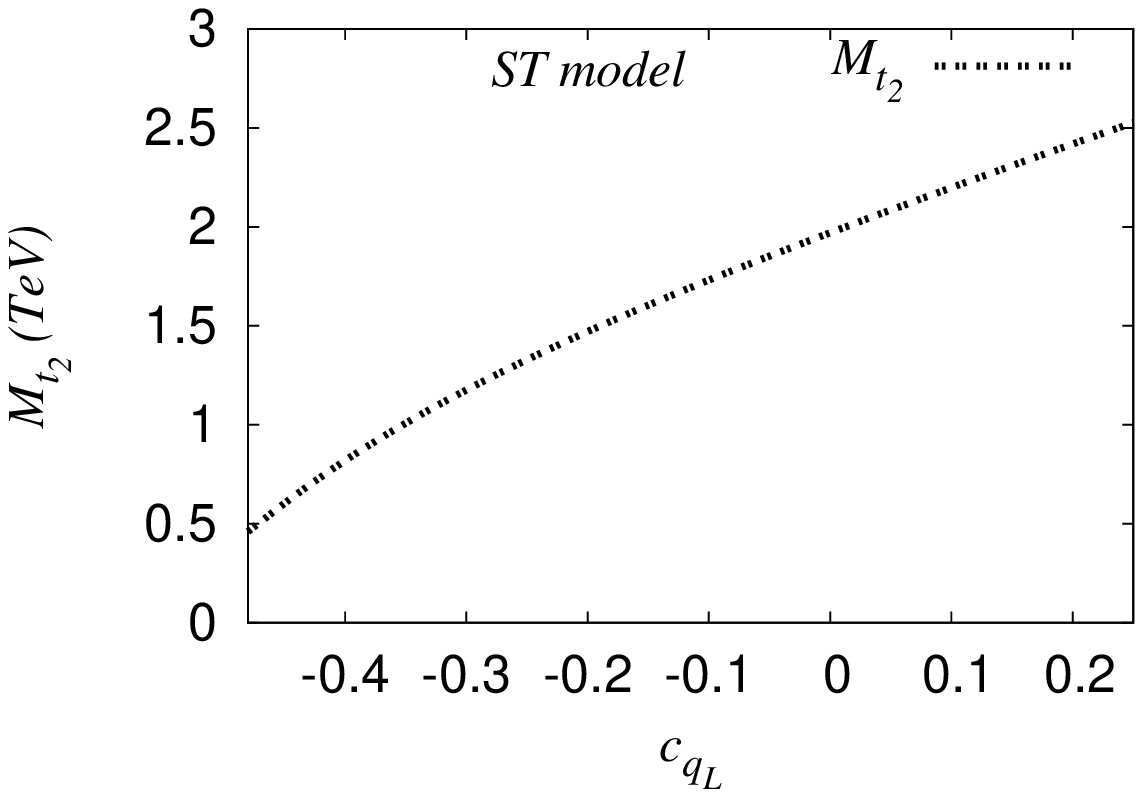}
\includegraphics[width=0.49\textwidth]{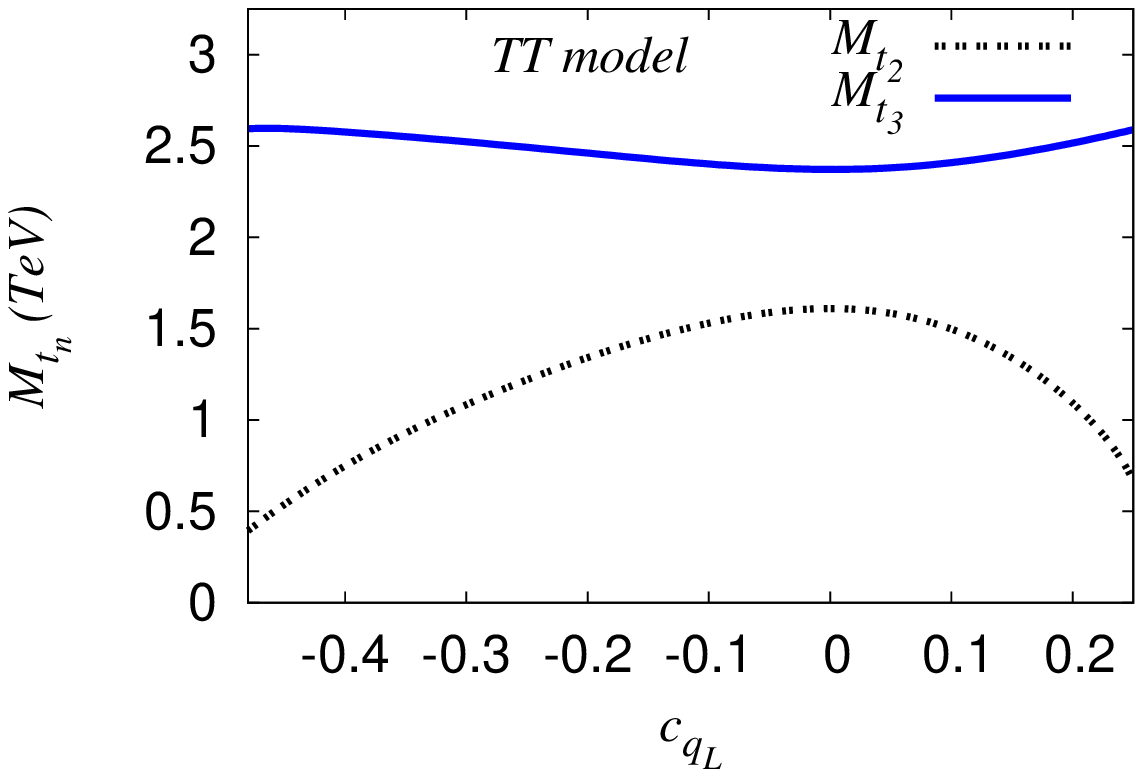}
\caption{$M_{t_2,t_3}$ as functions of $c_{q_L}$ in the ST and TT models with
$\tilde{\lambda}_t=1$, $\tilde{\lambda}_b=1$ and $M_{KK}=3$~TeV.
\label{cqL_Mtp.FIG}}
\end{center}
\end{figure}

\begin{figure}[]
\begin{center}
\includegraphics[width=0.49\textwidth]{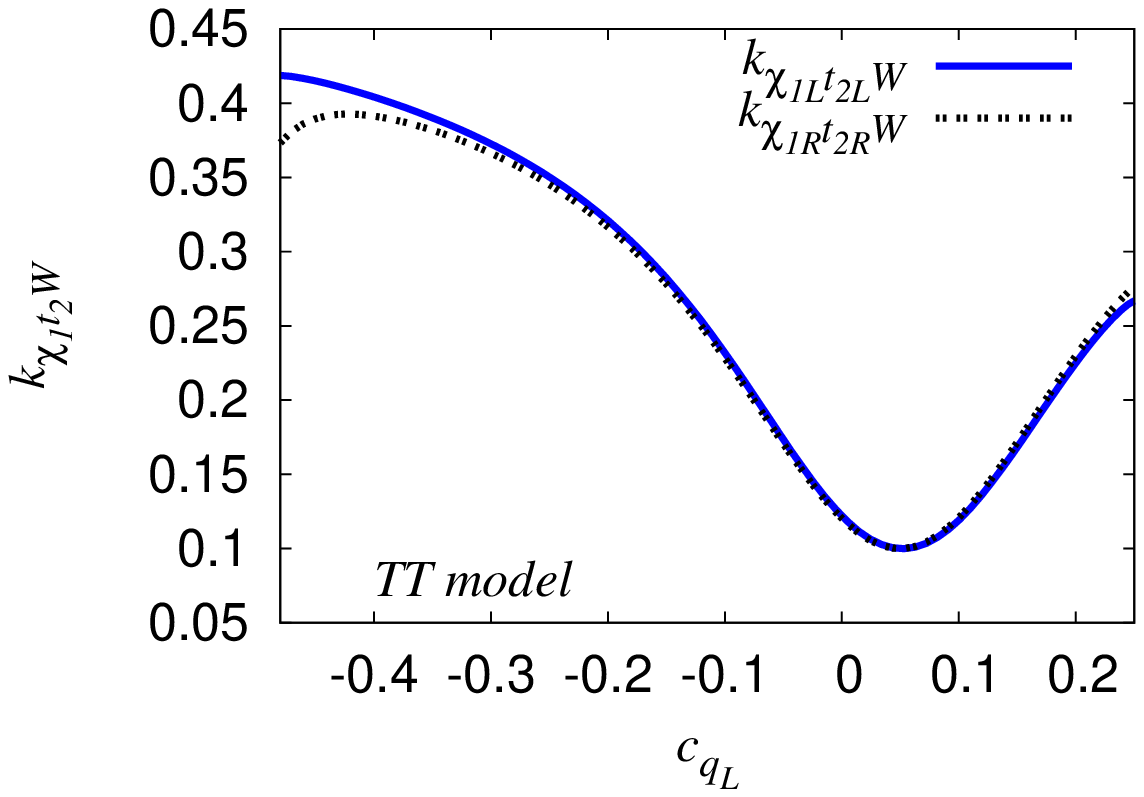}
\includegraphics[width=0.49\textwidth]{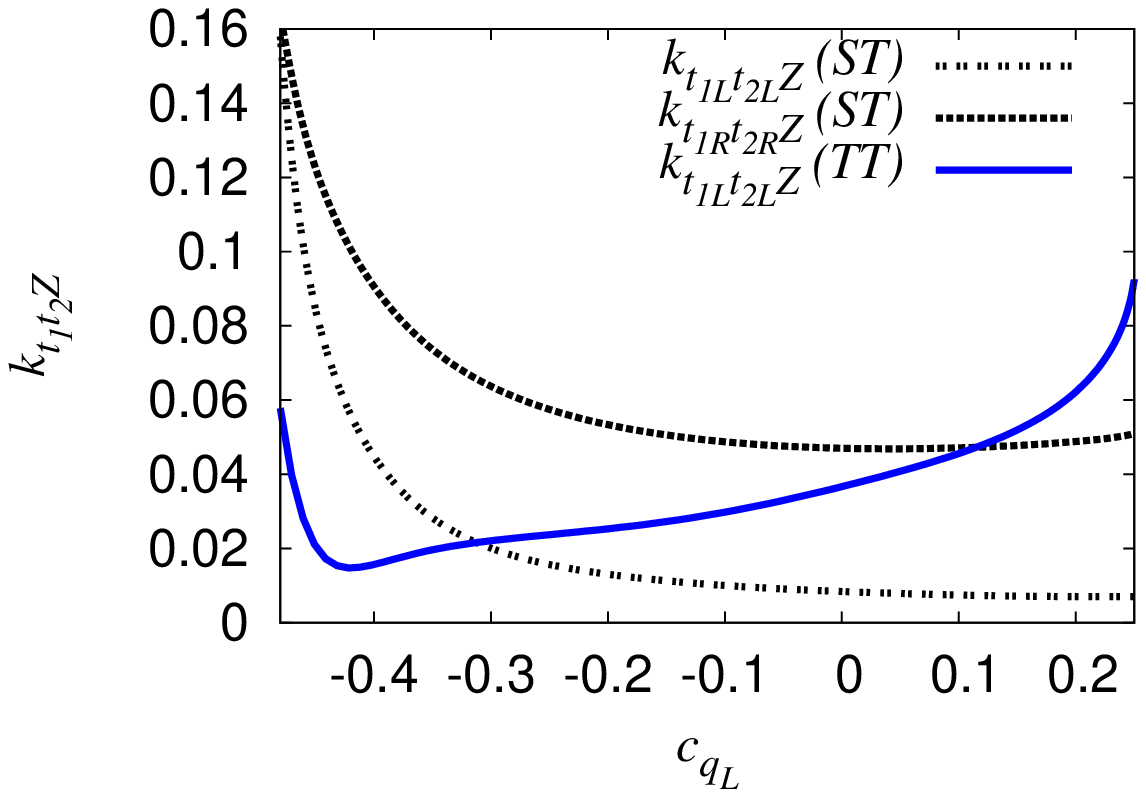}  
\includegraphics[width=0.49\textwidth]{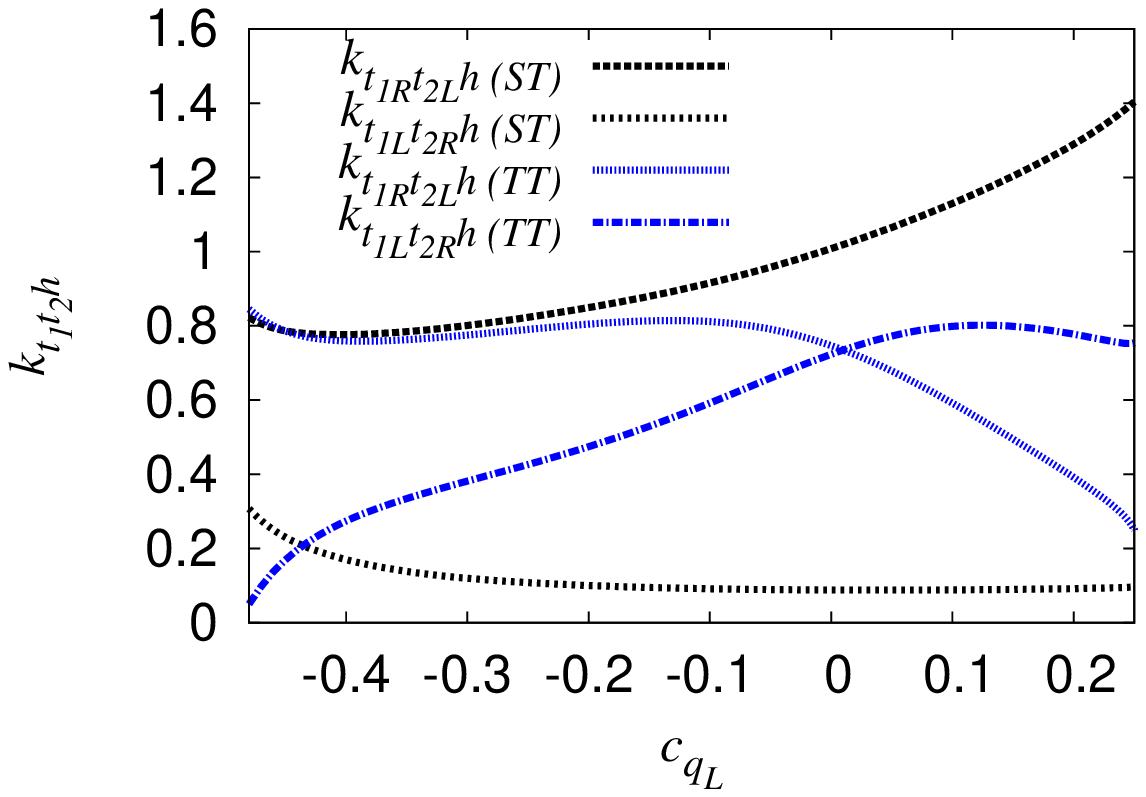}
\caption{The $\kappa$'s for $t_2$ as functions of $c_{q_L}$ in the ST and TT models, with 
$\tilde{\lambda}_t=1$, $\tilde{\lambda}_b=1$ and $M_{KK}=3$~TeV.
\label{t2para}}
\end{center}
\end{figure}

The $\kappa$ for the warped model are as detailed in Sec.~\ref{ThFrmWrk.SEC}. 
In the model with no $Zb\bar b$ protection (DT model), the $t'$ is quite heavy (above $3$~TeV) due to the choice of the 
$c_{b_R}$ required for the correct $b$-quark mass, making its LHC discovery challenging. 
We therefore will not discuss further the $t'$ in the DT model, and will restrict ourselves to the $Zb\bar b$ protected
ST and TT models. 
In Fig.~\ref{cqL_Mtp.FIG} we show the $M_{t'}$ as functions of $c_{q_L}$ in the ST and TT models. For the TT model, we note that 
the mass eigenvalue $M_{t_2}$ shows
a similar behavior as $M_{\chi_1}$, {\it i.e.}, with increasing $c_{Q_L}$, it first increases and then decreases.
We also find for the TT model that the $t_2$-$\chi_1$ mass-difference 
is larger than $m_W$ which allows the $t_2\to \chi_1 W$ decay mode.
We show the $t_2$ couplings 
in Fig.~\ref{t2para} for the various models as functions of $c_{q_L}$. 
$\kappa_{t_2 \chi_1 W}$ is large
since it is given by the $t'\chi W$ or $t^{\prime\prime}\chi^{\prime\prime}W$ couplings,
and is not proportional to any small off-diagonal mixing-matrix elements. 
In Table~\ref{t2STpara} we display the benchmark parameters and couplings in the ST model that are used for 
our numerical computations. 
\begin{table}[]
\centering
\begin{tabular}{|c|c|c|c|c|c|}
\hline $\mathcal T$ & $c_{q_L}$ & $c_{t_R}$  & $c_{b_R}$ & $\sin\theta_L$ & $\sin\theta_R$ \\ 
\hline 
 $\mathcal T_1$ & -0.471 & 0.196 & 0.586 & -0.167 & -0.442 \\
 $\mathcal T_2$ & -0.419 & 0.216 & 0.585 & -0.062 & -0.262 \\
 $\mathcal T_3$ & -0.356 & 0.204 & 0.584 & -0.034 & -0.195 \\
 $\mathcal T_4$ & -0.279 & 0.179 & 0.583 & -0.022 & -0.161 \\
 $\mathcal T_5$ & -0.191 & 0.140 & 0.581 & -0.016 & -0.141 \\
 $\mathcal T_6$ & -0.094 & 0.082 & 0.578 & -0.013 & -0.130 \\
\hline 
\hline $\mathcal T$ & $M_{t_2}$(GeV) & $\kappa_{t_{2L}t_{1R}h}$ & $\kappa_{t_{1L}t_{2R}h}$ & $\kappa_{t_{2R}t_{1R}Z}$ & $\kappa_{t_{2L}t_{1L}Z}$ \\ 
\hline 
 $\mathcal T_1$ &  500 & 0.806 & 0.277 & 0.148 & 0.123 \\
 $\mathcal T_2$ &  750 & 0.769 & 0.176 & 0.094 & 0.046 \\
 $\mathcal T_3$ & 1000 & 0.778 & 0.134 & 0.071 & 0.026 \\
 $\mathcal T_4$ & 1250 & 0.807 & 0.111 & 0.059 & 0.017 \\
 $\mathcal T_5$ & 1500 & 0.851 & 0.098 & 0.052 & 0.012 \\
 $\mathcal T_6$ & 1750 & 0.915 & 0.090 & 0.048 & 0.010 \\
 \hline 
\end{tabular} 
\caption{$t_2$ benchmark parameters (parameter set denoted by $\mathcal T$) and couplings obtained using $\tilde{\lambda}_t=1$, $\tilde{\lambda}_b=1$ and $M_{KK}=3$ TeV in the
ST model. The $c$ values for all the benchmark parameter sets reproduce correct top and bottom quark masses after mixing.
\label{t2STpara}}
\end{table}
%

%%%%%%%%%%%%%%%%%
\subsection{$b'$ Parameters and Couplings}
\begin{figure}[]
\begin{center}
\includegraphics[width=0.49\textwidth]{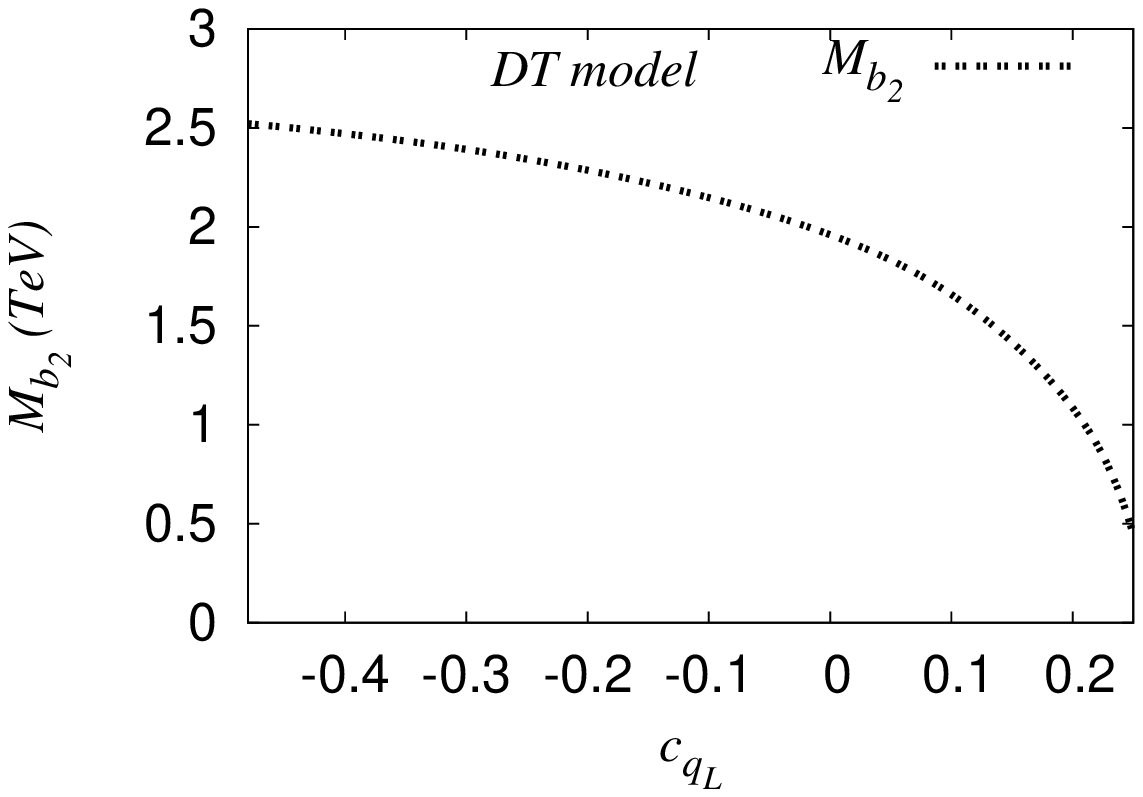}
\includegraphics[width=0.49\textwidth]{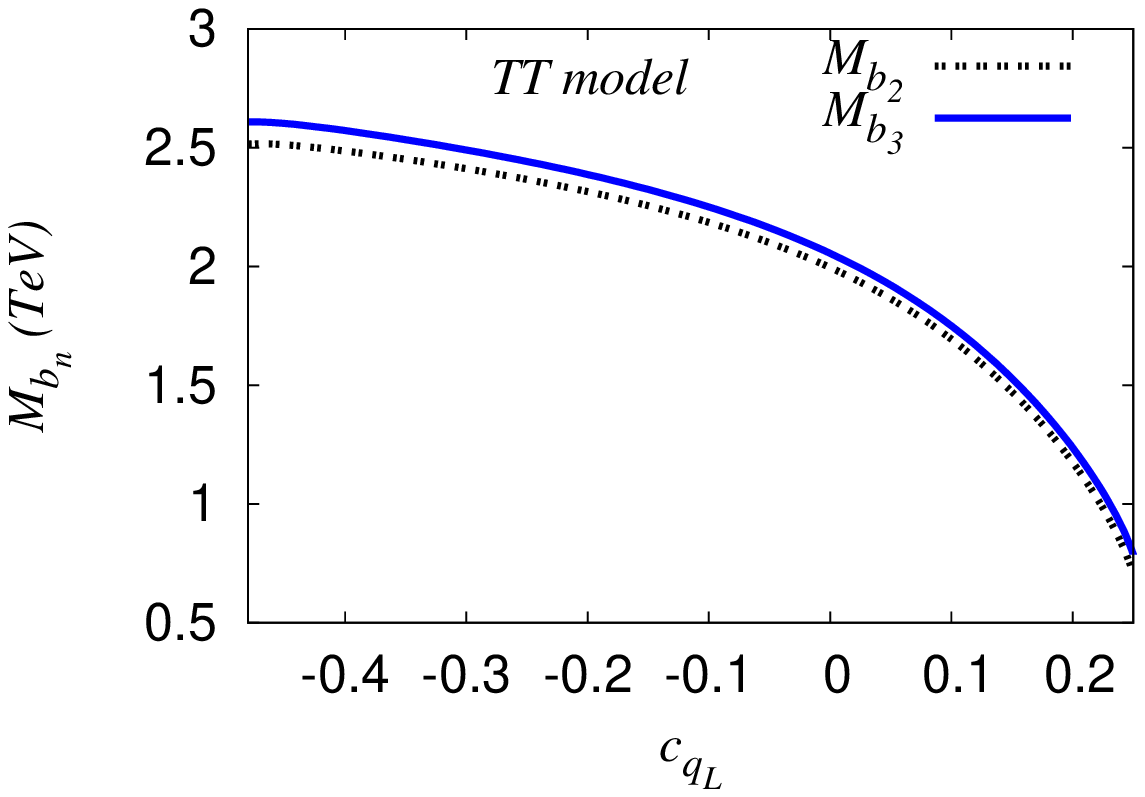}
\caption{$M_{b_2,b_3}$ as functions of $c_{q_L}$ in the DT and TT models, 
with $\tilde{\lambda}_t=1$, $\tilde{\lambda}_b=1$ and $M_{KK}=3$ TeV.
\label{b2mass}}
\end{center}
\end{figure}

\begin{figure}[]
\begin{center}
\includegraphics[width=0.49\textwidth]{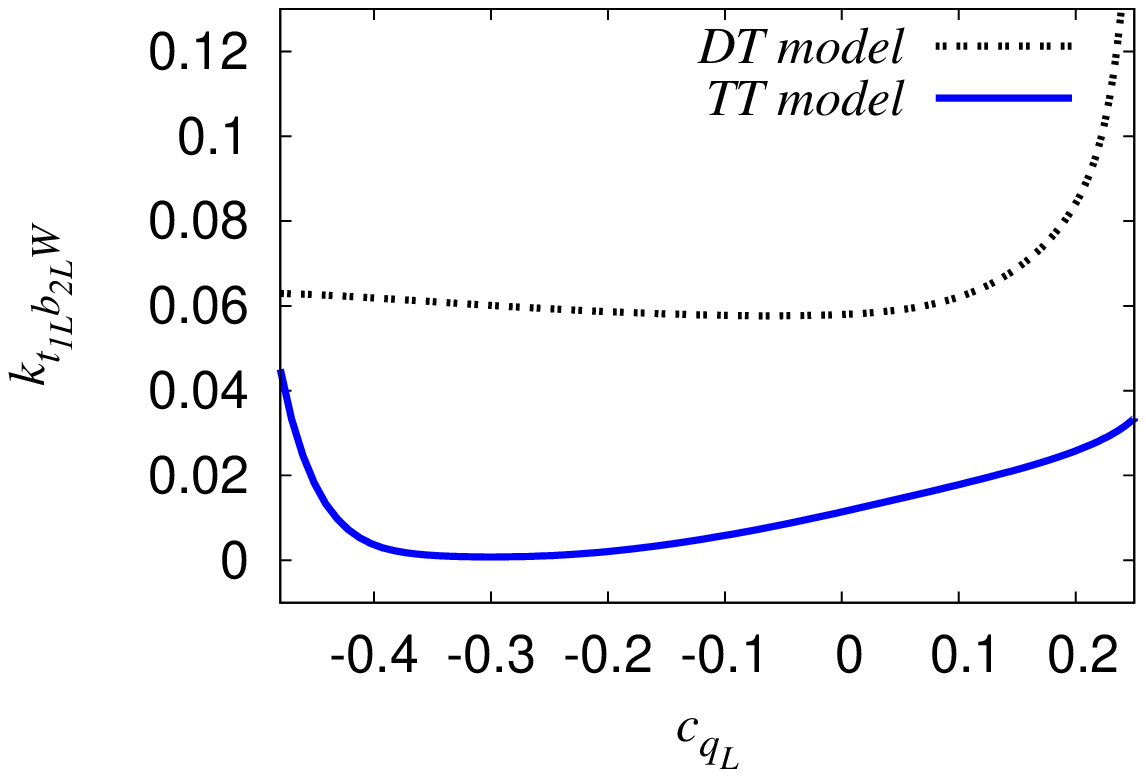}  
\includegraphics[width=0.49\textwidth]{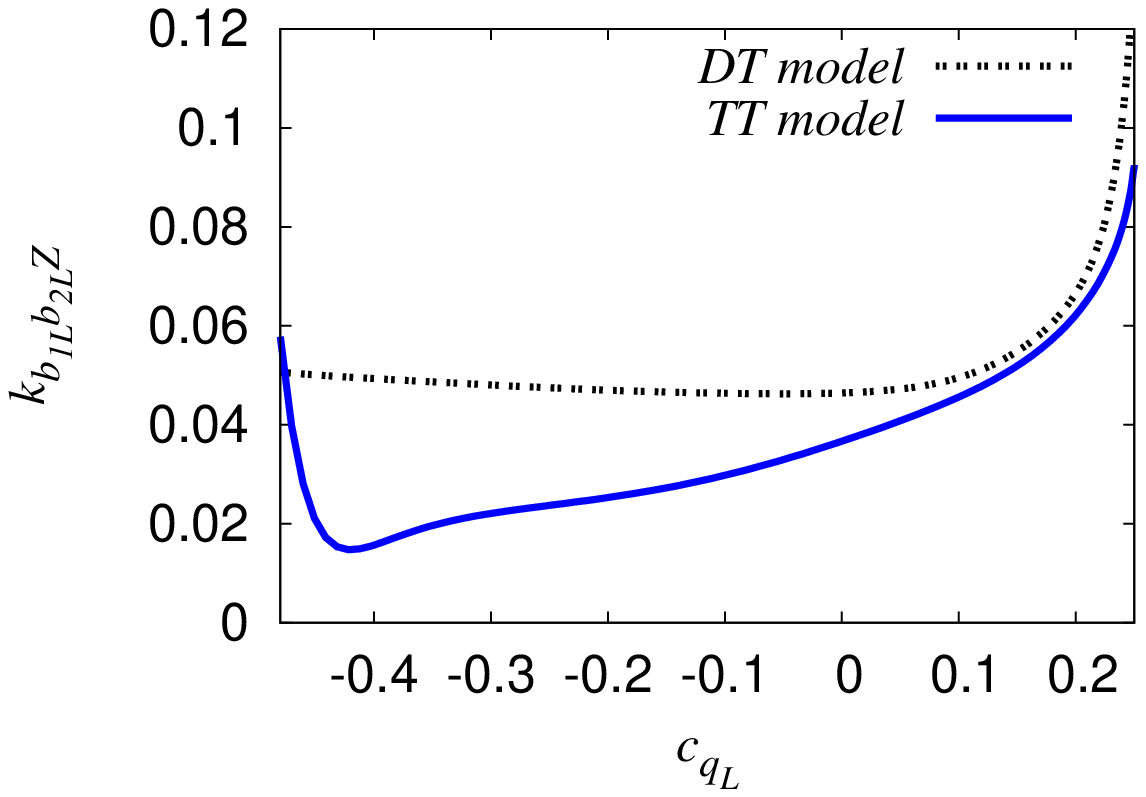}
\includegraphics[width=0.49\textwidth]{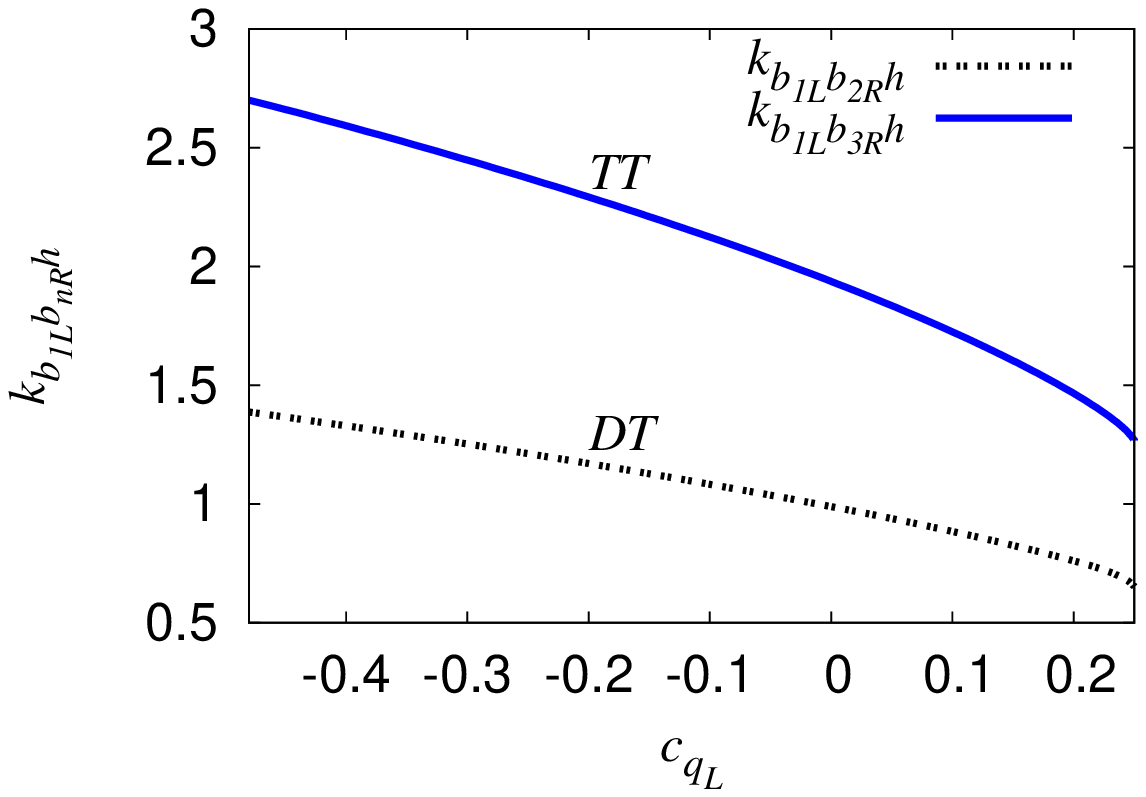}
\caption{The $\kappa$'s for $b_2$ as functions of $c_{q_L}$ in the DT and TT models, 
with $\tilde{\lambda}_t=1$, $\tilde{\lambda}_b=1$ and $M_{KK}=3$~TeV. 
\label{b2kappa}}
\end{center}
\end{figure}

The $\kappa$ for the warped model are as detailed in Sec.~\ref{ThFrmWrk.SEC}. 
As already mentioned, our convention of the Higgs coupling $\kappa$'s appearing in Eq.~(\ref{Lagb'H.EQ}) differ by a factor of $\sqrt{2}$ 
compared to that in Ref.~\cite{Gopalakrishna:2011ef}.
We display $M_{b_2}$, $\kappa_{b_2b_1h}$ and $\kappa_{b_2b_1Z}$ as functions of $c_{q_L}$ for the DT and TT models in 
Figs.~\ref{b2mass}~and~\ref{b2kappa}. 
In the TT model we have $M_{b^{\prime}}=M_{b^{\prime\prime}}$ due to the $P_{LR}$ symmetry of the theory
and we find that the $b_2 b_1 h$ couplings (both $L$ and $R$) to be zero as a consequence of this. 
The $b_2 b_3 h$ coupling is also zero.
Furthermore, the $P_{LR}$ symmetry also constrains $m_{bb'} =  m_{bb''}$ and as a result we find 
$b_3 b_1 Z$ (both $L$ and $R$) couplings to be zero. 
These are explicitly seen in the analytical formulas shown in \ref{tR3AnaDiag.APP} in the small mixing limit.
In Table~\ref{b'TTpara} we show the parameters for some benchmark points in the TT model.  
$R^{12}_{b_L}$ and $R^{12}_{b_R}$ are as defined in Sec.~\ref{tR3Defn.SEC}.   
\begin{table}[]
\centering
\begin{tabular}{|c|c|c|c|c|c|}
\hline $\mathcal B$ & $c_{q_L}$ & $c_{t_R}$  & $c_{b_R}$ & $R^{12}_{b_L}$ & $R^{12}_{b_R}$ \\ 
\hline 
 $\mathcal B_1$ & 0.259 & -0.464 & 0.562 & -0.400 & -0.0034 \\
 $\mathcal B_2$ & 0.247 & -0.414 & 0.566 & -0.299 & -0.0017 \\
 $\mathcal B_3$ & 0.226 & -0.350 & 0.569 & -0.242 & -0.0010 \\
 $\mathcal B_4$ & 0.197 & -0.274 & 0.571 & -0.207 & -0.0007 \\
 $\mathcal B_5$ & 0.156 & -0.186 & 0.574 & -0.186 & -0.0005 \\
 $\mathcal B_6$ & 0.098 & -0.088 & 0.577 & -0.173 & -0.0004 \\
\hline 
\hline $\mathcal B$ & $M_{b_2}$ (GeV) & $\kappa_{b_{2L}t_{1L}W}$ & $\kappa_{b_{2L}b_{1L}Z}$ & $\kappa_{b_{2L}t_{2L}W}$ & $\kappa_{b_{2R}t_{2R}W}$ \\ 
\hline 
 $\mathcal B_1$ &  500 & 0.118 & 0.210 & 0.300 & 0.322 \\
 $\mathcal B_2$ &  750 & 0.077 & 0.158 & 0.311 & 0.321 \\
 $\mathcal B_3$ & 1000 & 0.060 & 0.128 & 0.313 & 0.319 \\
 $\mathcal B_4$ & 1250 & 0.050 & 0.109 & 0.311 & 0.315 \\
 $\mathcal B_5$ & 1500 & 0.044 & 0.098 & 0.303 & 0.306 \\
 $\mathcal B_6$ & 1750 & 0.041 & 0.091 & 0.283 & 0.286 \\
 \hline 
\end{tabular} 
\caption{$b_2$ benchmark parameters (parameter set denoted by $\mathcal B$) and couplings obtained using $\tilde{\lambda}_t=1$, 
$\tilde{\lambda}_b=1$ and $M_{KK}=3$ TeV in the TT model.  
The $c$ values for all the benchmark parameter sets reproduce correct top and bottom quark masses after mixing.
\label{b'TTpara}}
\end{table}
We have 
$V_{tb} = R_{t_L}^{11*} R_{b_L}^{11}$,
and for the lower $b'$ masses this may be somewhat close to the 
experimental limit quoted earlier.
In \ref{tR3AnaDiag.APP} we give the analytical
expressions in the TT model in the small mixing limit
for illustration, and use exact numerical diagonalization in our results.  

\clearpage

%%%%%%%%%%%%%%%%%%%%%%%%%%%%%%%%%%%%%%%%%%%%%%%%%%%%
\section{Decay width and Branching Ratio}
\label{DecBR.SEC}

Here, we present the decay width and branching ratios (BRs) of vectorlike quarks. 
As concrete examples we take the different models detailed in Sec.~\ref{ThFrmWrk.SEC}, 
namely the DT, ST and TT models.

The analytical expressions for the vectorlike fermion partial decay widths are\footnote{
The Eqs.~(3)-(5) of Ref.~\cite{Gopalakrishna:2011ef} are special cases of these formulas. 
We point out a minor error in Eq.~(5) of Ref.~\cite{Gopalakrishna:2011ef} introduced by an ambiguity in 
specifying a number multiplying the $4\times 4$ identity in the program FORM. 
The decay width $\Gamma(b^\prime \rightarrow b h)$ shown in Eq.~(5) of Ref.~\cite{Gopalakrishna:2011ef} 
should read as shown here in Eq.~(\ref{Gamq2q1h.EQ}). 
Since the error is in terms suppressed as $m_b/M_{b^\prime}$, which is small, the error does not change any of the results of that paper. 
}
\begin{eqnarray}
\Gamma_{q_2\to q_1V} &= \frac{1}{32\pi}\frac{M_{q_2}^3}{M_V^2}
\left[\left(\kappa_L^2 + \kappa_R^2\right)
\left\{\left(1 - x_{q_1}^2\right)^2 + x_V^2\left(1 + x_{q_1}^2\right) - 2x_V^4\right\}\right. \nonumber \\ &- \left. 12\kappa_L\kappa_R x_{q_1}x_V^2 \right]\times 
\left(1+x_{q_1}^4+x_V^4-2x_{q_1}^2-2x_V^2 - 2x_{q_1}^2x_V^2 \right)^{\frac{1}{2}} 
\label{Gamq2q1V.EQ} \\
\Gamma_{q_2\to q_1h} &= \frac{1}{32\pi}M_{q_2}
\left[\left(\kappa_L^2 + \kappa_R^2\right)
\left\{\left(1 - x_{q_1}^2 - x_h^2\right)^2 \right\} + 4\kappa_L\kappa_R x_{q_1}\right]\nonumber \\ 
&\times \left(1+x_{q_1}^4+x_h^4-2x_{q_1}^2-2x_h^2 - 2x_{q_1}^2x_h^2\right)^{\frac{1}{2}} \ ,
\label{Gamq2q1h.EQ}
\end{eqnarray}
where the $\kappa_{L,R}$ are the couplings parametrized as in Sec.~\ref{parCoup.SEC}, and
$x_{q_1} \equiv M_{q_1}/M_{q_2}$, $x_V \equiv M_V/M_{q_2}$ and $x_h \equiv M_h/M_{q_2}$. 
We can obtain the total width and BRs in any model containing vectorlike fermions using the above equations.
Next, we present some results for the warped models.  

\begin{figure}[!h]
\begin{center}
\includegraphics[width=0.49\textwidth]{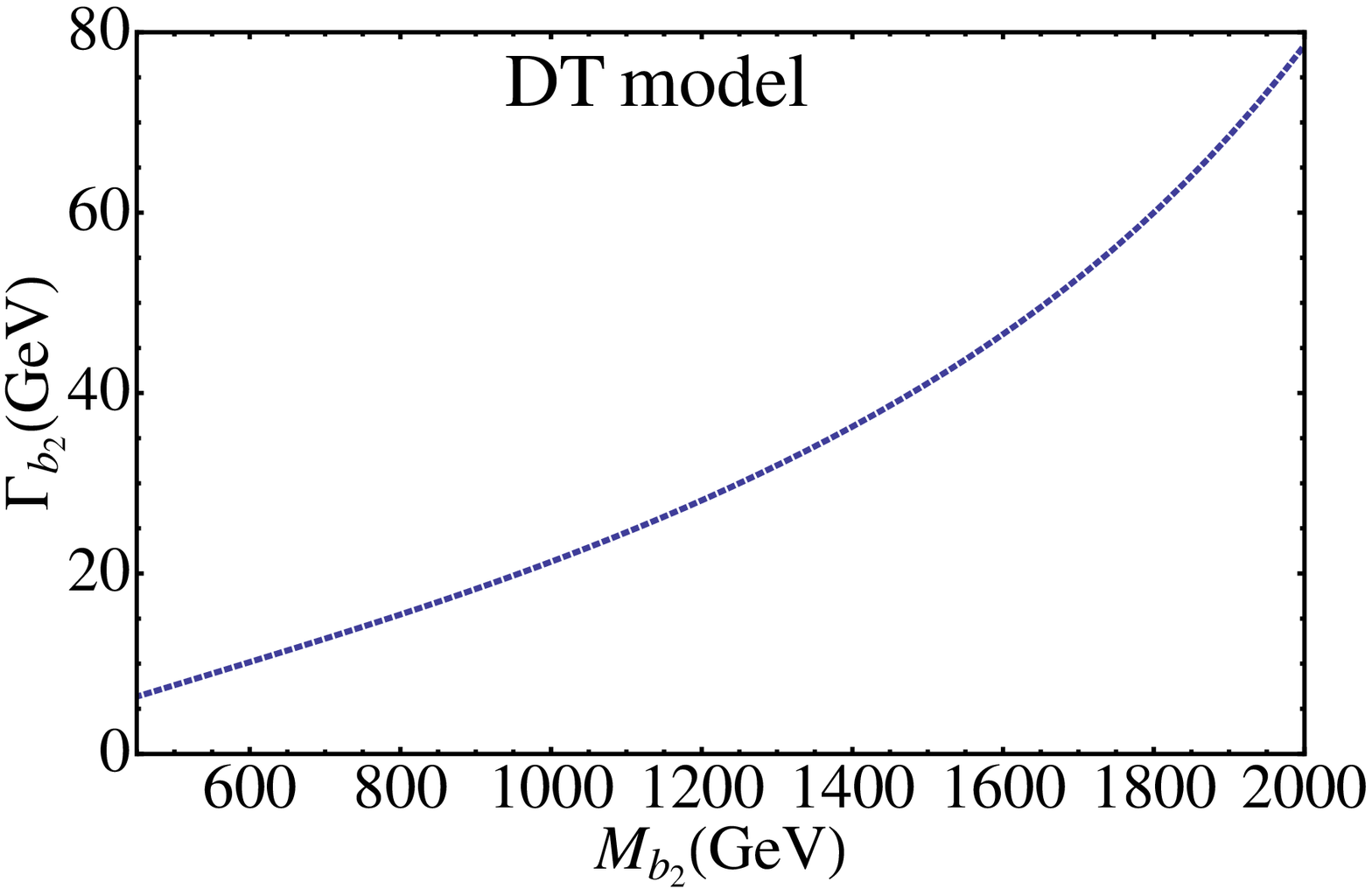}
\includegraphics[width=0.5\textwidth]{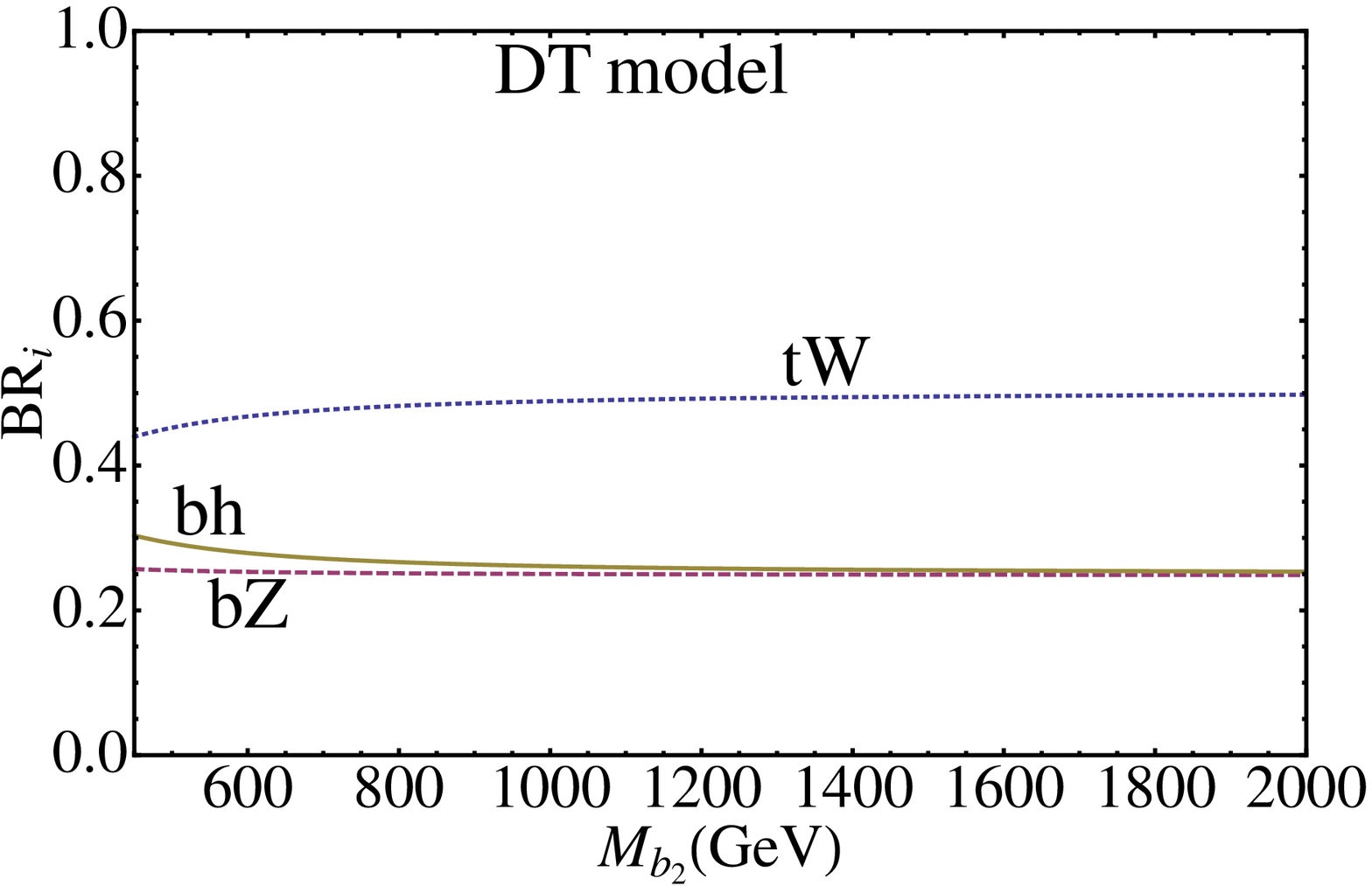}
\caption{Total decay width and branching ratios of $b_2$ as functions of $M_{b_2}$  in the model without $Zb\bar b$ protection for the DT model. 
\label{b2TWBR_DT.FIG}}
\end{center}
\end{figure}

\begin{figure}[!h]
\begin{center}
\includegraphics[width=0.49\textwidth]{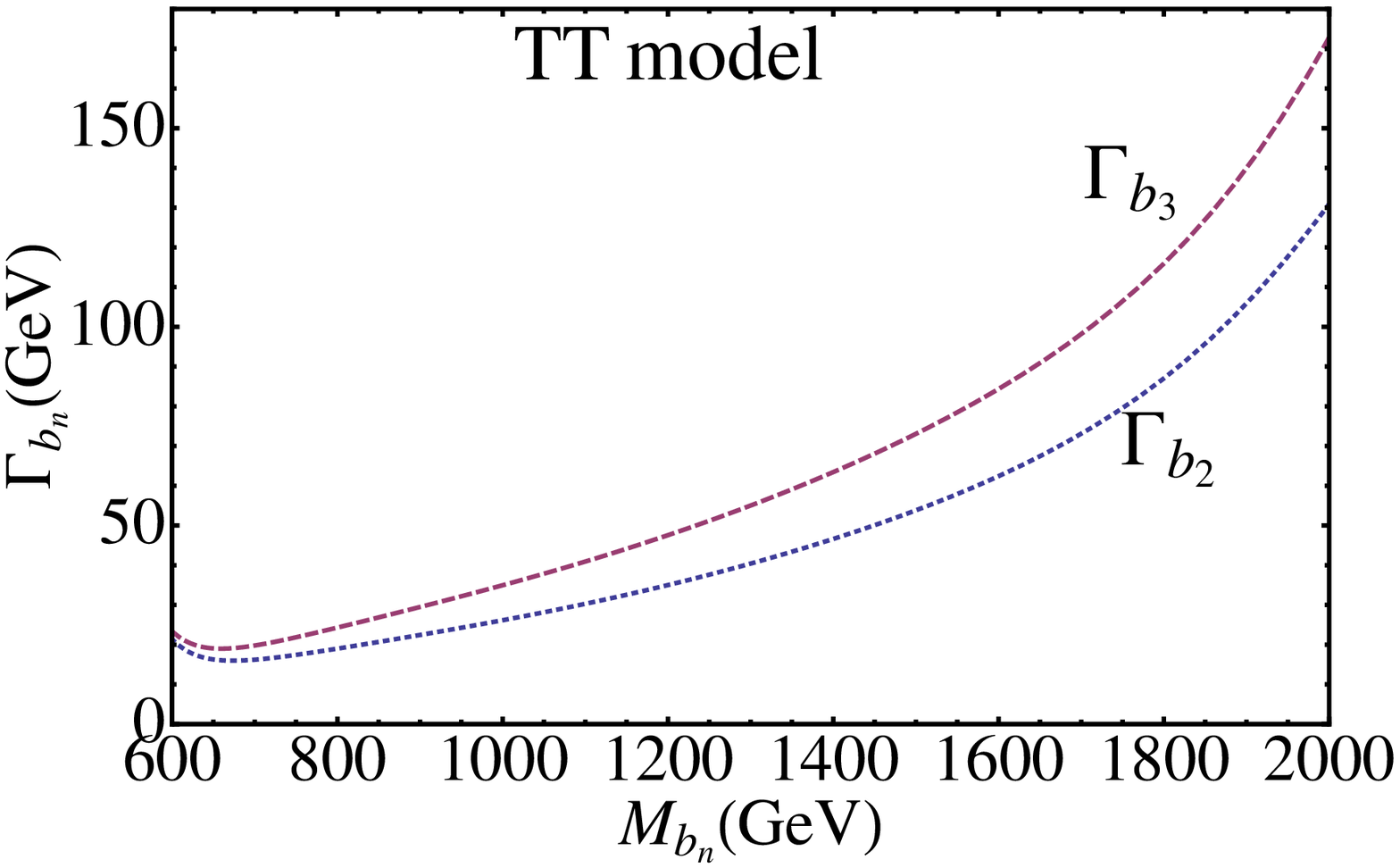}
\includegraphics[width=0.49\textwidth]{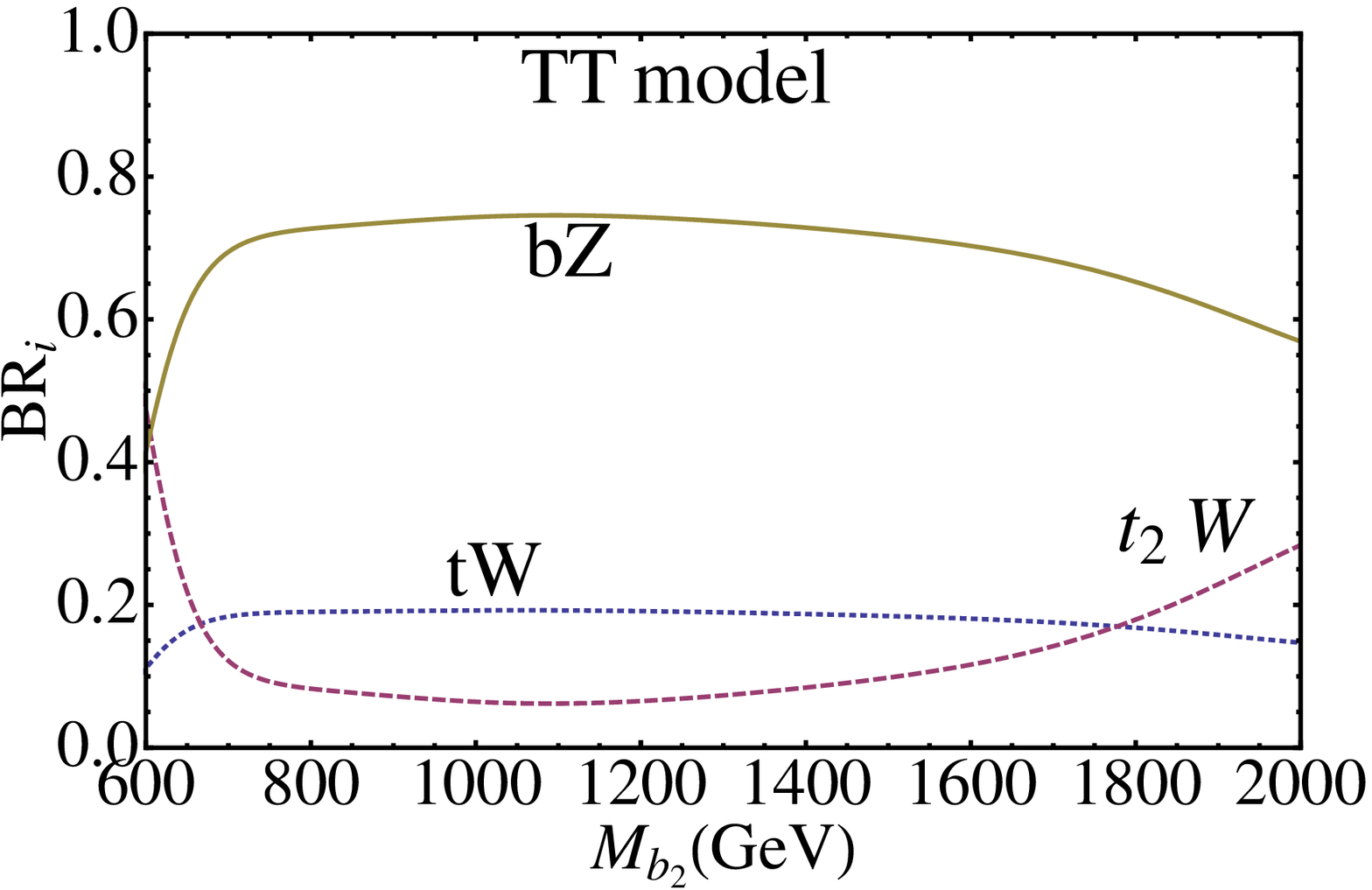}
\includegraphics[width=0.49\textwidth]{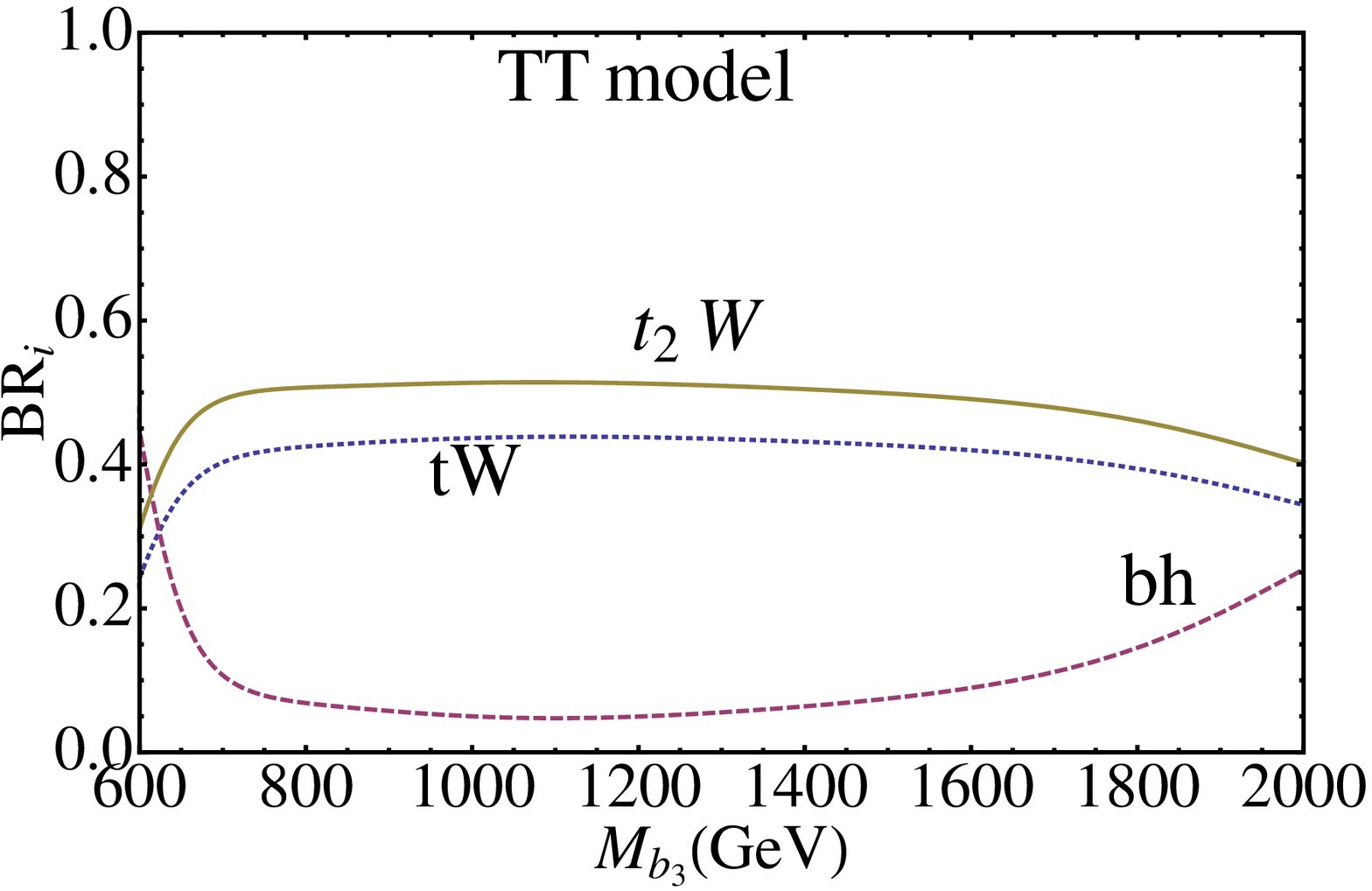}
\caption{Total decay width (left) and branching ratios of $b_2$ (center) and $b_3$ (right) as functions of their masses in the model with 
$Zb\bar b$ protection for the TT model. 
\label{b2TWBR_TT.FIG}}
\end{center}
\end{figure}

In the warped model without custodial protection of the $Zb\bar b$ coupling, presented in Sec.~\ref{ThNoZbb.SEC}
(DT model), the new vectorlike fermions are the $b'$ and $t'$.
We first focus on the on the $b'$ here, and will present the $t'$ decay width and BRs in the context of the
ST and TT models later.    
In Fig.~\ref{b2TWBR_DT.FIG} we show the total decay width (left) and BRs (right)
as functions of $M_{b_2}$ for the $b'$ in the DT model. 
The total width is a few percent of the mass.  
Its roughly linear dependence on $M_{b_2}$ can be understood by noting from 
Eqs.~(\ref{Gamq2q1V.EQ})~and~(\ref{Gamq2q1h.EQ}) that (in the
large $M_{b_2}$ limit) $s_\theta^L \propto 1/M_{b_2}, \ c_\theta^L \approx 1$, 
leaving a $\Gamma_i \sim M_{b_2}$ behavior for all the partial widths. 
All three modes have comparable branching ratios. For the $t W$ channel, for $M_{b_2}$ not too much
bigger than $m_t$, the phase space suppression due to the large top mass is significant, 
but is overcome for large $M_{b_2}$.
The $b Z$ and $b h$ BR curves are quite similar, particularly for large $M_{b_2}$,  
since, neglecting the (small) $x_b$, the $b_2 b_1 Z$ and $b_2 b_1 h$ couplings are proportional to 
$g_Z c^L_\theta s^L_\theta$ and $c^L_\theta \lambda_{Q_L b'_R}$ respectively. 
Since $s^L_\theta \propto \lambda_{Q_L b'_R}$ and in the 
$b Z$ partial width the factor of $g_Z^2$ cancels against the $1/m_Z^2$, the 
two BRs end up being equal as can be shown using 
Eq.~(\ref{Gamq2q1V.EQ}). 
In Table~\ref{bpBR.TAB} we give the $b_2$ branching ratio for each of the three channels as a function of its mass
in the DT model. 
\begin{table}
\centering
\begin{tabular}{|c|c|c|c|c|c|c|c|} \hline
$M_{b_2}$ (GeV) &  500  &  750  & 1000  & 1250  & 1500  & 1750 & 2000\\ \hline
$b_2\to t_1W$   & 0.452 & 0.480 & 0.489 & 0.493 & 0.495 & 0.497 & 0.498\\ %\hline
$b_2\to b_1Z$   & 0.292 & 0.269 & 0.261 & 0.257 & 0.255 & 0.254 & 0.253\\ %\hline
$b_2\to b_1h$   & 0.255 & 0.251 & 0.250 & 0.250 & 0.249 & 0.249 & 0.249\\ \hline
\end{tabular}
\caption{$b_2$ branching ratios for the warped-space DT model.
\label{bpBR.TAB}}
\end{table}
In Fig.~\ref{b2TWBR_TT.FIG} we show the total decay width and BRs of the $b_2$ and $b_3$ for the TT model. 
An additional decay mode $b_2 \to t_2 W$ opens up at large $M_{b_2}$. 
Since this BR is not too big for the masses of interest, we do not consider this mode further.  

\begin{figure}[!h]
\begin{center}
\includegraphics[width=0.49\textwidth]{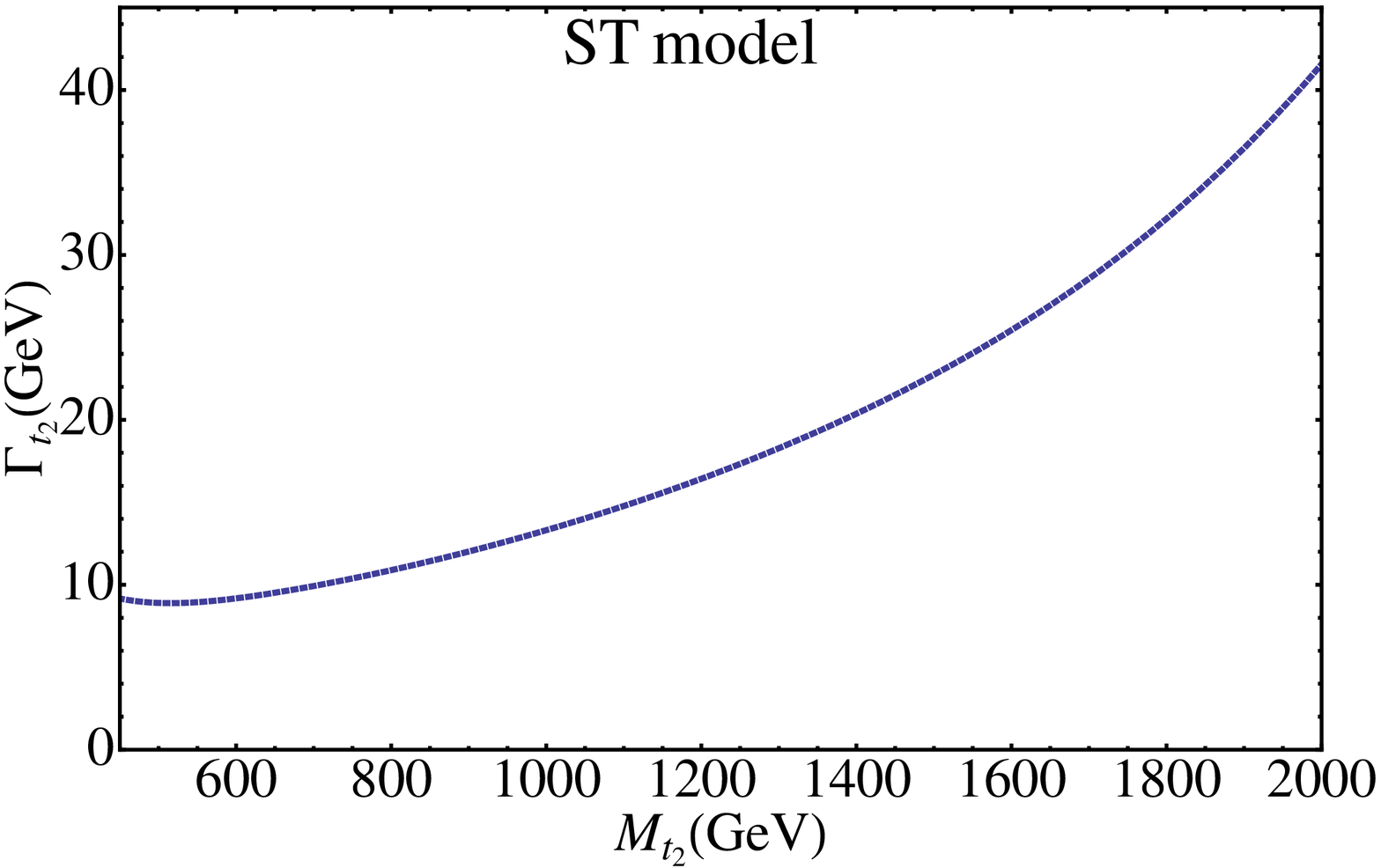}
\includegraphics[width=0.5\textwidth]{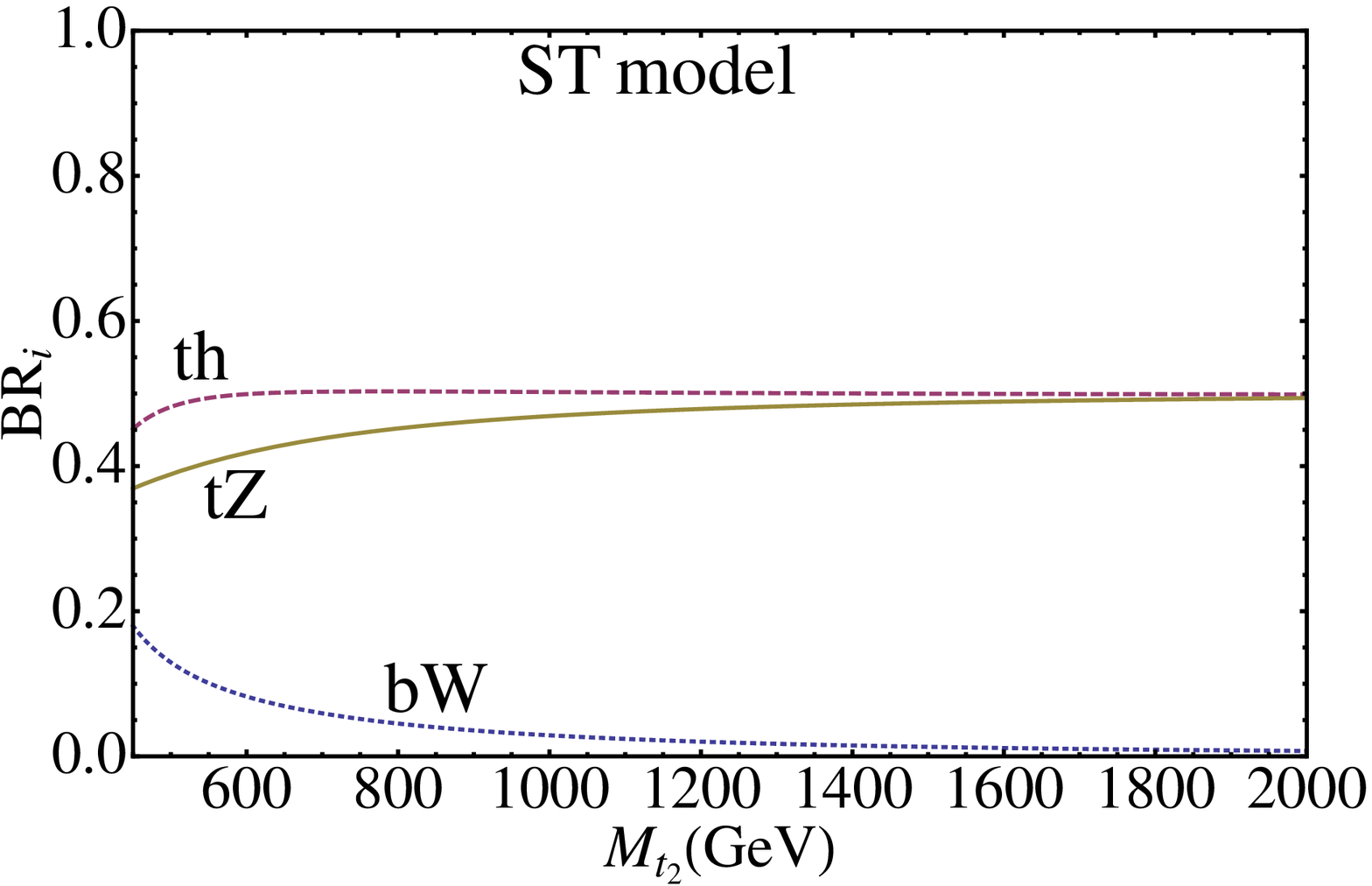}
\caption{Total decay width and branching ratios of $t_2$  as functions of $M_{t_2}$ in the model with $Zb\bar b$ protection 
for the ST model. 
\label{t2TWBR_ST.FIG}}
\end{center}
\end{figure}

\begin{figure}[!h]
\begin{center}
\includegraphics[width=0.49\textwidth]{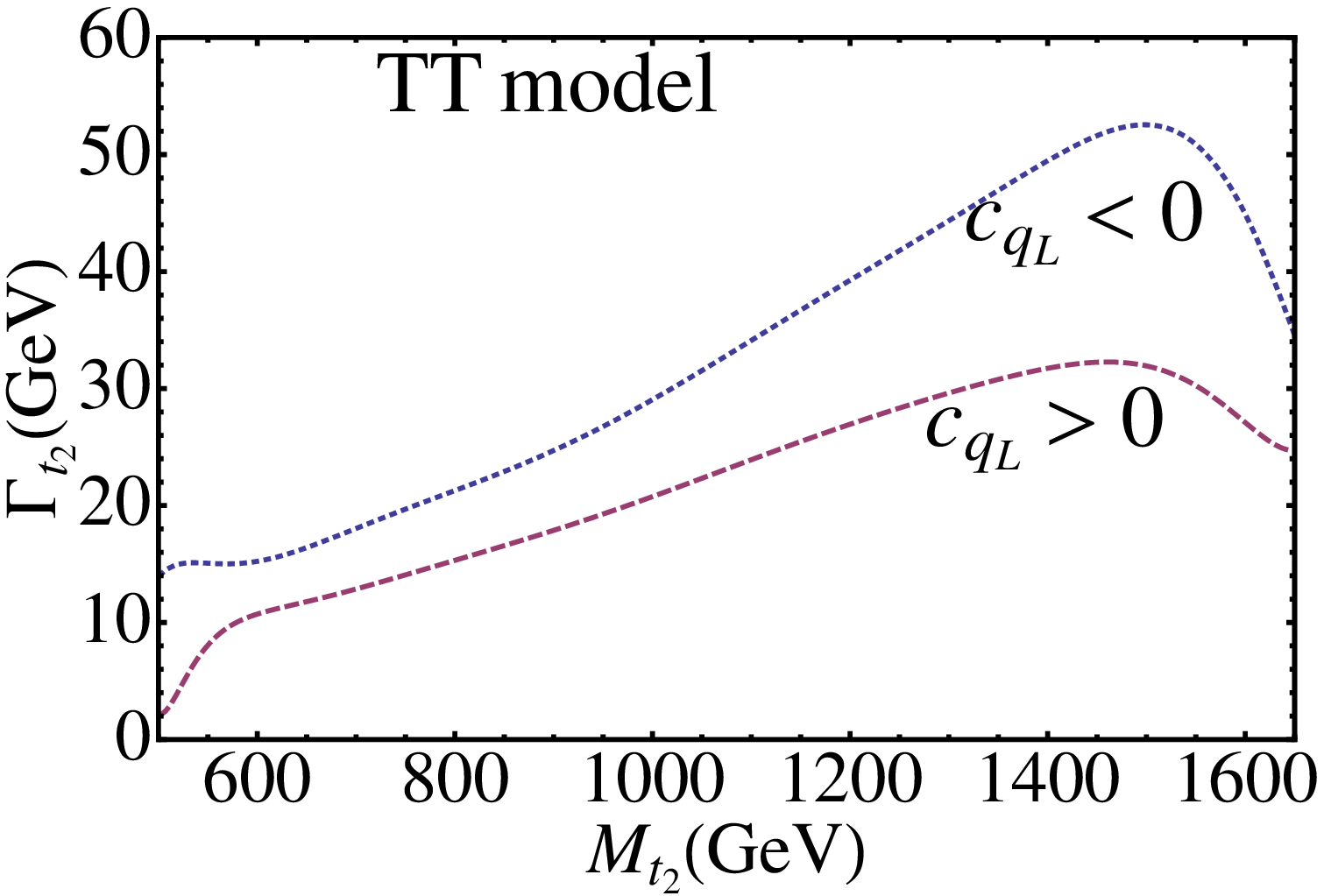}
\includegraphics[width=0.49\textwidth]{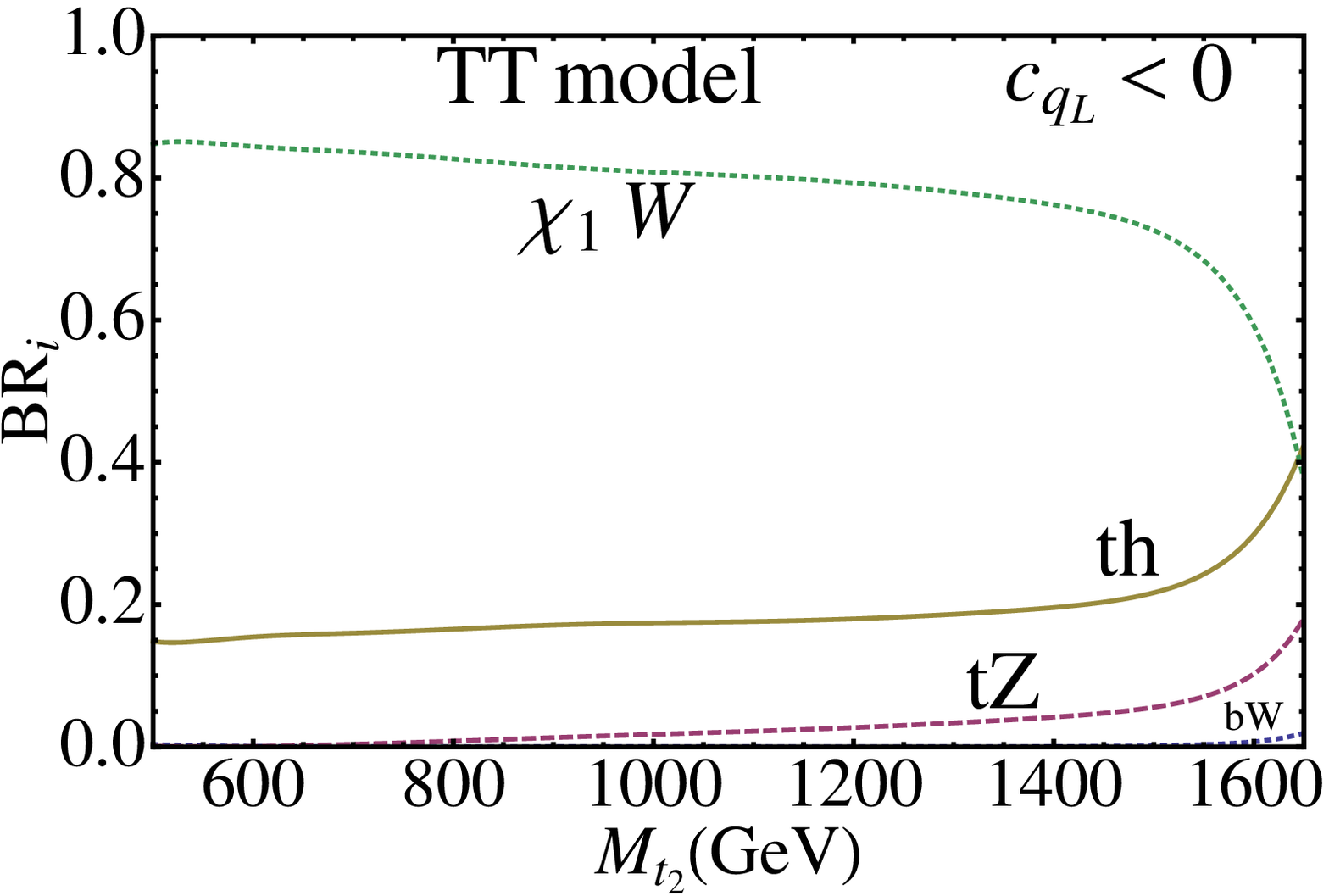}
\includegraphics[width=0.49\textwidth]{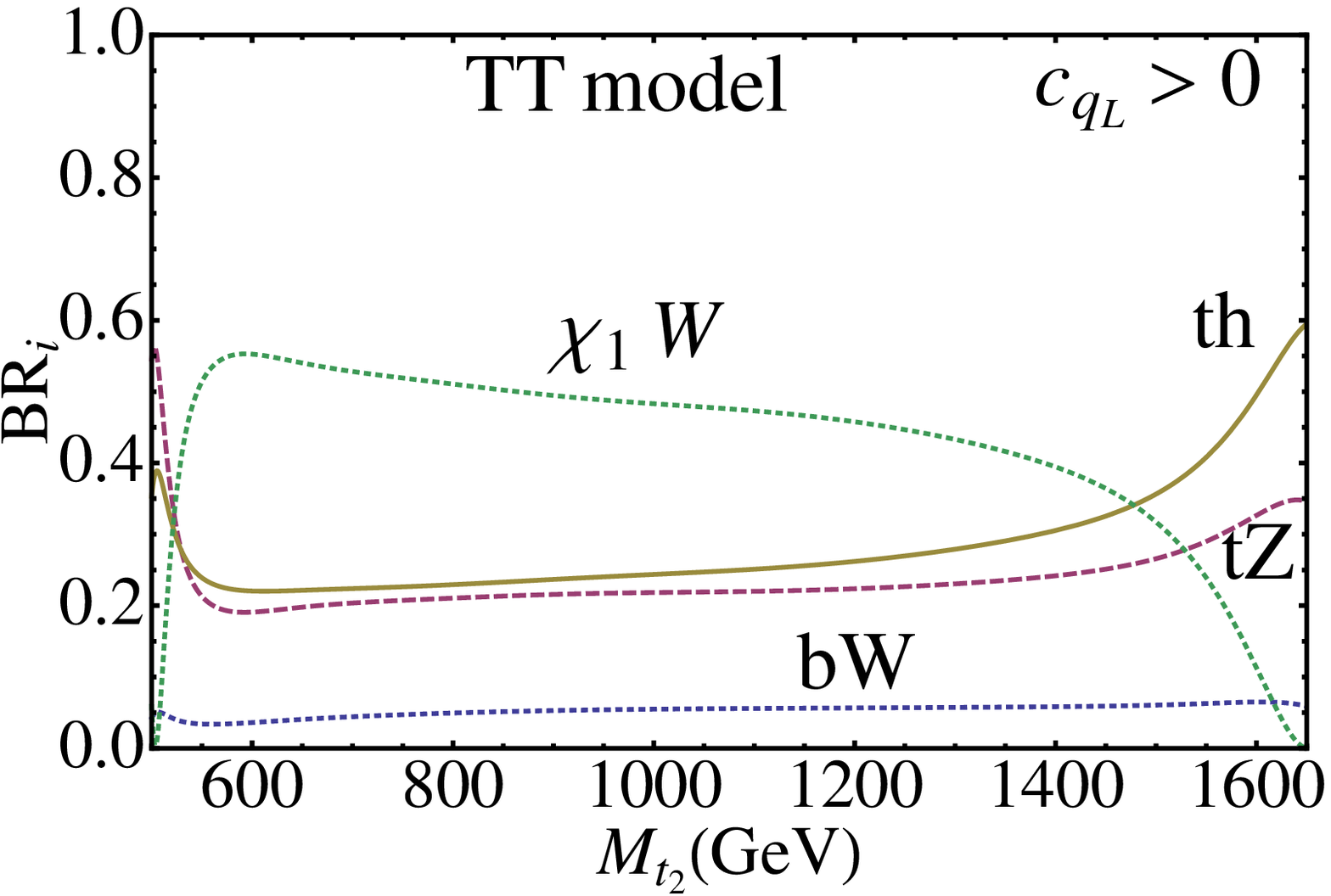}
\caption{Total decay width and branching ratios of $t_2$  as functions of $M_{t_2}$ in the model with $Zb\bar b$ protection 
for the TT model. 
For the TT model, we show different plots for $c_{q_L} < 0$ and $c_{q_L} > 0$ becuase $M_{t_2}$ is two-fold degenerate 
for different $c_{q_L}$ as can be seen from Fig~\ref{cqL_Mtp.FIG}. 
\label{t2TWBR_TT.FIG}}
\end{center}
\end{figure}

In Fig.~\ref{t2TWBR_ST.FIG} we present the $t_2$ decay width and branching ratio for the ST model, 
and in Fig.~\ref{t2TWBR_TT.FIG} for the TT model. 
We notice that the $t_2\to bW$ decay width becomes small at large $m_2$.  
The reason for this is that there is no $Tb\phi^{+}$ coupling in Eq.~(\ref{tR1-LYuk.EQ}) and it will be generated after mixing as a 
$t_2b\phi^{+}$ term. This is of $\mathcal{O}(\tilde{x})$ and is negligible in the large $m_2$ limit. 
In the TT model, the additional decay mode $t_2 \to \chi_1 W$ is present, and ends up being the 
dominant decay mode. 
The reason for this is the large coupling relevant here for the reason mentioned in Sec.~\ref{tpParamCoup.SEC}.
For the TT model, we show different plots for $c_{q_L} < 0$ and $c_{q_L} > 0$,
because $M_{t_2}$ is two-fold degenerate for different $c_{q_L}$ as can be seeen from Fig~\ref{cqL_Mtp.FIG}. 
For $c_{q_L} < 0$ the $t_2 \to tZ$ BR is quite small while for $c_{q_L} > 0$ it increases to about 0.2.

\begin{figure}[!h]
\begin{center}
\includegraphics[width=0.49\textwidth]{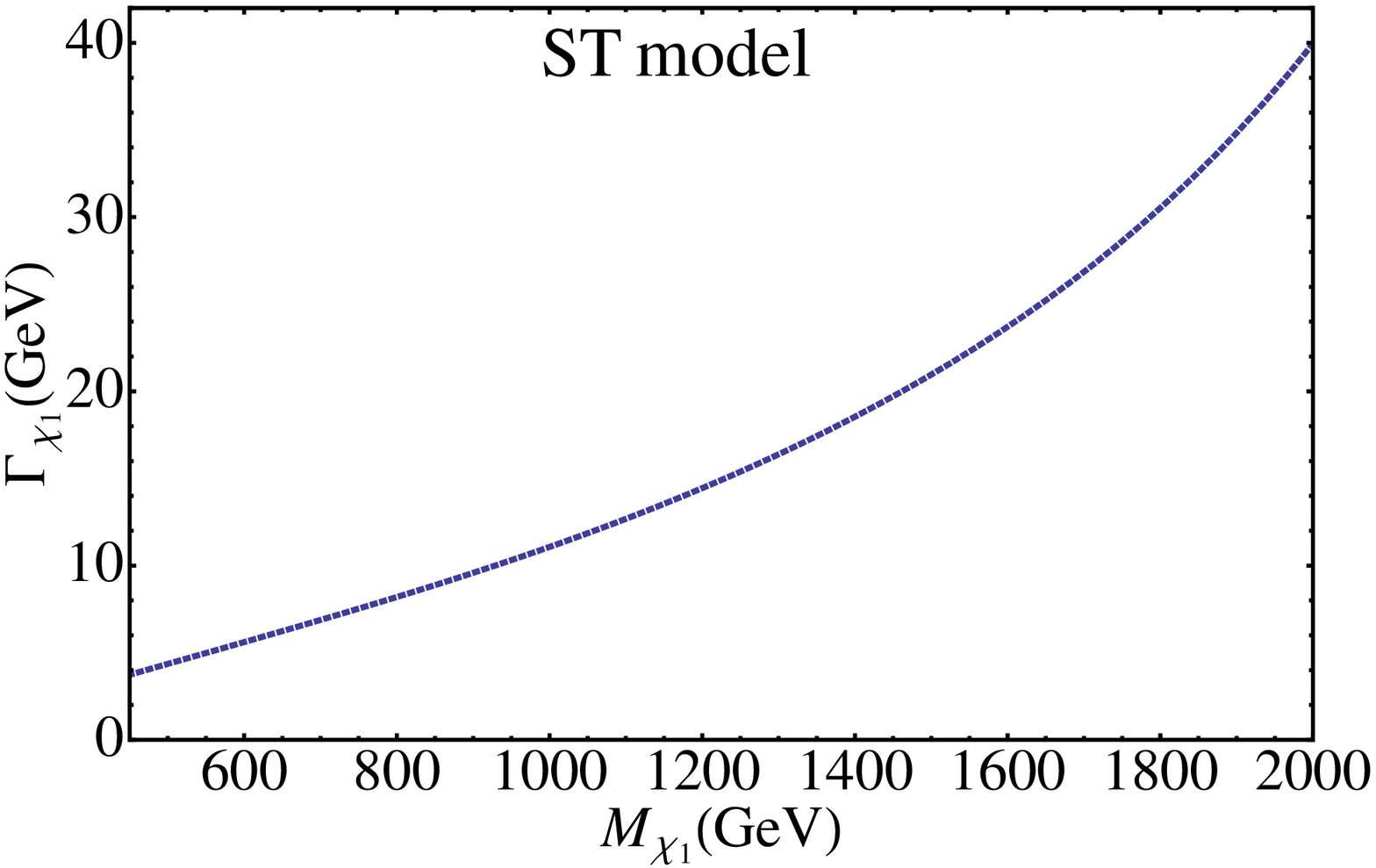}
\includegraphics[width=0.45\textwidth]{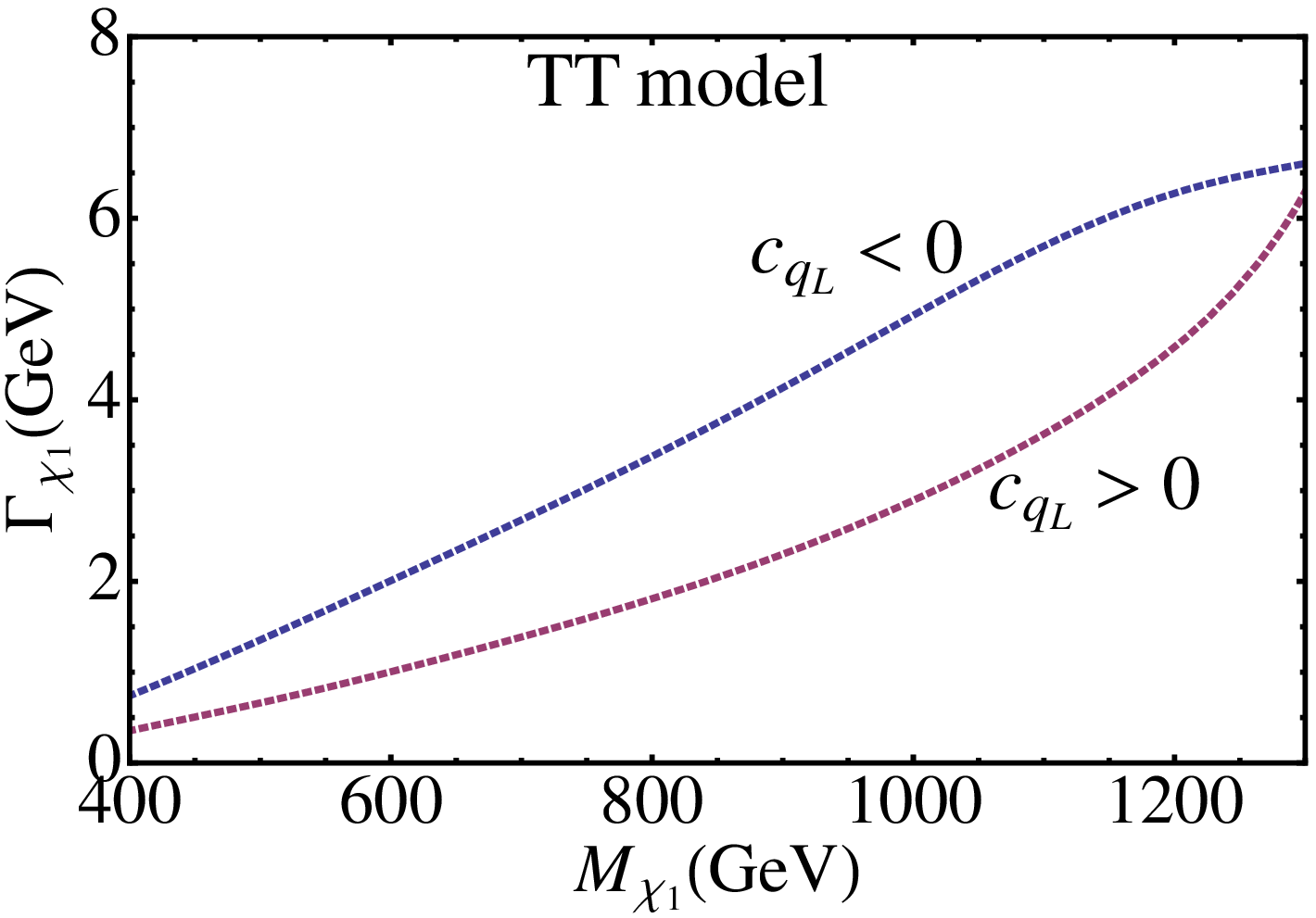}
\caption{Total decay widths of $\chi_1$  as functions of $M_{\chi_1}$ in the ST and  TT models.  
For the TT model, we show different plots for $c_{q_L} < 0$ and $c_{q_L} > 0$ 
because $M_{\chi_1}$ is two-fold degenerate for different $c_{q_L}$ as can be seen from Fig.~\ref{chimass}. 
\label{XTW_Zbb}}
\end{center}
\end{figure}

In Fig.~\ref{XTW_Zbb} we show the $\chi_1$ total decay width for the ST and TT models. 
The $\chi_1$ BR is 100~\% into the $tW$ mode as this is the only channel accessible. For the TT model, we show different plots for $c_{q_L} < 0$ and $c_{q_L} > 0$ as, like $M_{t_2}$, $M_{\chi_1}$ also shows degeneracy as a function of $c_{q_L}$ (see Fig.~\ref{chimass}). 
In the TT model, the additional decay mode $\chi_2 \to \chi_1 Z$ is present, and ends up being the 
dominant decay mode (with BR about 0.8). 
The reason for this is the large coupling. 
Interestingly, $\chi_2$ has many more decay modes, namely $tW$ (with BR of about 0.2), $\chi_1 Z$, $\chi_1 h$, and $t_2 W$, 
but we do not consider the $\chi_2$ as we expect its production c.s. to be smaller owing to its 
larger mass.

%%%%%%%%%%%%%%%%%%%%%%%%%%%%%
\section{LHC signatures}
\label{LHCSign.SEC}
In this section we study the LHC signatures of the $\chi$ (EM charge 5/3), $t'$ (charge 2/3) and $b'$ (charge -1/3) 
vectorlike quarks.
We present many of our results model-independently and also show specific signatures and the reach 
for the different warped models detailed in Sec.~\ref{ThFrmWrk.SEC},
namely, the model without custodial protection of $Zb\bar b$ (DT model), 
and the two cases with custodial protection, singlet $t_R$ (ST model) and triplet $t_R$ (TT model).
The warped model parameter choices we use for our numerical studies are given in Sec.~\ref{parCoup.SEC}. 

Generally, at the LHC, the dominant production channel of these quarks is their pair production. However in this paper,
in addition to the pair productions, we also look into some of their important single production
channels. The single production channels can give useful information about model dependent weak coupling
parameters and thus, help us to identify the underlying model at colliders. 
Single production can also have less complications from combinatorics compared to pair-production.
Moreover, in general, depending on the coupling, some single production channel can
even be the dominant production channel if the vectorlike quark is too heavy due to the phase-space suppression 
in pair-production.
For instance, for electroweak size couplings, the single production starts to dominate for masses roughly 
above $700$~GeV. 

Due to mixing of the SM top and bottom quarks with the $t'$ and $b'$ respectively, $V_{tb}$ can be shifted. 
The current measured value of $|V_{tb}|$ from the direct measurement of the single top production cross section at the 
Tevatron with $\sqrt{s}=1.96$ TeV is $|V_{tb}|=0.88\pm 0.07$ 
with a limit~\cite{Beringer:1900zz} of $|V_{tb}|>0.77$ at the 95\% C.L. assuming a top quark mass $m_t = 170$ GeV. 
While presenting the results for the warped models, the parameters we use 
for numerical computations satisfy the above $|V_{tb}|$ constraint.

For each of the $\chi$, $t'$, and $b'$ we identify promising pair and single production channels, 
compute the signal cross-section and dominant SM backgrounds, and compute the luminosity required 
($\mathcal L_5$) for 5$\sigma$ significance, {\it i.e.} $S/\sqrt{B} = 5$, 
and additionally ($\mathcal L_{10}$) for obtaining 10 signal events. 
We take the larger of $\mathcal L_5$ and $\mathcal L_{10}$ as the luminosity for discovery. 

We have implemented the warped model Lagrangian in FeynRules version~1.6.0~\cite{Christensen:2008py} 
and generated the model files for the Monte-Carlo event-generator MadGraph5~\cite{Alwall:2011uj}, 
using which we obtain the signal cross-sections. 
We use CTEQ6L1 Parton Distribution Functions (PDFs)~\cite{Pumplin:2002vw}. 
We perform a patron-level study, and do not include hadronization and detector resolution effects in this first level of study. 

%%%%%%%%%%%%    
\subsection{$\chi$ LHC Signatures}
\label{ChiLHCSign.SEC}

\begin{figure}[!h]
\centering
\subfloat{
\begin{tabular}{ccc}
\resizebox{40mm}{!}{\includegraphics{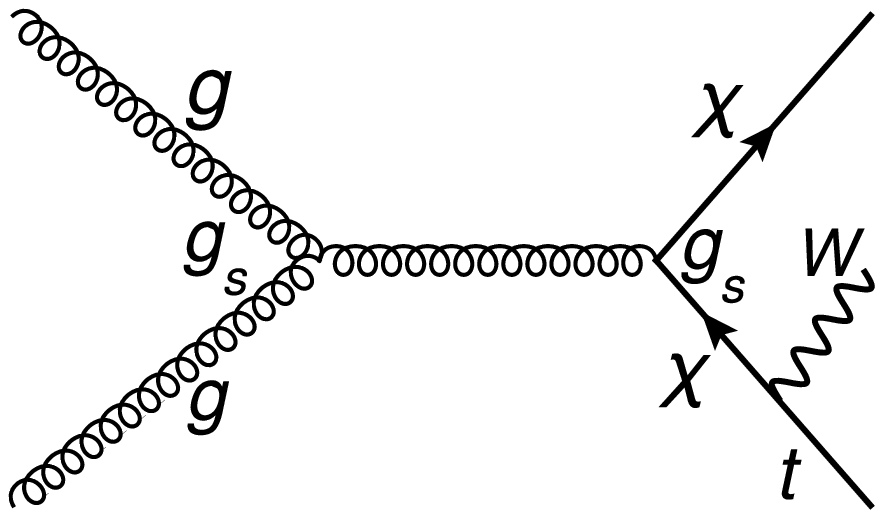}} &&
\resizebox{40mm}{!}{\includegraphics{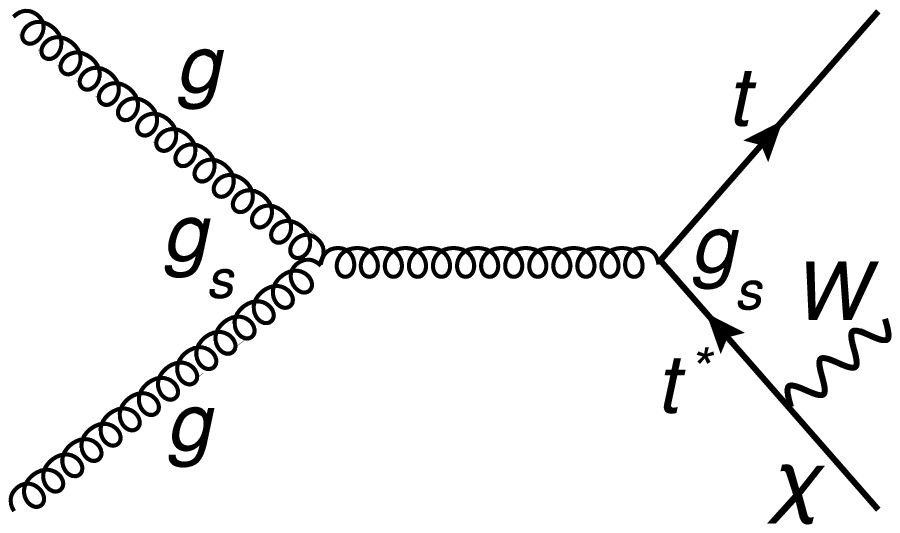}} \\
\footnotesize{(a)}&&\footnotesize{(b)}
\end{tabular}}\\
\caption{Sample Feynman diagrams contributing to the $pp\to\chi_1 tW $ process. In (a) when both the $\chi$'s are on-shell, we have a DR contribution, while when
one of them is off-shell we have the SR process; the other contribution to SR comes from strict single production diagrams like the one shown in (b).
\label{gg2XtW.FIG}}
\end{figure}

We assume that the only decay is $\chi\to t W$, which is the case in many BSM scenarios. 
We parametrize the $\chi$ couplings model-independently as shown in Eq.~(\ref{kapChiMI.EQ}).    
At the LHC, we consider the $\chi t W$ production process as we find this to be the dominant $\chi$ production channel. 
As shown in Fig.~\ref{gg2XtW.FIG}, this includes 
(i) the double resonant (DR) pair-production $\chi_1 \bar\chi_1$ (both on-shell) followed by the decay of one of the on-shell
$\chi$ to $tW$, and, 
(ii) the single resonant (SR) channel including $\chi_1 \bar\chi_1^*$ (one of the $\chi$ off-shell), and in addition,
the strict single-production of $\chi_1$ shown in (b). 
\vspace{0.5cm}
We include both DR and SR and focus on the channel  
\begin{eqnarray}
pp \rightarrow \chi_1 tW \rightarrow tWtW \rightarrow tWt\ell\nu\,.
\end{eqnarray}
If the $W$'s (including the ones coming from the tops) decay hadronically, then the signature for this channel would be 
$bb\ell E\!\!\!/_{T}+\textrm{jets}$. If the tops can be reconstructed, then 
the main SM background for this signature would be $pp\to tt+\textrm{jets}$, $ttV+\textrm{jets}$,
$ttVV+\textrm{jets}$ (where $V=\{W,Z\}$), $tth+\textrm{jets}$ etc. In addition to
the tops, if the hadronically decaying $W$ is also reconstructed, then $pp\to tWtW$ becomes the
dominant background. Therefore, for the background we consider the SM process $pp \rightarrow tWtW \rightarrow tWt\ell\nu$.
We consider it at the $tWtW$ level keeping in mind that the top-jets
can be tagged with high efficiency using advanced top-tagging algorithms. We have discussed
this issue in~\ref{chiSophist.APP} in more detail.  
We obtain the signal and background cross-sections at the $ttW\ell\nu$ level
{\it i.e.}, only one $W$ decays leptonically. We perform our analysis at this
level because for the signal we expect the lepton coming from the $W$ to have large $p_T$,
whereas it is less probable for the background to have a high $p_T$ lepton. This 
feature of the lepton can be used to isolate the signal from the background. The lepton can be used as a trigger. 
We consider the $bb~6j~\ell E\!\!\!/_{T}$ final state where $j$ includes only ``light'' jets ($u$, $d$, $c$, $s$) 
and $\ell$ includes $e$ and $\mu$. 
From the $tWt\ell\nu$ level cross-section, we compute the rate for the final-state of interest 
by multiplying with appropriate branching ratios. 

In order to select the signal while suppressing the background, 
we apply the following ``basic'' and ``discovery'' cuts and 
present the signal and the background cross sections 
in Table \ref{chipro} (Table \ref{chipro8}) for the 14 TeV (8 TeV) LHC: 
\begin{enumerate}
\item {\bf Basic}
\begin{enumerate}
\item
$|y(\ell)| \leq 2.5$
\item
$p_T(\ell)\geq 10$ GeV
\end{enumerate}
\item {\bf Discovery}
\begin{enumerate}
\item
$|y(\ell)| \leq 2.5$
\item
$p_T(\ell) \geq 125$ GeV
\item
$p_T(W) \geq 250$ GeV.
\end{enumerate}
\end{enumerate}
The second set of cuts is chosen to optimize the signal over background ratio. It is our ``discovery cut'' motivated by the fact that in the signal,
there are two high-$p_T$ $W$'s present at the $ttWW$ level and one of them decays to a high-$p_T$ lepton.
To account for the various efficiencies we
multiply both signal and background cross sections with a factor  
\begin{eqnarray}
\label{factors}
\eta_{\chi_1} = (\epsilon^b_{tag})^2\times(\epsilon^W_{rec})^3\times {(\epsilon^t_{rec})}^2\times (BR_{W\rightarrow jj})^3\approx 0.082 \ ,
\end{eqnarray}
where $\epsilon^b_{tag}$ is the $b$-tagging efficiency, $\epsilon^W_{rec}$ is the $W$ reconstruction efficiency from $jj$,
$\epsilon^t_{rec}$ is the $t$ reconstruction efficiency from $bW$. Combinatorics might be an important issue for reconstruction but at our level of analysis we ignore this complication.  We take $\epsilon^b_{tag}=0.5$, $\epsilon^t_{rec}=1$,
$\epsilon^W_{rec}=1$ and $W\rightarrow jj$ branching ratio $BR_{W\rightarrow jj}=0.69$.
As explained earlier, we then compute $\mathcal L_5$ for 5$\sigma$ significance
and $\mathcal L_{10}$ for obtaining 10 signal events, 
and the larger of $\mathcal L_5$ and $\mathcal L_{10}$ is the discovery luminosity.
In~\ref{chiSophist.APP} we present a more sophisticated analysis by including additional 2-jets background
and identify cuts that can bring them under control without sacrificing the signal much. 
The $\kappa$ can be probed by isolating the SR contribution.
Typically, for the range of the coupling arising in warped models, 
the contribution of the second type of diagrams shown in Fig.~\ref{gg2XtW.FIG}(b) to the total cross-section is very small.
This means the $tW$ pair in the SR production of $\chi_1$ is 
dominantly coming from an off-shell heavy quark - $\chi_1^{*}$.
At the $\chi_1tW$ level we isolate the SR contribution by applying only the kinematical cut on the 
invariant mass $M(tW)$,
\begin{equation}
|M(tW)-M_{\chi_1}| \geq \alpha_{cut} M_{\chi_1} ; \; \alpha_{cut} = 0.05,
\label{invmasscut}
\end{equation}
which ensures that the $t$ quark and the $W$ do not reconstruct to an on-shell $\chi_1$, 
{\it i.e.} this cut removes the DR contribution. 
To understand why the cross-section after the $\alpha_{cut}$ scales as $\kappa^2$, let us consider the $\kappa$
dependent part of the cross-section (from the type of diagram in Fig.~\ref{gg2XtW.FIG}(a)),
\begin{equation}
\sigma_{\chi_1tW} \propto \frac{\kappa^2}{(p^2-M_{\chi_1}^2)^2+\Gamma_{\chi_1}^2M_{\chi_1}^2} \,,
\end{equation}
where $p$ is the momentum carried by the internal $\chi_1$.
The cut of Eq.~(\ref{invmasscut}) is chosen such that $|p^2-M_{\chi_1}^2|$ 
dominates over $\Gamma_{\chi_1}M_{\chi_1}$, and one can neglect 
$\Gamma_{\chi_1}M_{\chi_1}$ compared to $|p^2-M_{\chi_1}^2|$,  
ensuring that $\sigma_{\chi_1tW}$ scales as $\kappa^2$.
\begin{figure}[!h]
\begin{center}
\includegraphics[width=0.55\textwidth]{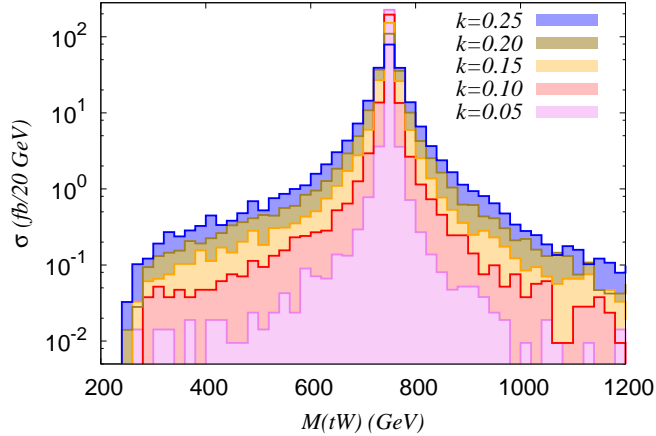}
\caption{The $tW$ invariant mass distributions for the $pp \to \chi_1 tW$ process for different $\kappa_{\chi_{1R}t_{1R}W}$ (denoted as $\kappa$), for 
$M_{\chi_1}=750$~GeV at the 14~TeV LHC.
\label{Mtw}}
\end{center}
\end{figure}
\begin{table}[!h]
\centering
\begin{tabular}{|c|c|c|}
\hline $\kappa_{\chi_{1R}t_{1R}W}$ & $\sigma(pp\to \chi_1 tW)$ & $\sigma(pp\to \chi_1 tW)$ \\ 
  & (fb) before cut & (fb) after cut \\ 
\hline 0.05 & 239.37 & 4.945 \\ 
\hline 0.10 & 238.91 & 21.09 \\ 
\hline 0.15 & 236.31 & 45.92 \\ 
\hline 0.20 & 233.52 & 79.71 \\ 
\hline 0.25 & 229.40 & 118.71 \\
\hline 
\end{tabular}
\caption{Scaling behavior of $pp \to \chi_1 tW$ single production cross-sections after the invariant mass cut defined in Eq.~(\ref{invmasscut}), for $M_{\chi_1}=750$ GeV at the 14 TeV LHC. 
\label{scaling}} 
\end{table}
In Fig.~\ref{Mtw} we show the $tW$ invariant mass distribution for the $pp \to \chi_1 tW$ process, and in Table~\ref{scaling} the cross section before and after the $\alpha_{cut}$. 
We observe that the total cross section before the cut is almost constant but decrease slightly with increasing 
$\kappa$ due to finite width effects.
The contribution in the off-shell region increases with $\kappa$ since the total width grows as $\kappa^2$ which makes the Breit-Wigner distributions wider.\footnote{A 
similar plot for a $b'$ is shown in Ref.~\cite{{Coleppa:2014qja}}.}
This is seen more quantitatively in Table~\ref{scaling}, where the cross section after the $\alpha_{cut}$ scales as $\kappa^2$ (for $\kappa$ not too large). 
%The $pp\to \chi_1tW$ cross section is mainly due to DR contribution which decreases slightly for larger
%$\kappa$ mainly because of the large width.  
As $\kappa$ increases, 
the $\sigma_{\chi_1tW}$ value after the $\alpha_{cut}$ cannot increase arbitrarily as it remains bounded by the 
$\sigma_{\chi_1tW}$ value before the cut.
This can be seen by keeping in mind that $\Gamma_{\chi_1}$ depends on $\kappa$, 
and from the fact that for a fixed value of $\alpha_{cut}$, 
the $\kappa^2$ scaling behavior of $\sigma$ breaks down as $\Gamma_{\chi_1}$ increases with increasing $\kappa$
and at some point the $\Gamma_{\chi_1}M_{\chi_1}$ term in the denominator starts dominating again.
Therefore, to be sensitive to $\sigma_{SR}$, the choice of $\alpha_{cut}$ is crucial 
(see Ref.~\cite{Coleppa:2014qja} for more details). 
%It is dictated by the fact that we expect $\sigma_{SR} $ to scale as 
%$\kappa^2_{\chi_1tW}$ whereas $\sigma_{DR}$ is set by $g_s$. 
Taking $\alpha_{cut}$ too small will spoil the scaling because of the contamination 
from the pair production, but it should not be too large either as that  will make the cross-section very small. 
%In Table~\ref{scaling} we explicitly demonstrate that our choice of $\alpha_{cut}$ retains the $\kappa^2_{\chi_1tW}$ scaling.
%
\begin{table}[t]
\centering
\begin{tabular}{|c|c|c|c|c|c|c|c|} \hline
\multicolumn{1}{|c|}{$\mathcal X$} &\multicolumn{1}{|c|}{$M_{\chi}$} & \multicolumn{1}{|c|}{$\sigma_{tot}$} & \multicolumn{1}{|c|}{$\sigma_{SR}$} & \multicolumn{1}{|c|}{cuts} & \multicolumn{1}{|c|}{S} & \multicolumn{1}{c|}{BG} & \multicolumn{1}{|c|}{$\mathcal{L}$} \\ 
&\multicolumn{1}{|c|}{(GeV)} & \multicolumn{1}{|c|}{$(fb)$} & \multicolumn{1}{|c|}{$(fb)$} & \multicolumn{1}{|c|}{} & \multicolumn{1}{|c|}{$(fb)$} & \multicolumn{1}{c|}{$(fb)$} & \multicolumn{1}{|c|}{$(fb^{-1})$} \\ \cline{1-8} 
$\mathcal X_1$& 500  & 2566  & 261.5  & Basic  & 977.5  & 3.257  & -        \\ \cline{5-8} 
   & &       &        & Disc.  & 146.1  & 0.115  & 0.826    \\ \hline
$\mathcal X_2$&750  & 260.0 & 29.31  & Basic  & 99.99  & 3.257  & -        \\ \cline{5-8} 
    & &       &        & Disc.  & 42.74  & 0.115  & 2.824    \\ \hline
$\mathcal X_3$&1000 & 46.47 & 5.198  & Basic  & 17.92  & 3.257  & -        \\ \cline{5-8} 
    & &       &        & Disc.  & 11.36  & 0.115  & 10.63    \\ \hline
$\mathcal X_4$&1250 & 11.22 & 1.231  & Basic  & 4.305  & 3.257  & -        \\ \cline{5-8} 
     &&       &        & Disc.  & 3.226  & 0.115  & 37.42    \\ \hline     
$\mathcal X_5$&1500 & 3.242 & 0.364  & Basic  & 1.235  & 3.257  & -        \\ \cline{5-8} 
     &&       &        & Disc.  & 1.010  & 0.115  & 119.5    \\ \hline
$\mathcal X_6$&1750 & 1.040 & 0.121  & Basic  & 0.393  & 3.257  & -        \\ \cline{5-8} 
     &&       &        & Disc.  & 0.339  & 0.115  & 355.8    \\ \hline	 
\end{tabular}
\caption{Signal (S) and background (BG) cross sections (in $fb$) for $pp \rightarrow \chi tW\rightarrow ttW\ell\nu$ channel
at the 14 TeV LHC for the ST model. For the BG we have considered $pp \to ttW\ell\nu$ process within the SM. The $\mathcal X_i$'s correspond to the parameter sets detailed in Table \ref{chiSTpara}. The luminosity requirement ($\mathcal{L}$) is computed using $\sigma_{tot}$ after including the factor $\eta_{\chi_1}$ defined in Eq.~(\ref{factors}). 
The $\sigma_{tot}$ is computed at the $\chi_1tW$ level with no cut applied. 
$\sigma_{SR}$ is computed at the $\chi tW$ level with only an invariant mass cut applied on 
$tW$ as defined in Eq.~(\ref{invmasscut}). 
\label{chipro}}
\end{table}
\begin{table}[t]
\centering
\begin{tabular}{|c|c|c|c|c|c|c|c|} \hline
\multicolumn{1}{|c|}{$\mathcal X$} &\multicolumn{1}{|c|}{$M_{\chi}$} & \multicolumn{1}{|c|}{$\sigma_{tot}$} & \multicolumn{1}{|c|}{$\sigma_{SR}$} & \multicolumn{1}{|c|}{cuts} & \multicolumn{1}{|c|}{S} & \multicolumn{1}{c|}{BG} & \multicolumn{1}{|c|}{$\mathcal{L}$} \\ 
&\multicolumn{1}{|c|}{(GeV)} & \multicolumn{1}{|c|}{$(fb)$} & \multicolumn{1}{|c|}{$(fb)$} & \multicolumn{1}{|c|}{} & \multicolumn{1}{|c|}{$(fb)$} & \multicolumn{1}{c|}{$(fb)$} & \multicolumn{1}{|c|}{$(fb^{-1})$} \\ \cline{1-8} 
$\mathcal X_1$&500  & 374.2 & 36.63 & Basic & 144.0  & 0.622  & -       \\ \cline{5-8}
	 &&       &       & Disc. & 18.40  & 0.011  & 6.560   \\ \hline
$\mathcal X_2$&750  & 25.61 & 2.741 & Basic & 9.927  & 0.622  & -       \\ \cline{5-8}
	 &&       &       & Disc. & 4.103  & 0.011  & 29.42   \\ \hline
$\mathcal X_3$&1000 & 2.817 & 0.315 & Basic & 1.092  & 0.622  & -       \\ \cline{5-8}
	 &&       &       & Disc. & 0.680  & 0.011  & 177.5   \\ \hline 
$\mathcal X_4$&1250 & 0.381 & 0.042 & Basic & 0.147  & 0.622  & -       \\ \cline{5-8}
	 &&       &       & Disc. & 0.109  & 0.011  & 1105    \\ \hline 	 
\end{tabular}
\caption{Same as in Table~\ref{chipro} for the 8~TeV LHC. 
\label{chipro8}}
\end{table}
In the warped $Zb\bar b$ protected model (ST and TT models), 
the $\kappa$ of Eq.~(\ref{kapChiMI.EQ}) are given in Eqs.~(\ref{tR1-CC.EQ})~and~(\ref{tR3-CC.EQ}) 
and shown in Fig.~\ref{chikappa} and Table~\ref{chiSTpara} respectively. 
For all $M_\chi$ considered here, we find $\mathcal{L}_5 < \mathcal L_{10}$, 
and therefore in Table~\ref{chipro} we present only $\mathcal L_{10}$. 
From Table~\ref{chipro} we find that using $\sigma_{tot}$, {\it i.e.} including both SR and DR, 
the 14 TeV LHC can probe $M_{\chi_1}$ up to 1.5 TeV (1.75 TeV) with 100 $fb^{-1}$ (300 $fb^{-1}$) of integrated luminosity 
for the ST model. 
The numbers in Table~\ref{chipro} show that for the parameter ranges we are interested in, the $pp\to \chi_1tW$ process is dominated 
by the DR production. Hence, we do not display the cross sections and discovery luminosity separately for the TT model 
as the difference between them is only due the SR production (which depends on the $\kappa_{\chi_1tW}$ coupling). 
\begin{figure}[!h]
\begin{center}
\includegraphics[width=0.5\textwidth]{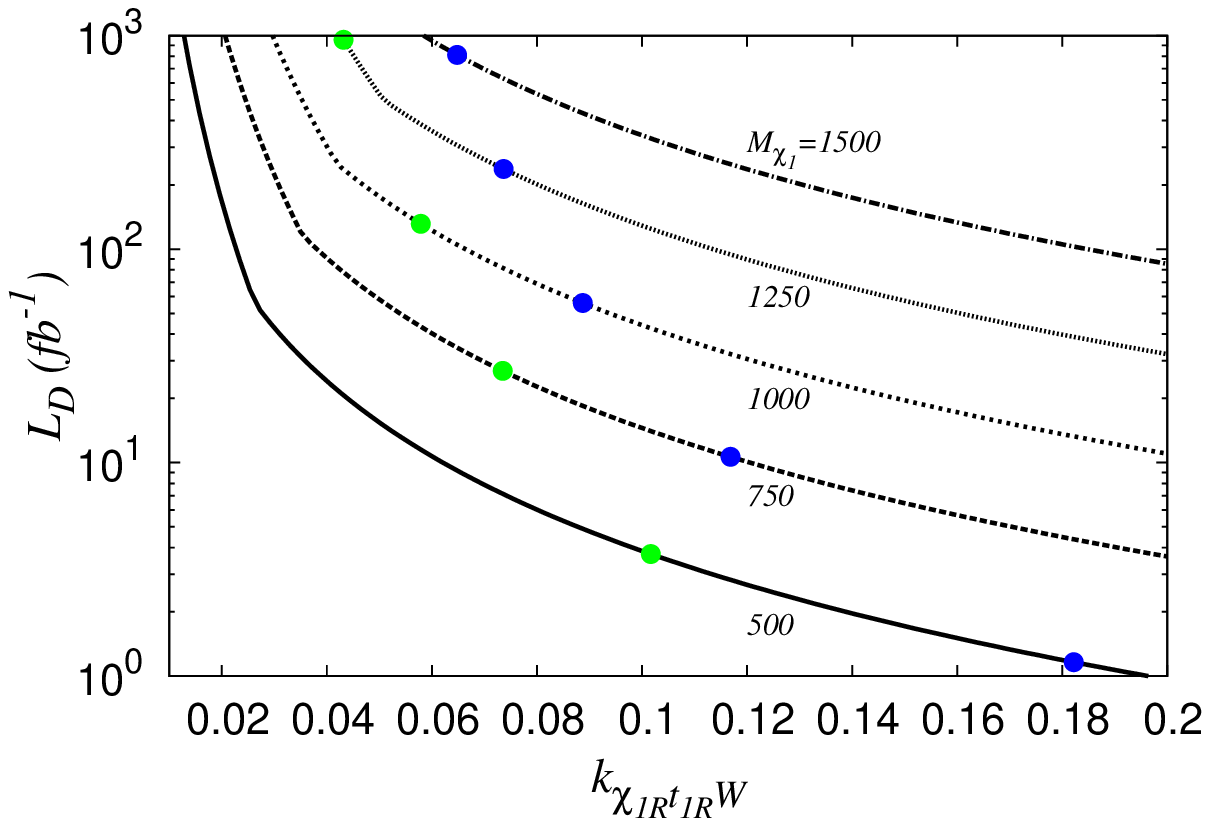}
\caption{Luminosity requirements ($\mathcal{L}_D$, in $fb^{-1}$) for observing the $pp\to\chi_1tW$ single resonant (SR) channel 
as functions of $\kappa_{\chi_{1R}t_{1R}W}$ for different $M_{\chi_{1}}$ (in GeV) at the 14 TeV LHC. 
$\mathcal{L}_D$ is computed after including all BRs and $b$-tagging efficiency. 
The blue and green dots correspond to the ST and TT models respectively. 
\label{lumchi}}
\end{center}
\end{figure}

As mentioned, the $\kappa$ can be probed by isolating the SR contribution.
To present our results model-independently such that it is useful for other models with a $\chi tW$ coupling,
we show in Fig.~\ref{lumchi} the luminosity requirement ($\mathcal{L}_D$) to observe the $pp\to\chi_1tW$ 
SR production process assuming the $\chi_1 \to tW$ BR to be 100\%, where,
$\mathcal{L}_{D} = {\rm Max}(\mathcal{L}_5,\mathcal{L}_{10})$.
%In Fig.~\ref{lumchi} 
The blue and green dots show the reach for the SR process for the warped ST and TT models respectively.
Although we compute $\mathcal{L}_{D}$ at the $\chi t W$ level multiplied by the appropriate BRs, 
with only the invariant mass cut of Eq.~(\ref{invmasscut}), 
we expect that the inclusion of the full decays and the basic and discovery cuts should change
$\mathcal{L}_{D}$ only by a small amount.  
Here we vary $\kappa_{\chi_{1R}t_{1R}W}$ keeping the other coupling $\kappa_{\chi_{1L}t_{1L}W}$ zero
(since this is the case in the ST and TT models).
The plot will look identical if we instead vary $\kappa_{\chi_{1L}t_{1L}W}$ keeping $\kappa_{\chi_{1R}t_{1R}W} = 0$.
The background for the $\chi_1tW$ SR production is computed at the $tWtW$ level after demanding that any one of the $tW$ pair satisfies the cut defined in Eq.~(\ref{invmasscut}). 
This can be expressed as 
\begin{equation}
|M(t_1W_i) - M_{\chi_1}| \geq \alpha_{cut} M_{\chi_1}~~~\textrm{AND}~~~
|M(t_2W_j) - M_{\chi_1}| \leq \alpha_{cut} M_{\chi_1}
\end{equation} 
where $t$'s and $W$'s are $p_T$-ordered and $i,j=\{1,2\}$ with $i\neq j$.
The kinks in the graphs appear because of the
transition from $\mathcal{L}_5$ to $\mathcal{L}_{10}$ along the increasing values of
the coupling. 
For getting the SR reach in the warped model, Tables~\ref{chipro}~and~\ref{chipro8} give the SR cross-section $\sigma_{SR}$ for the ST model. 

Finally, we note that there are other single production channels for $\chi_1$ at the LHC like the $W^\pm$ mediated $pp\to\chi_1t$ or $pp\to \chi_1 tq$ (studied in Ref.~\cite{Contino:2008hi} in the context of composite Higgs models). 
However, unlike the $pp\to\chi_1tW$ process, these are electroweak processes 
due to which we find their cross-sections to be much smaller.
Also, we expect $\sigma(\chi_2 \chi_2) < \sigma(\chi_1 \chi_1)$ due to the larger $M_{\chi_2}$, 
and since already the $\chi_1$ pair-production is signal rate limited, we do not 
explore the $\chi_2$ production and the subsequent $\chi_2 \to \chi_1 h$ or $\chi_2 \to \chi_1 Z$ channels.

%%%%%%%%%%%%%%%%%%%%%%%%%%%%%%%%%%%%%%%%%%%%%%%%%%%%%%%%%%%%%%%%%%%%%%%%%%%%%%%%%%%%
\subsection{$t'$ LHC Signatures}

At the LHC, apart from the usual pair production channel, a charge 2/3 vectorlike $t_2$ can be produced through the following single production channels
\begin{equation}
pp \to t_2W,t_2b,t_2t,t_2bW,t_2tZ, t_2th \ .
\end{equation} 
In models where the $t_2 b W$ coupling is much smaller than the others (as for instance in the warped ST and TT models), 
we can ignore the single production channels $t_2W,t_2b, t_2bW $ channels.
We parametrize the $t_2 t Z$ and $t_2th$ interaction terms model-independently as shown in Eq.~(\ref{t'kapMI.EQ}).

\begin{figure}[!h]
\centering
\subfloat{
\begin{tabular}{ccc}
\resizebox{40mm}{!}{\includegraphics{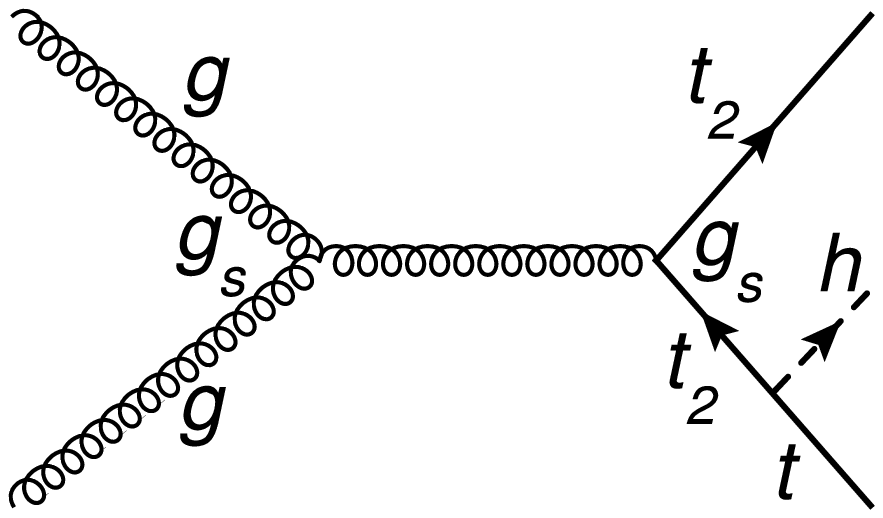}} &&
\resizebox{40mm}{!}{\includegraphics{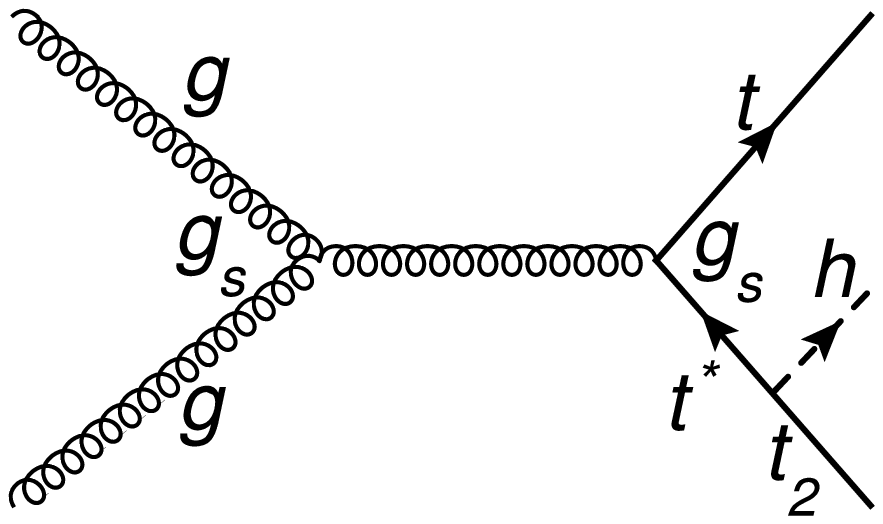}} \\
\footnotesize{(a)}&&\footnotesize{(b)}
\end{tabular}}\\
\caption{Sample Feynman diagrams contributing to the $pp\to t_2 th$ process.  In (a) when both the $\chi$'s are on-shell, we have a DR contribution, while when
one of them is off-shell we have the SR process; the other contribution to SR comes from the strict single production diagrams like the one shown in (b).
\label{t2_single}}
\end{figure}

Similar to the discussion for the $\chi$ in Sec.~\ref{ChiLHCSign.SEC}, here too we identify the 
double resonant (DR) and single resonant (SR) channels, and consider the $thth$ final state.
As shown in Fig.~\ref{t2_single}, this includes 
(i) the double resonant (DR) pair-production $t_2 \bar t_2$ (both on-shell) followed by the decay of one of the on-shell
$t_2 \to th$, 
(ii) the single resonant (SR) channel including $t_2 \bar t_2^*$ (one of the $t_2$ off-shell), and in addition, 
the single-production of $t_2$. 
\vspace{0.5cm}
We therefore include DR and SR and consider the process 
\begin{eqnarray}
pp \rightarrow t_2th\rightarrow thth\rightarrow tbbtbb \ ,
\end{eqnarray}
and focus on the $6~b+4~j$ final-state, where $j$ includes only ``light'' jets ($u, d, c, s$). 
We obtain the signal cross-sections at the $tbbtbb$ level and multiply by appropriate branching ratios 
relevant to the above final state. 
We take the Higgs boson mass to be $125$~GeV in all our computations. 
We assume a $b$-tagging efficiency $\epsilon^b_{tag}=0.5$, and demand only four of the six $b$-jets to be $b$-tagged 
(Ref.~\cite{Girdhar:2012vn} also follows a similar approach) to get a better signal rate.
We require the two top-quarks to be reconstructed from two $b$-tagged jets and four $J$
(where $J$ stands for either a light-jet or an untagged $b$-jet) 
and then the two $h$ to be reconstructed from the remaining two $b$-tagged jets and two $J$.
Here we do not deal with any complications of combinatorics. 
For this channel the main SM background come from $pp\to tt V_h + \textrm{jets}$,
$tt V_h V_h +\textrm{jets}$ (where $V_h=\{W,Z,h\}$) processes.
We compute the background cross-sections at the $ttbbJJ$ level, {\it i.e.} $pp\to ttbbJJ$
process which includes all tree level SM processes leading to $ttbbJJ$ final state. 
%since there could be potentially other sources of background. 
However, due to requiring the four jets to reconstruct to the two $h$ by applying the invariant mass cuts, 
the SM QCD contribution to the $pp\to ttbbJJ$ process becomes negligible and the dominant SM background contribution comes from the $pp\to tthh$ process.
%
%We require a minimum angular separation between any two jets
%\begin{equation}
%\Delta R(ij) = \sqrt{\Delta\phi^2_{ij} + \Delta\eta^2_{ij}}\, , 
%\end{equation}
%where $\phi$ is the azimuthal angle and $\eta$ is the pseudo-rapidity. 
%
%To optimize the signal and get rid of the background, we identify the following cuts:
To keep most of the signal events while suppressing the background, 
we apply the following ``basic'' and ``discovery'' cuts on the $ttbbJJ$ events:
\begin{enumerate}
\item {\bf Basic}
\begin{enumerate}
\item
$|y(J)| \leq 2.5$
\item
$\Delta R(JJ) \geq 0.4$
\item
$p_T(J) \geq 25$ GeV 
\end{enumerate}
\item {\bf Discovery} 
\begin{enumerate}
\item
$|y(J)| \leq 2.5$
\item
$\Delta R(JJ) \geq 0.4$
\item
For $p_T$ ordered jets: \\$p^{1st}_T(J),p^{2nd}_T(J)  \geq 175$ GeV and $p^{3rd}_T(J),p^{4th}_T(J)  \geq 25$ GeV
\item
$|M(J_i,J_j)-m_h| \leq 10$ GeV  and  $~|M(J_k,J_l)-m_h| \leq 10$ GeV
where $i\neq j\neq k\neq l$.
\end{enumerate}
\end{enumerate}
where $\Delta R(ij) = \sqrt{\Delta\phi^2_{ij} + \Delta\eta^2_{ij}}$ is the 
angular separation between any two jets, $\phi$ is the azimuthal angle and $\eta$ is the pseudo-rapidity.
The ``discovery cut'' is motivated by the fact that 
for the signal, there is at least one high-$p_T$ Higgs coming from the heavy $t_2$ decay, 
and we expect the $b$-quarks coming from the Higgs decay to have a large $p_T$. 
We multiply both signal and background cross sections with a factor
\begin{eqnarray}
\label{ft2b2j2}
\eta_{t_2} = (\epsilon^b_{tag})^4\times(\epsilon^W_{rec})^2\times {(\epsilon^t_{rec})}^2\times(BR_{W\rightarrow jj})^2\approx 0.0299 \ . 
\end{eqnarray}
where we take $\epsilon^b_{tag}=0.5$, $\epsilon^t_{rec}=1$,
$\epsilon^W_{rec}=1$ and $W\rightarrow jj$ branching ratio $BR_{W\rightarrow jj}=0.69$.

In the warped models detailed in Sec.~\ref{ThFrmWrk.SEC}, the $t_2bW$ coupling 
({\it i.e.} $\kappa_{t_2bW}$) becomes very small for heavy $t_2$ as explained in Sec.~\ref{parCoup.SEC}. 
As a result, the production cross sections for the $pp \to t_2W,t_2b,t_2bW$ 
channels are small compared to the rest of the single production channels. 
Among the other channels, the $pp\to t_2t$ channel is weak interaction mediated~\footnote{However, this could also arize from the 
decay of the KK Gluon; see Ref.~\cite{Barcelo:2011wu}.} 
(the $t_2t$ pair actually comes from an off-shell $Z$ or $h$) and so is less significant
than the $pp\to t_2tZ$ or $pp\to t_2th$ channels, and we do not consider the former due to the small $BR_{Z\to \ell\ell}$. 
Thus in the warped models, the $pp\to t_2th$ channel that we have focused on is a promising channel. 
As already mentioned, the $t_2$ in the warped model without $Zb\bar b$ protection (DT model) is very heavy making its
discovery very challenging. We therefore do not consider further the $t'$ in the DT model.  
The $\kappa$ in the warped models with $Zb\bar b$ protection (ST and TT models) are given in Sec.~\ref{ThFrmWrk.SEC}. 
We present our results for the ST model at the 14 TeV (8 TeV) LHC in Table~\ref{tthjj} (Table \ref{tthjj8}) 
after the cuts shown above. 
\begin{table}[!h]
\centering
\begin{tabular}{|c|c|c|c|c|c|c|c|} \hline
$\mathcal T$&\multicolumn{1}{|c|}{$M_{t_2}$} & \multicolumn{1}{|c|}{$\sigma_{tot}$} & \multicolumn{1}{|c|}{$\sigma_{SR}$} & \multicolumn{1}{|c|}{cuts} & \multicolumn{1}{|c|}{S} & \multicolumn{1}{c|}{BG} & \multicolumn{1}{|c|}{$\mathcal{L}$} \\ 
&\multicolumn{1}{|c|}{(GeV)} & \multicolumn{1}{|c|}{$(fb)$} & \multicolumn{1}{|c|}{$(fb)$} & \multicolumn{1}{|c|}{} & \multicolumn{1}{|c|}{$(fb)$} & \multicolumn{1}{c|}{$(fb)$} & \multicolumn{1}{|c|}{$(fb^{-1})$} \\ \cline{1-8} 

$\mathcal T_1$&500  & 1247  & 223.0 & Basic  & 237.4  & 102.7  & -       \\ \cline{5-8}
	 &&       &       & Disc.  & 52.38  & 0.389  & 6.379   \\ \hline
$\mathcal T_2$&750  & 122.3 & 18.30 & Basic  & 22.67  & 102.7  & -       \\ \cline{5-8}
	 &&       &       & Disc.  & 13.25  & 0.389  & 25.22   \\ \hline
$\mathcal T_3$&1000 & 20.33 & 2.715 & Basic  & 3.088  & 102.7  & -       \\ \cline{5-8}
	 &&       &       & Disc.  & 2.421  & 0.389  & 138.0   \\ \hline
$\mathcal T_4$&1250 & 4.444 & 0.590 & Basic  & 0.477  & 102.7  & -       \\ \cline{5-8}
	 &&       &       & Disc.  & 0.415  & 0.389  & 1889.2   \\ \hline
\end{tabular}
\caption{\label{tthjj} Signal (S) and background (BG) cross sections (in $fb$) for 
$pp\rightarrow t_2th\rightarrow ttbbbb$ channel at the 14 TeV LHC for the ST model. For the BG we have considered the SM $pp \to ttbbJJ$ process where the dominant contribution comes from $pp\to tthh$. The $\mathcal T_i$'s correspond to the parameter sets detailed in Table \ref{t2STpara}. The luminosity
requirement $\mathcal{L}$ is computed using $\sigma_{tot}$ after
including the factor $\eta_{t_2}$ defined in Eq.~(\ref{ft2b2j2}). 
These numbers are obtained using $BR_{h\rightarrow bb}=0.8$. 
The $\sigma_{tot} = \sigma_{DR} + \sigma_{SR}$ is computed at the $t_2th$ level with no cut applied, 
whereas $\sigma_{SR}$ is computed at the $t_2 th$ level with only the $tW$ invariant mass cut of Eq.~(\ref{invmasst2})
applied.}
\end{table}
\begin{table}[!h]
\centering
\begin{tabular}{|c|c|c|c|c|c|c|c|} \hline
$\mathcal T$&\multicolumn{1}{|c|}{$M_{t_2}$} & \multicolumn{1}{|c|}{$\sigma_{tot}$} & \multicolumn{1}{|c|}{$\sigma_{SR}$} & \multicolumn{1}{|c|}{cuts} & \multicolumn{1}{|c|}{S} & \multicolumn{1}{c|}{BG} & \multicolumn{1}{|c|}{$\mathcal{L}$} \\ 
&\multicolumn{1}{|c|}{(GeV)} & \multicolumn{1}{|c|}{$(fb)$} & \multicolumn{1}{|c|}{$(fb)$} & \multicolumn{1}{|c|}{} & \multicolumn{1}{|c|}{$(fb)$} & \multicolumn{1}{c|}{$(fb)$} & \multicolumn{1}{|c|}{$(fb^{-1})$} \\ \cline{1-8} 

$\mathcal T_1$&500  & 181.3 & 32.48 & Basic  & 35.83  & 16.43  & -       \\ \cline{5-8}
	 &&       &       & Disc.  & 6.702  & 0.035  & 49.85   \\ \hline
$\mathcal T_2$&750  & 11.96 & 1.690 & Basic  & 2.353  & 16.43  & -       \\ \cline{5-8}
	 &&       &       & Disc.  & 1.325  & 0.035  & 252.3   \\ \hline
$\mathcal T_3$&1000 & 1.222 & 0.168 & Basic  & 0.206  & 16.43  & -       \\ \cline{5-8}
	 &&       &       & Disc.  & 0.162  & 0.035  & 2056.8  \\ \hline 
\end{tabular}
\caption{\label{tthjj8}
Same as in Table~\ref{tthjj} for the 8~TeV LHC.}
\end{table}
Defining as before, $\mathcal{L}_{5\sigma}$ as the Luminosity for $S/\sqrt{B} = 5$ and $\mathcal{L}_{10}$ that for 10 events, 
we find that $\mathcal{L}_{5\sigma} < \mathcal{L}_{10}$ in most of parameter-space, except for $M_{t_2}=1250$ GeV for 14 TeV LHC, 
and we present the maximum of $\mathcal{L}_{5\sigma}$ and $\mathcal{L}_{10}$ in Table~\ref{tthjj}.
From $\sigma_{tot} = \sigma_{DR} + \sigma_{SR}$, we find that the 14 TeV LHC can probe $M_{t_2}$ of the order of 1 TeV with 
100 $fb^{-1}$ of integrated luminosity in the ST model.  

\begin{figure}[!h]
\begin{center}
\includegraphics[width=0.5\textwidth]{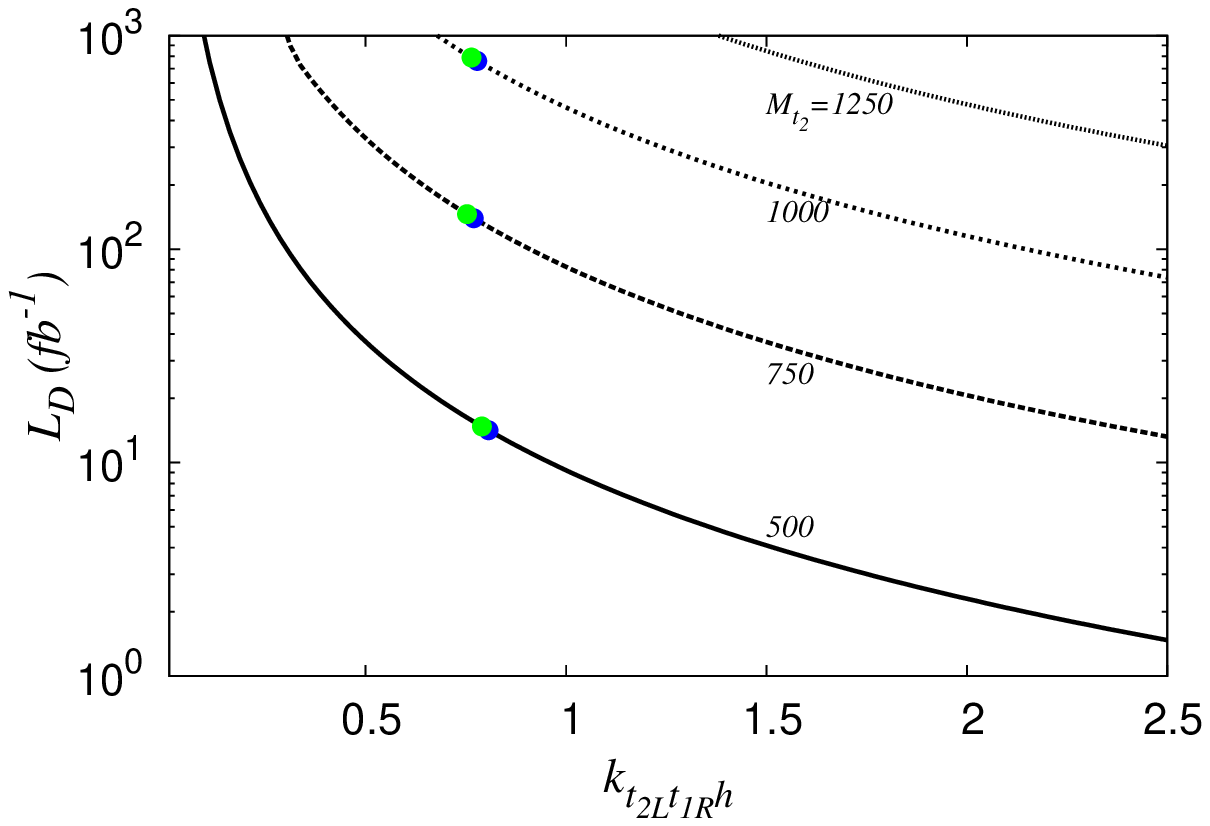}
\caption{Luminosity requirements ($\mathcal{L}_D$, in $fb^{-1}$) for observing the $pp\to t_2 t h$ SR process
as functions of $\kappa_{t_{2L}t_{1R}h}$ for different $M_{t_{2}}$ (in GeV) at the 14 TeV LHC. 
The luminosity is computed after including all BRs and $b$-tagging efficiency. 
The blue and green dots correspond to the ST and TT models respectively.
\label{lumt2}}
\end{center}
\end{figure}

As mentioned earlier, the SR process can give important information on the electroweak couplings $\kappa$ 
(while the DR depends dominantly on $g_S$). To explore this aspect, 
we compute the $pp\to t_2th$ SR production cross-sections from the $pp\to t_2th$ signal events by applying the kinematical cut
\begin{equation}
\label{invmasst2}
|M(th)-M_{t_2}| \geq \alpha_{cut}M_{t_2}; \; \alpha_{cut} = 0.05 \ .
\end{equation}
The background for the $t_2th$ SR production is computed at the $thth$ level after demanding that any one of the $th$
pairs satisfies the invariant mass cut defined in Eq.~(\ref{invmasst2}). This cut can be expressed as
\begin{equation}
|M(t_1h_i) - M_{t_2}| \geq \alpha_{cut} M_{t_2}~~~\textrm{AND}~~~
|M(t_2h_j) - M_{t_2}| \leq \alpha_{cut} M_{t_2}
\end{equation} 
where $t$'s and $h$'s are $p_T$-ordered and $i,j=\{1,2\}$ with $i\neq j$.
Just as in the case of $\chi_1$ production, for the parameter ranges we are interested in, 
$pp\to t_2th$ process is dominated by the DR production. We have  also verified that 
with our choice of $\alpha_{cut}$ the $\sigma_{SR}$ scales as $\kappa_{t_2th}^2$.
Since the SR production can give us information about the off-diagonal $t_2th$ coupling, in Fig.~\ref{lumt2}
we present model-independently the luminosity required for $pp\to t_2th$ SR production channel assuming $BR_{t_2 \to th}$ to be 100\%. 
In doing this we vary $\kappa_{t_{2L}t_{1R}h}$ keeping the other coupling $\kappa_{t_{1L}t_{2R}h}$ to zero 
(as is the case for instance in the warped model). 
We find that $pp \to t_2th$ events are signal rate limited ({\it i.e.,} $\mathcal{L}_{10} > \mathcal{L}_5$) in the parameter range we have considered.
In Fig.~\ref{lumt2} we show the luminosity required for the warped ST model as blue dots and the TT model as green dots. 

In the ST or TT models, for heavy $t_2$,  the branching ratios for $t_2\to th$ and $t_2\to tZ$ are comparable, {\it i.e.}, 
\begin{equation}
 BR_{t_2\to th} \approx BR_{t_2\to tZ} .%\gg BR_{t_2\to bW}.
\end{equation} 
Hence one could as well study the following processes:
\begin{eqnarray}
&&pp \to t_2\bar{t}h\rightarrow (tZ)th\rightarrow bWZbWh,\\
&&pp \to t_2\bar{t}Z\rightarrow (th)tZ\rightarrow bWhbWZ,\\
&&pp \to t_2\bar{t}Z\rightarrow (tZ)tZ\rightarrow bWZbWZ.
\end{eqnarray}
Of these the first two can even lead to $4b+6j$ final states which is exactly what we have used for our analysis by demanding only 4~$b$-tagged jets. We don't expect the LHC reach to be very different for these two channels from what we have estimated. This is because, the main difference between these two channels and what we have considered comes from the facts that the Higgs boson is a bit heavier than the $Z$ and $BR_{h\to bb} > BR_{Z\to JJ}$. However for the last
process, {\it i.e.} $pp \to t_2\bar{t}Z\to (tZ)tZ$, we cannot demand 4~$b$-tagged jets anymore and as a result we must consider one of the $Z$ decaying leptonically to act as the trigger. Since $BR_{Z\to \ell\ell} < BR_{Z\to JJ}$, in this case the signal rate will be quite small.

%%%%%%%%%%%%%%%%%%%%%%%%%%%%%%%%%%%%%%%%%%%%%%%%%%%%%%%%%%%%%%%%%%%%%%%%%%%%%%%%%%%
\subsection{$b'$ LHC Signatures}

\begin{figure}[!h]
\centering
\subfloat{
\begin{tabular}{ccc}
\resizebox{40mm}{!}{\includegraphics{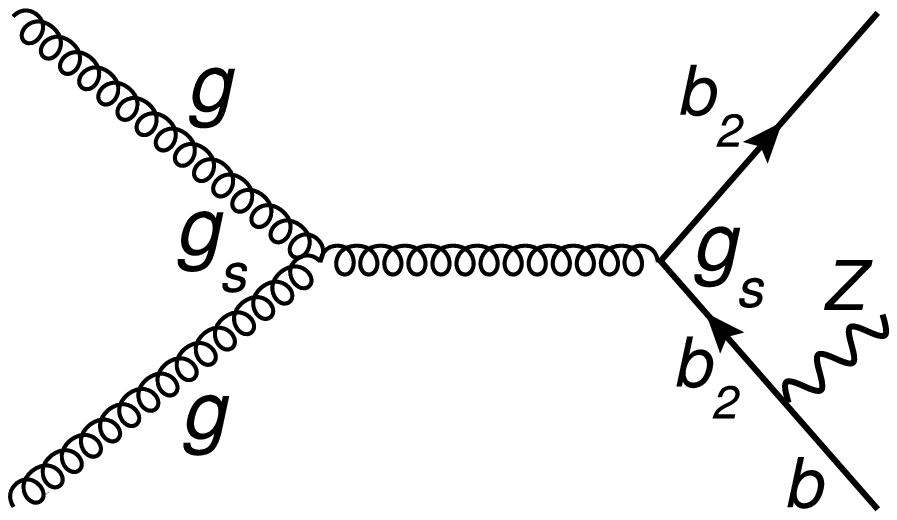}} &&
\resizebox{40mm}{!}{\includegraphics{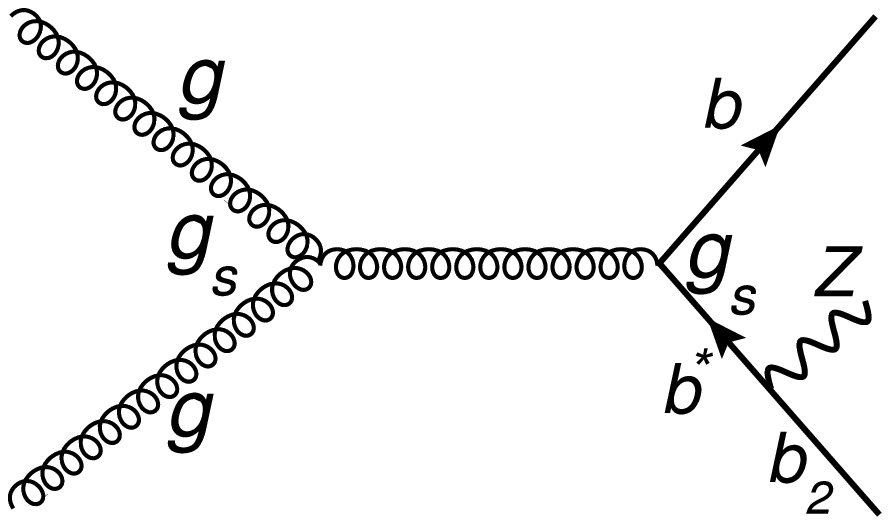}} \\
\footnotesize{(a)}&&\footnotesize{(b)}
\end{tabular}}\\
\caption{ In (a) when both the $b_2$ are on-shell, we have a double resonant (DR) contribution, while when
one of them is off-shell we have the single resonant (SR) process; the other contribution to SR coming from 
the strict single production diagram is shown in (b).
\label{b2_single}}
\end{figure}

\begin{figure}[!h]
\begin{center}
\includegraphics[width=0.5\textwidth]{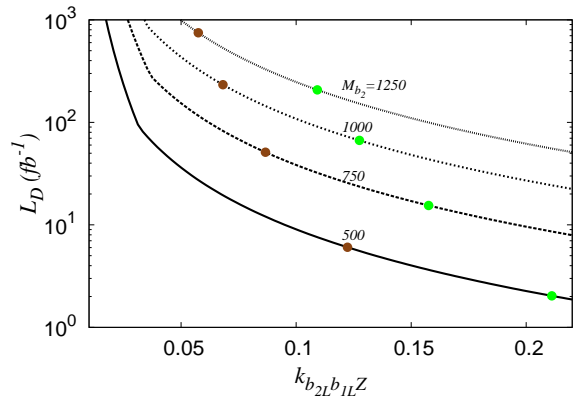}
\caption{Luminosity requirements ($\mathcal{L}_D$, in $fb^{-1}$) for observing the $pp\to b_2bZ$ SR process as functions of 
$\kappa_{b_{2L}b_{1L}Z}$ for different $M_{b_2}$ (in GeV) at the 14 TeV LHC. $\mathcal{L}_D$ is computed after including all BRs and $b$-tagging efficiency as
shown in Eq.~(\ref{factorb2}). The brown and green dots correspond to the DT and TT models respectively. 
\label{lumb2}}
\end{center}
\end{figure}

The important single production channels of a vectorlike $b'$ were explored in Ref.~\cite{Brooijmans:2010tn},
which included $tb'$, $bb'$, $b'h$, $b'Z$, $qtb'$, $qbb'$, $bb'Z$, $bb'h$, $qb'Z$, $qb'h$, $tb'W$ and $qb'W$ processes.  
As mentioned earlier, Ref.~\cite{Bini:2011zb} studies the $bb'$ production via KK-gluon. 
The $qtb'$ process has been studied in Ref.~\cite{Contino:2008hi} in the context of composite Higgs models. 
A detailed study of the collider signatures and discovery reach for $b'$ pair production and $b'Z$ and $b'h$ single production channels
is already presented in a model independent manner in Ref.~\cite{Gopalakrishna:2011ef}. 
Here we consider another $b'$ single production process, 
thus adding to the study of Ref.~\cite{Gopalakrishna:2011ef}. 
The process we consider is shown in Fig.~\ref{b2_single}, namely
\begin{eqnarray}
p p \to b_2 b Z \to bZbZ \ ,
\end{eqnarray}
and select the $bb\ell\ell JJ$ channel.  
\vspace{0.5cm}
To obtain the luminosity requirements, we multiply the cross-section obtained at the $bZbZ$ level by the factor
\begin{eqnarray}
\label{factorb2}
\eta_{b_2} = 2\times(\epsilon^b_{tag})^2\times\epsilon^{(\ell\ell\to Z)}_{rec}\times \epsilon^{(JJ\to Z)}_{rec}\times (BR_{Z\to JJ})\times (BR_{Z\to \ell\ell})\approx 0.023 \ ,
\end{eqnarray}
to take into account the various BR and efficiencies. 
Here $\epsilon^{(\ell\ell\to Z)}_{rec}$ and $\epsilon^{(JJ\to Z)}_{rec}$ stand
for reconstruction efficiency of $Z$ from $\ell\ell$ and $JJ$ respectively.
We take $\epsilon^b_{tag}=0.5$, $\epsilon^{(\ell\ell\to Z)}_{rec}=1$ and
$\epsilon^{(JJ\to Z)}_{rec}=1$ and the branching ratios $BR_{Z\to JJ}=0.69$ and $BR_{Z\to \ell\ell}=0.068$. The extra 2 factor appears because
either of the $Z$ can decay to the $\ell\ell$ pair.
We parametrize the $b_2bZ$ interaction terms model-independently as shown in Eq.~(\ref{b'kapMI.EQ}).
Analogous to the previous subsections, we have both double resonant (DR) and single resonant (SR) contributions to the
$b_2 b Z$ final state.
Isolating the SR contribution can give us information about the off-diagonal $b_2bZ$ couplings.
To this end, we compute the $pp\to b_2bZ$ SR production cross-section from the $pp\to b_2bZ$ 
cross-section by applying the kinematical cut
\begin{equation}
\label{invmassb2}
|M(bZ)-M_{b_2}| \geq \alpha_{cut}M_{b_2}; \quad \alpha_{cut} = 0.05 \ .
\end{equation}
We have also verified that with our choice of $\alpha_{cut}$ the $\sigma_{SR}$ scales as $\kappa_{b_2bZ}^2$.
The main SM backgrounds for the $b_2bZ$ SR production come from $pp\to bbZ+\textrm{jets}$, $bbZV$ 
(where $V=\{W,Z\}$) processes. Applying invariant mass-cut around $Z$-mass one can
significantly reduce $bbZ+\textrm{jets}$ and $bbZW$ contributions. Therefore, we compute the background for the $b_2bZ$ SR production 
at the $bZbZ$ level. 
We demand that any one of the $bZ$ pairs satisfies the invariant mass cut of Eq.~(\ref{invmassb2}) {\it i.e.}
\begin{equation}
|M(b_1Z_i) - M_{b_2}| \geq \alpha_{cut} M_{b_2}~~~\textrm{AND}~~~
|M(b_2Z_j) - M_{b_2}| \leq \alpha_{cut} M_{b_2}
\end{equation} 
where $b$'s and $Z$'s are $p_T$-ordered and $i,j=\{1,2\}$ with $i\neq j$.
In Fig.~\ref{lumb2} we present the luminosity requirement for $pp\to b_2bZ$ SR production channel in a model-independent manner assuming $BR_{b_2 \to bZ}$ to be 100\%. 
The kinks in the graphs appear because of the transition from $\mathcal{L}_5$ to $\mathcal{L}_{10}$ along the 
increasing values of the coupling. 
In doing this we  vary $\kappa_{b_{2L}b_{1L}Z}$ keeping the other coupling $\kappa_{b_{2R}b_{1R}Z}$ zero. 
(This is the case in the warped models we have considered.)
\begin{table}[!h]
\begin{center}
\begin{tabular}{|c|c|c|c|}
\hline $M_{b_2}$ (GeV)  & $\sigma(pp\to b_2 Z)$ (fb) & $\sigma(pp\to b_2 b)$ (fb) & $\sigma(pp\to b_2 bZ)$ (fb)\\ 
\hline 500  & 81.50 & 15.86 & 47.12 \\ 
\hline 750  & 16.67 & 3.910 & 11.10 \\ 
\hline 1000 & 4.630 & 1.256 & 3.933 \\ 
\hline 1250 & 1.534 & 0.472 & 1.722 \\ 
\hline 1500 & 0.565 & 0.193 & 0.804 \\ 
\hline 
\end{tabular} 
\caption{SR production cross-sections of $b_2$ for different $M_{b_2}$ with $\kappa_{b_2bZ}=0.1$.
The $b_2 b Z$ cross-section is after applying the invariant mass cut of Eq.~(\ref{invmassb2}), 
while the others are without any cuts.
\label{b2-SR-cs.TAB}}
\end{center}
\end{table}
In Table~\ref{b2-SR-cs.TAB} we compare the various SR channel cross-sections model-independently. 
The $b_2 b Z$ cross-section is after applying the invariant mass cut of Eq.~(\ref{invmassb2}), 
while the others are without any cuts.
We see that the $b_2 Z$ channel studied in Ref.~\cite{Gopalakrishna:2011ef} and the $b_2 b Z$ SR process studied here
are comparable in signal cross-section; however the latter case requires larger luminosity since the background is larger.

In the warped models, a $b'$ is present in the DT and TT models, and
the $\kappa$ of Eq.~(\ref{b'kapMI.EQ}) are given in Eqs.~(\ref{kapNoZbb.EQ})~and in Sec.~\ref{tR3Defn.SEC} respectively.  
The $\kappa$ for the DT and TT models are shown in Fig.~\ref{b2kappa}, and for the TT model in Table~\ref{b'TTpara}. 
We can infer the luminosity required for the DR process from Ref.~\cite{Gopalakrishna:2011ef}. 
For the DT model in the $pp\to b' \bar{b'}\to bZ \bar{b}Z \to b\ell\ell b jj$ channel, the 14~TeV LHC reach is about $1250$~GeV with about 500~${\rm fb}^{-1}$. 
For the TT model, the $BR(b' \to b Z)$ is about a factor of two bigger compared to the DT model; hence the luminosity being signal-rate limited, is about half. 
Turning next to the SR process, the brown and green dots in Fig.~\ref{lumb2} are for the DT and TT warped models respectively. 
The corresponding signal cross-sections are shown in Table~\ref{b2bZDTTT.TAB}.
In the TT model, for simplicity, we have focused only on the $b_2$ signatures, although the $b_3$ is quite close in mass;
a more complete analysis can include the $b_3$ contributions also. 
Analogously, one can also look at the $bhbh$ channel which we have not explored in this work.  
In the DT model, for the choice of benchmark parameters discussed in Sec.~\ref{parCoup.SEC}, we have a reach of 
$M_{b_2} = 1000~$GeV with about 250~$fb^{-1}$, and in the TT model it is about $M_{b_2} = 1250~$GeV with about 250~$fb^{-1}$.
\begin{table}
\begin{center}
\begin{tabular}{|c|c|c|}
\hline  \multicolumn{3}{|c|}{DT model}\\
\cline{1-3} $M_{b_2}$ (GeV)& $\kappa_{b_{1L}b_{2L}Z}$ & $\sigma_{b_2 b Z}$ (fb) \\ 
\hline  500  & 0.122 & 70.49 \\ 
\hline  750  & 0.087 & 8.341 \\ 
\hline  1000 & 0.068 & 1.829 \\ 
\hline  1250 & 0.057 & 0.569 \\ 
\hline 
\end{tabular} 
\begin{tabular}{|c|c|c|}
\hline  \multicolumn{3}{|c|}{TT model}\\
\cline{1-3} $\mathcal B$ & $M_{b_2}$ (GeV) & $\sigma_{b_2 b Z}$ (fb) \\ 
\hline  $\mathcal{B}_1$ &  500 & 210.05 \\ 
\hline  $\mathcal{B}_2$ &  750 & 27.56 \\ 
\hline  $\mathcal{B}_3$ & 1000 & 6.394 \\ 
\hline  $\mathcal{B}_4$ & 1250 & 2.054 \\ 
\hline 
\end{tabular}
\caption{Cross-sections for the process $pp\to b_2 b Z$ in the DT and TT models for different choices of $M_{b_2}$. 
The  cross-sections are obtained after applying the invariant mass cut of Eq.~(\ref{invmassb2}).
The couplings for the TT model corresponding to the parameter sets labelled by ${\cal B}_i$ are shown in Table~\ref{b'TTpara}.
\label{b2bZDTTT.TAB}}
\end{center} 
\end{table}

%%%%%%%%%%%%%%%%%%%%%%%%%%%%%%%%%%%%%%%%%%%%%%%%%%%%%%%%%%%
\section{Conclusions}
\label{CONCL.SEC}

We present the phenomenology and LHC Signatures of colored vectorlike fermions 
$\chi$ (EM charge 5/3), $t'$ (EM charge 2/3) and $b'$ (EM charge -1/3). 
Such fermions appear in many BSM extensions. 
We take warped extra-dimensional models as the motivating framework for our analysis. 
However, our analysis applies to other models
that have such fermions, and we present our results model-independently wherever possible. 
Our focus is the phenomenology due to the mixing of SM fermions with the new vectorlike fermions
induced by EWSB. 

We identify the allowed decay modes of the vectorlike quarks, compute their partial widths
and branching ratios. This guides us in identifying promising channels for discovery of these 
vectorlike quarks at the LHC.  
While pair production via the gluon coupling usually has the largest cross-section at the LHC
for the range of parameters we consider,
a particular focus is single production channels of these vectorlike quarks, which although challenging,
can probe the EW structure of the BSM model. 

We consider three different cases of warped models as motivating examples, 
differing in the fermion representations under $SU(2)_L \otimes SU(2)_R \otimes U(1)_X$ gauge group.  
We label them by the representation $t_R$ appears in, namely, Doublet Top (DT), Singlet Top (ST) and 
Triplet Top (TT) models. 
The first, the DT model, does not have the $Z b\bar b$ coupling protected and has stronger 
constraints on it,
while the ST and TT models have custodial protection of the $Zb\bar b$ coupling and have less 
severe constraints on them. 
More than one $\chi$, $t'$ or $b'$ can be present depending on the model,
and they can mix among themselves and the SM quarks as a result of off-diagonal EWSB induced mass mixing terms.  
We identify the mass eigenstates by diagonalizing the mass matrix, and work out the couplings that are
relevant to the LHC phenomenology we discuss. 

At the LHC we have computed the signal cross-sections and the dominant SM background to $\chi$, $t'$ and $b'$ productions,
and find the 8~TeV and 14~TeV LHC discovery reach. 
For the $\chi$, we identify $pp \rightarrow \chi tW \rightarrow tWtW$ 
in the $2b~ 6j ~ \ell E\!\!\!/_{T}$ channel as a promising one. 
The $pp \to \chi tW$ process has contributions from: 
(a) double resonant (DR) process $pp \to \chi \bar\chi$ followed by $\chi\to t W$ decay where both $\chi$ are on-shell, 
and, 
(b) the single resonant (SR) process $pp \to \chi tW$ where only one $\chi$ is on-shell.
The DR process 
dominantly depends only on the strong coupling $g_s$, 
while the SR process is directly sensitive to the EW couplings and mixing effects,
and its measurement would give valuable information on the EW structure of the underlying BSM theory.   
We show that by applying an invariant mass cut to remove one on-shell $\chi$, we can get sensitivity to the
EW couplings.
Including both SR and DR, we find that at the 14~TeV LHC the reach is about $m_\chi=1750$~GeV with about $350~{\rm fb}^{-1}$.  
In the same vein, for the $t'$, we study the process $pp \to t_2th \to thth$ in the $2j~ 6b~ \ell E\!\!\!/_{T}$ 
channel as a promising one, and find that the 14~TeV LHC can probe of the order of 1~TeV mass with about 150~${\rm fb}^{-1}$. 
For the $b'$ we discuss the process $p p \to b_2 b Z \to bZbZ$ in the $2j~ 2b~ \ell^+ \ell^-$ channel, 
and infer that the 14~TeV LHC reach is about $1250$~GeV with about 250~${\rm fb}^{-1}$ for the TT model. 

%%%%%%%%%%%%%%%%%%%%%%%%%
\medskip
\noindent {\it Acknowledgements:} 
We thank K.~Agashe and A.~Pomarol for valuable discussions.

%%%%%%%%%%%%%%%%%%%%%%%%%%%%%%%%%%%%%%%%%%%%%%%%%%%%%%%%%%%%%
\appendix
%%%%%%%%%%%%%%%%%%%%%%%%%%%%%%%%%%%%%%%%%%%%%
% Put the following in place of the \appendix command
%%%%%%%%%%%%%%%%%%%%%%%%%%%%%%%%%%%%%%%%%%%%%
\setcounter{section}{0}
\setcounter{figure}{0}
\setcounter{table}{0}

\renewcommand\thesection{Appendix \Alph{section}}               % use this to modify the section numbering style
\renewcommand\thesubsection{\Alph{section}.\arabic{subsection}}
\renewcommand\thesubsubsection{\Alph{section}.\arabic{subsection}.\arabic{subsubsection}}

\renewcommand{\theequation}{\Alph{section}.\arabic{equation}}    % use this to modify the equation numbers
\renewcommand{\thetable}{\Alph{section}.\arabic{table}}          % use this to modify the table numbers
\renewcommand{\thefigure}{\Alph{section}.\arabic{figure}}        % use this to modify the figure numbers

%%%%%%%%%%%%%%%%%%%%%%%%%%%%%%%%%%%%%%%%%%%%
\section{Fermion Profiles}
\label{fermProf.APP}
\setcounter{equation}{0}
\setcounter{figure}{0}
\setcounter{table}{0}

\begin{figure}[!h]
\begin{center}
\includegraphics[width=0.5\textwidth]{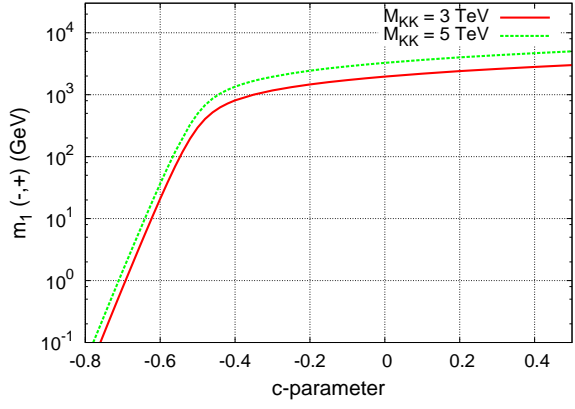}
\caption{\label{m1KKmp}Masses of the first KK fermion with $(-,+)$ BC as functions of $c$ for different values of the KK gauge boson masses.}
\end{center}
\end{figure}

The fermion KK mode profiles read as \cite{Gherghetta:2000qt}
\begin{eqnarray}
f^{(0)}(y) &=& \sqrt{\frac{(1-2c)k\pi R}{e^{(1-2c)k\pi R}-1}}e^{-cky}\\
f^{(n)}(y) &=& \frac{e^{ky/2}}{N_n}\left[J_{\alpha}\left(\frac{m_n}{k}e^{ky}\right)
+ b_{\alpha}(m_n)Y_{\alpha}\left(\frac{m_n}{k}e^{ky}\right)\right]~~(n=1,2,...)
\end{eqnarray}
where $\alpha=|c+1/2|$. $J_{\alpha}$ and $Y_{\alpha}$ are the Bessel functions of
order $\alpha$ of the first and the second kind respectively. 
These profiles satisfy the following orthonormality condition,
\begin{eqnarray}
\label{ortho}
\frac{1}{\pi R}\int^{\pi R}_{0} dy e^{ky}f^{(m)}(y)f^{(n)}(y) = \delta_{mn},
\end{eqnarray}
from which one can determine the normalization, $N_n$. 
$b_\alpha(m_n)$ and $m_n$ are determined through the BC on the branes.
For fermions obeying $(-,+)$ BC, which means
\begin{eqnarray}
f^{(n)}(y)|_{y=0}=0~~\textrm{and}~~(\partial_y + ck)f^{(n)}(y)|_{y=\pi R}=0
\end{eqnarray}
From these two equations one obtains the following condition
\begin{eqnarray}
b_{\alpha}(m_n) = - \frac{J_{\alpha}\left(\frac{m_n}{k}\right)}{Y_{\alpha}\left(\frac{m_n}{k}\right)} = -\frac{\left(c + \frac12\right)J_{\alpha}\left(\frac{m_n}{k}e^{\pi kR}\right) + \left(\frac{m_n}{k}e^{\pi kR}\right) J^{\prime}_{\alpha}\left(\frac{m_n}{k}e^{\pi kR}\right)}{\left(c + \frac12\right)Y_{\alpha}\left(\frac{m_n}{k}e^{\pi kR}\right) + \left(\frac{m_n}{k}e^{\pi kR}\right) Y^{\prime}_{\alpha}\left(\frac{m_n}{k}e^{\pi kR}\right)}
\end{eqnarray}
This condition can be solved numerically for $m_n$ and $b_{\alpha}(m_n)$. 
The first fermion KK mass $m_1$ for $(-,+)$ BC as a function of $c$ is
shown in Fig. \ref{m1KKmp}.

%%%%%%%%%%%%%%%%%%%%%%%%%%%%%%%%%%%%%%%%%%%
\section{$t_R$ Triplet Case Diagonalization}
\label{tR3AnaDiag.APP}
\setcounter{equation}{0}
\setcounter{figure}{0}
\setcounter{table}{0}

Here we present analytical results for the mass matrix diagonalization
and the resulting couplings in the mass basis 
in the limit of $m_{ij}/M_{\psi^\prime} \ll 1$ 
for the $t_R$ Triplet case (TT model) detailed in Sec.~\ref{tR3Defn.SEC}.  
 
The $m_{ij}$ in the charge $-1/3$ mass matrix in Eq.~(\ref{tR3Mmat.EQ}) are the same,
and defining $r_b \equiv m/M$, we find
\beq
R_L^T = \frac{1}{\sqrt{1+2 r_b^2}} \bmat -1 & r_b & r_b \\ 0 & -\frac{\sqrt{1+2r_b^2}}{\sqrt{2}} & \frac{\sqrt{1+2r_b^2}}{\sqrt{2}} \\ \sqrt{2}r_b & \frac{1}{\sqrt{2}} & \frac{1}{\sqrt{2}} \emat \ ; \quad 
R_R = \bmat 1 & 0 & 0 \\ 0 & -\frac{1}{\sqrt{2}} & \frac{1}{\sqrt{2}} \\ 
0 & \frac{1}{\sqrt{2}} & \frac{1}{\sqrt{2}} \emat 
\eeq
with the mass eigenvalues $0, M, M\sqrt{1+2r_b^2}$. The $b_1$ is identified as the SM b-quark, 
and the zero eigenvalue will be lifted when $\lambda_b$ terms are included.

The $Z$ boson neutral current interactions 
are 
(although not shown, the vector index on the gauge fields and the $\gamma^\mu$ between the fermion fields are implied)
\bea
{\cal L}_{NC}^Z &\supset& g_Z \left\{ 
 \bar{b_1}_L \left[-\frac{1}{2} - s_W^2 Q_b \right] {b_1}_L  + 
 \bar{b_2}_L \left[-\frac{1}{2} - s_W^2 Q_b \right] {b_2}_L  + 
 \bar{b_3}_L \left[-\frac{1}{2} - s_W^2 Q_b \right] {b_3}_L + 
\right. \nonumber \\ &&\left.
\left[ 
 \bar{b_1}_L \left(\frac{\sqrt{2} r_b}{\sqrt{1+2r_b^2}} \right) \left(-\frac{1}{2} \right) {b_2}_L + 
 \bar{b_2}_L \left(\frac{1}{\sqrt{1+2r_b^2}} \right) \left(-\frac{1}{2} \right)  {b_3}_L + h.c. \right] 
\right. \nonumber \\ &&\left.
 \bar{b_1}_R \left(- s_W^2 Q_b \right) {b_1}_R  + 
 \bar{b_2}_R \left(-\frac{1}{2} - s_W^2 Q_b \right) {b_2}_R  + 
 \bar{b_3}_R \left(-\frac{1}{2} - s_W^2 Q_b \right) {b_3}_R  +
\right. \nonumber \\ &&\left.
\left[
 \bar{b_2}_R \left(-\frac{1}{2} \right) {b_3}_R  +  h.c. \right]
\right\} Z
\eea
where $g_Z \equiv \sqrt{g_L^2 + {g^\prime}^2}$, $Q_b=-1/3$. 
Note that the $b_1 b_1 Z$ interactions come out standard due to the custodial protection.  
The photon couplings are not shown and as usual has vectorlike couplings to the fermions
given by their electromagnetic charge. 
We have taken all ${\cal I}_{\psi\psi V} = 1$ as earlier, ignoring corrections to this due to EWSB $(0)-(1)$ gauge boson mixing which are
at most a few percent. 
The Higgs interactions are got by $v\rightarrow v(1+h/v)$ 
and are
\beq
{\cal L}^h \supset \frac{1}{\sqrt{1+2r_b^2}} \left(\frac{2m^{bb'}}{v}\right) \left[ \bar{b_1}_L  {b_3}_R - \sqrt{2} r_b \bar{b_3}_L {b_3}_R \right] h + h.c. 
\eeq

%%%%%%%%%%%%%%%%%%%%%%%%%%%%%%%%%%%%%%%%%%%
\section{$\chi$ Signature in More Detail}
\label{chiSophist.APP}
\setcounter{equation}{0}
\setcounter{figure}{0}
\setcounter{table}{0}

%So far in this paper, we have used parton level estimations for both signal and backround cross-sections 
%with simple kinematical cuts. 
In this section, we perform a more detailed analysis of the $pp \to \chi_1 tW \to tWtW$ channel that we discussed
in Sec.~\ref{ChiLHCSign.SEC}. 
Our aim is to show that the discovery luminosity estimates that we obtained there stand up to a more detailed analysis. 
In Sec.~\ref{ChiLHCSign.SEC}, to estimate the LHC discovery reach of $\chi$, 
we compute the $pp\to ttWW\to ttW\ell\nu$ as the SM background for $pp\to \chi_1tW \to tWt\ell\nu$. 
For $M_\chi \gtrsim 750$~GeV, the top quarks will be quite boosted and so, instead of using conventional
top reconstruction algorithm with $b$-tagging, one could use modern top-tagging 
algorithms~\cite{Plehn:2010st,Plehn:2009rk,Plehn:2011tg} like HEPTopTagger~\cite{Plehn:2010st} 
which has much higher top-tagging efficiency. 
These advanced algorithms can achieve a reconstruction efficiency $\epsilon_t\sim 40-50$\%
(mistag rate is only a few percent and can even be reduced further)
in the top-$p_T$ ranging from 200 GeV to 600 GeV. %For higher $p_T$ the efficiency can even increase. 
With HEPTopTagger, $b$-tagging is not necessary and combinatorics issues are automatically
resolved by the algorithm.
We note that the hadronic $W$-tagging efficiency is also quite high. It is around 70-80\% 
for moderately boosted $W$~\cite{Thaler:2010tr,CMS:2013uea}. 

With these in mind, after reconstruction of the two high $p_T$ tops ($p_T\geq 200$ GeV), 
for the $pp\to\chi_1tW\to ttW\ell\nu$ signal process, a problematic background can be the SM $pp\to ttjj\ell\nu$. 
The main contribution for this background will come from the processes where the jets are from 
the decay of $Z$ or $W$, or two QCD jets. 
%To obtain Table~\ref{chipro}, we assumed $W$ reconstruction and hence, 
%considered only $pp\to ttWW\to ttW\ell\nu$ as the dominant background process. 
We demonstrate here that these extra backgrounds can be brought under control, 
for example by using the following set of cuts on the $ttjj\ell\nu$ final state,
\begin{itemize}
\item Cut-II:
\begin{enumerate}
\item
$|y(l)|,|y(j)| \leq 2.5$, \quad $p_T(l),p_T(j)\geq 25$ GeV
\item
$p_T(t) \geq 200$ GeV, 
\item
$|M(jj)-M_W|\leq 15$ GeV, 
\item
$\big(|M(t_1jj)-M_{\chi_1}|$  or $|M(t_2jj)-M_{\chi_1}| \big)\leq 0.2M_{\chi_1}$\\ where
$t_1$ and $t_2$ are the two $p_T$-ordered tops.
\end{enumerate}
\end{itemize}
In Table \ref{cut2} we display the signal and total background cross-sections with Cut-II for the $\chi$ benchmark points. 
Here the background includes all the processes where the jets are coming from a EW vector boson or are QCD jets.
\begin{table}[t]
\begin{center}
\begin{tabular}{|c|c|c|c|c|}
\hline 
\multicolumn{1}{|c|}{$M_{\chi}$} & \multicolumn{3}{|c|}{$\sigma$ (fb) after Cut-II} & \multicolumn{1}{|c|}{$\mathcal{L}_D$}\\ \cline{2-4}
(GeV) & Signal & $ttjj\ell\nu$ (EW BG)  & $ttjj\ell\nu$ (QCD BG)  & fb$^{-1}$\\ 
\hline 
500 & 136.33 & 0.18 & 0.41  & 0.654\\ 
\hline 
750 & 33.66 & 0.16 & 0.29   & 2.647\\ 
\hline 
1000 & 8.006 & 0.09 & 0.18 & 11.13\\ 
\hline 
1250 & 2.173 & 0.05 & 0.10 & 41.01\\ 
\hline
1500 & 0.660 & 0.03 & 0.05 & 135.0\\ 
\hline 
1750 & 0.217 & 0.02 & 0.03 & 410.6\\ 
\hline 
\end{tabular}
\caption{\label{cut2}We display the signal and background (EW and QCD) c.s. at the $ttjj\ell\nu$ level at the
14 TeV LHC after Cut-II as defined in the text. While computing 
$\mathcal{L}_D$ we multiply both signal and background by a factor 
$\eta = (\epsilon_t)^2\times (BR_{W\to jj})^2$. We use $BR_{W\to jj}=0.67$ and, take
$\epsilon_t=0.5$.}
\end{center}
\end{table}
From the Table~\ref{cut2} we can see that Cut-II is very effective to reduce 
background for higher $M_{\chi}$ values and thus, making the $\chi$ discovery channel 
signal rate limited for all benchmark $M_{\chi}$ values we have considered.  
The luminosity requirements obtained here differ from the ones shown in Table~\ref{chipro}
by about 10-15\% only.

%%%%%%%%%%%%%%%%%%%%%%%%%%%%%%%%%%%%%%%%%%%%%%%%%%%%

\end{document}